\documentclass[aps,prl,reprint,superscriptaddress]{revtex4-1}

\usepackage{graphicx}
\usepackage{amsmath}
\usepackage{amssymb}
\usepackage{esvect}
\usepackage{braket}
\usepackage{physics}
\usepackage{setspace}
\usepackage{cancel}

\begin{document}    

\title{Witnessing quantum correlations in a nuclear ensemble via an electron spin qubit}

\author{Dorian A.\,Gangloff\textsuperscript{1,*}}
\author{Leon Zaporski\textsuperscript{1,*}}
\author{Jonathan H.\,Bodey\textsuperscript{1,*}}
\author{Clara Bachorz\textsuperscript{1}}
\author{Daniel M.\,Jackson\textsuperscript{1}}
\author{Gabriel \'Ethier-Majcher\textsuperscript{1}}
\author{Constantin Lang\textsuperscript{1}}
\author{Edmund Clarke\textsuperscript{2}}
\author{Maxime Hugues\textsuperscript{3}}
\author{Claire Le Gall\textsuperscript{1,$\dagger$}}
\author{Mete Atat\"ure\textsuperscript{1,$\dagger$}}

\noaffiliation

\affiliation{Cavendish Laboratory, University of Cambridge, JJ Thomson Avenue, Cambridge, CB3 0HE, UK}
\affiliation{EPSRC National Epitaxy Facility, University of Sheffield, Broad Lane, Sheffield, S3 7HQ, UK}
\affiliation{Universit\'e C\^ote d'Azur, CNRS, CRHEA, rue Bernard Gregory, 06560 Valbonne, France
\\ \ \\
\textsuperscript{*}\,These authors contributed equally to this work.
\\
\textsuperscript{$\dagger$}\,Correspondence should be addressed to: cl538@cam.ac.uk; ma424@cam.ac.uk.
\\ \ \\
}

\begin{abstract}
A coherent ensemble of spins interfaced with a proxy qubit is an attractive platform to create many-body coherences and probe the regime of collective excitations. An electron spin qubit in a semiconductor quantum dot can act as such an interface to the dense nuclear spin ensemble within the quantum dot consisting of multiple high-spin atomic species. Earlier work has shown that the electron can relay properties of its nuclear environment through the statistics of its mean-field interaction with the total nuclear polarisation, namely its mean and variance. Here, we demonstrate a method to probe the spin state of a nuclear ensemble that exploits its response to collective spin excitations, enabling a species-selective reconstruction beyond the mean field. For the accessible range of optically prepared mean fields, the reconstructed populations indicate that the ensemble is in a non-thermal, correlated nuclear state. The sum over reconstructed species-resolved polarisations exceeds the classical prediction threefold. This stark deviation follows from a spin ensemble that contains inter-particle coherences, and serves as an entanglement witness that confirms the formation of a dark many-body state.
\end{abstract}

\maketitle

Reconstructing the internal state of an ensemble of interacting particles can reveal many-body correlations at the heart of non-equilibrium or quantum phases of matter \cite{Abanin2019a,Kaufman2016,Shimazaki2020}. In the weakly interacting regime of identical particles, the ensemble can be seen as a single entity whose state can be modified via global controls and measured with a particle-summed signal power, as for example in nuclear magnetic resonance (NMR) \cite{Vandersypen2005}. When interactions are significant this view breaks down and the internal state of the ensemble contains correlations that only multi-particle measurements can extract. Intuitive realisations of such measurements are obtained through single-particle spatial resolution in atomic quantum gas microscopes \cite{Bakr2009,Haller2015}, or through site selectivity in dilute central-spin systems \cite{Childress2006,Pla2013,Taminiau2014,Car2018,Metsch2019,Bourassa2020,Hensen2020}. A single collective excitation of an ensemble \cite{Dicke1954}, e.g. a polariton \cite{Fleischhauer2005} or a magnon \cite{Nikuni2000}, is sensitive to inter-particle phase coherences and can be used to reveal information about the underlying state of interacting particles \cite{Munoz2018}. In fact, in cases where an ensemble is particularly dense, collective modes may be the only gateway to access many-body coherences.

In this context, a proxy qubit coupled uniformly and strongly to a dense ensemble becomes the ideal interface to a many-body system. Examples include a single photon coupled to an ensemble of (artificial) atoms in cavity quantum electrodynamics \cite{Simon2007,Mlynek2014,Casabone2015}, a Rydberg polariton in a cloud of atoms \cite{Heidemann2007,Dudin2012}, as well as a single electron spin coupled via the hyperfine interaction to a dense nuclear spin ensemble \cite{Hanson2007,Urbaszek2013}. Together with the intrinsic coherence and control demonstrated for nuclear spins in semiconductor nanostructures \cite{Wust2016,Waeber2019,Chekhovich2020,Coish2009}, the latter is a system of great interest in engineering quantum correlated states of a mesoscopic spin ensemble in the solid state \cite{Imamoglu2003,Taylor2003,Taylor2003a,Giedke2006}. To date, treating the nuclear ensemble interacting with this electron spin qubit as a mean field -- the Overhauser field -- has been sufficient to account for numerous insights into nuclear spin dynamics \cite{Dyakonov1973,Khaetskii2002,Urbaszek2013}. The electron-mediated back-action on the Overhauser field has been employed
to create nuclear states with controlled mean \cite{Braun2006,Eble2006,Maletinsky2007,Tartakovskii2007,Latta2009,Hogele2012,Puebla2013} and sub-thermal variance \cite{Greilich2007,Vink2009,Bluhm2010,Sun2012,Chow2016,Ethier-Majcher2017}. Meanwhile, Hahn-echo spectroscopy has provided a species-resolved decomposition of the mean-field dynamics \cite{Bechtold2015,Stockill2016,Botzem2016} and, recently, advanced interferometry measurements have captured the quantum back-action of single electron spins on the nuclear dynamics \cite{Bethke2020}, as well as the Overhauser shift arising from a single electron-mediated nuclear spin-flip \cite{Jackson2020}. A reconstruction that goes beyond the mean-field treatment, and with sensitivity to inter-particle coherences, is the next challenge.

\begin{figure*}
\centering
\includegraphics[width=2\columnwidth]{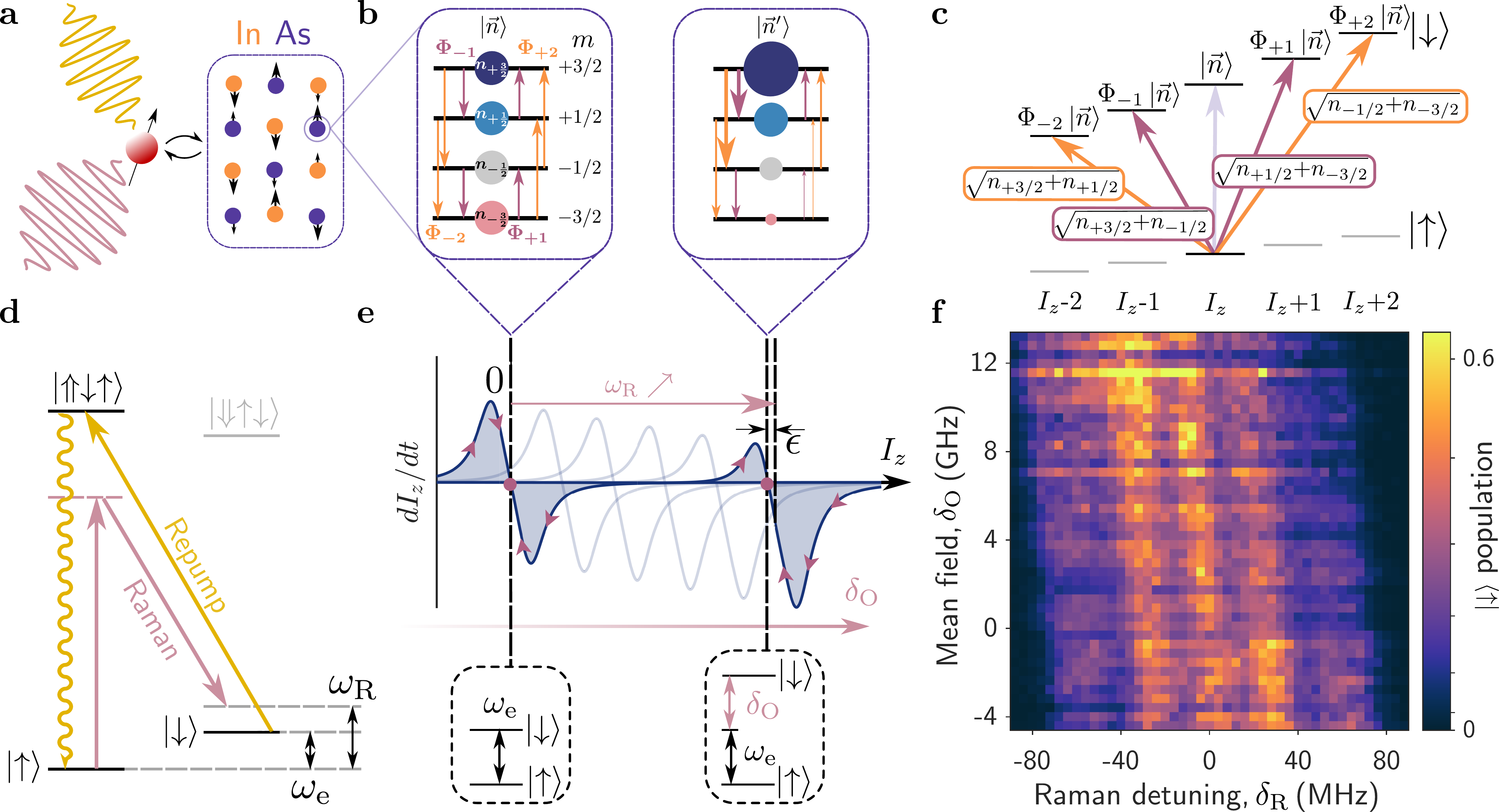}
\caption{\textbf{Proxy qubit interface to spin states of a nuclear ensemble. a,} An optically addressed electron spin interacts with an ensemble of $N$ nuclear spins composed of arsenic and indium (gallium not shown). \textbf{b,} An  external magnetic field of $3.5$\,T Zeeman-splits nuclear energy levels according to their spin projection $m$. Each level has fractional population $n_{m}$. $\Phi_{\pm1,2}$ are the magnon raising and lowering operators. Left: at $4.2$\,K, the system's thermal state is unpolarised, $I_z=0$. Changing $I_{z}$ changes populations $n_m$, altering the relative strengths of the $\Phi$ transitions. \textbf{c,} In the collective basis, the electron spin-state manifold $\{\ket{\uparrow},\ket{\downarrow}\}$ is dressed by nuclear spin-state population configuration $\ket{\vec{n}}$. Starting in the state $\ket{\uparrow,\vec{n}}$, magnon transitions whose strength depends on $n_m$ change the ensemble nuclear polarisation $I_z$ by one or two units.  \textbf{d,} QD level scheme in Voigt geometry. The cooling cycle pumps the electron resonantly to spin $\ket{\uparrow}$ via a charged exciton state, while a two-photon Raman process at frequency $\omega_\text{R}$ drives the electron spin, whose splitting at zero nuclear polarisation, set by $g_{\mathrm{e}}=0.52$, is $\omega_\text{e} = 25.3$\,GHz. \textbf{e,} This cooling cycle is equivalent to a feedback function $dI_{z}/dt$ on the nuclear polarisation $I_{z}$ (blue curves) via the mean field $\delta_\text{O} = \omega_\text{R} - \omega_\text{e}$. The stable mean-field polarisation (violet dot) can be dragged from zero by translating the Raman frequency $\omega_\text{R}$. At finite polarisation, there is a MHz-scale ($\sim0.1\%$) feedback-induced correction $\epsilon$ to the mean field. \textbf{f,} Population in the $\ket{\downarrow}$ manifold after a $1$-$\mu$s Raman drive, as a function of mean-field shift $\delta_\text{O}$ and probe Raman detuning $\delta_\text{R} = \omega_\text{R}^{\text{(probe)}} - \omega_\text{R}$, where $\omega_\text{R}^{\text{(probe)}}$ is the Raman frequency during the probing step.}
\label{fig:fig1}
\end{figure*}

In this work, we use a proxy qubit to detect interparticle coherences in a nuclear ensemble. To do so, we benchmark a technique that performs species-selective spin-state readout of a nuclear ensemble using a single electron spin as a proxy qubit. We first inject collective excitations -- nuclear magnons \cite{Gangloff2019,Bodey2019} -- into an optically cooled nuclear ensemble to probe the asymmetry of polarisation-increasing versus polarisation-decreasing spin-flip processes. From this magnon asymmetry alone, we reconstruct the classical-equivalent spin distribution of the nuclear state without requiring any knowledge of the mean field. Up to a mean-field polarisation of $\sim$\,$25\%$, our magnon spectroscopy  evidences correlated spin-state distributions and the non-thermal nature of our reduced-variance nuclear state \cite{Gangloff2019}. Finally, at all mean-field polarisations, we reveal that the measured magnon asymmetry far exceeds that of a classical nuclear state, which provides a witness for nuclear coherences akin to a sub-radiant many-body state. This entanglement witness and spin-state reconstruction constitute a precursor to quantifying many-body coherences in a dense ensemble. 

\begin{figure}[t!]
\centering
\includegraphics[width=1\columnwidth]{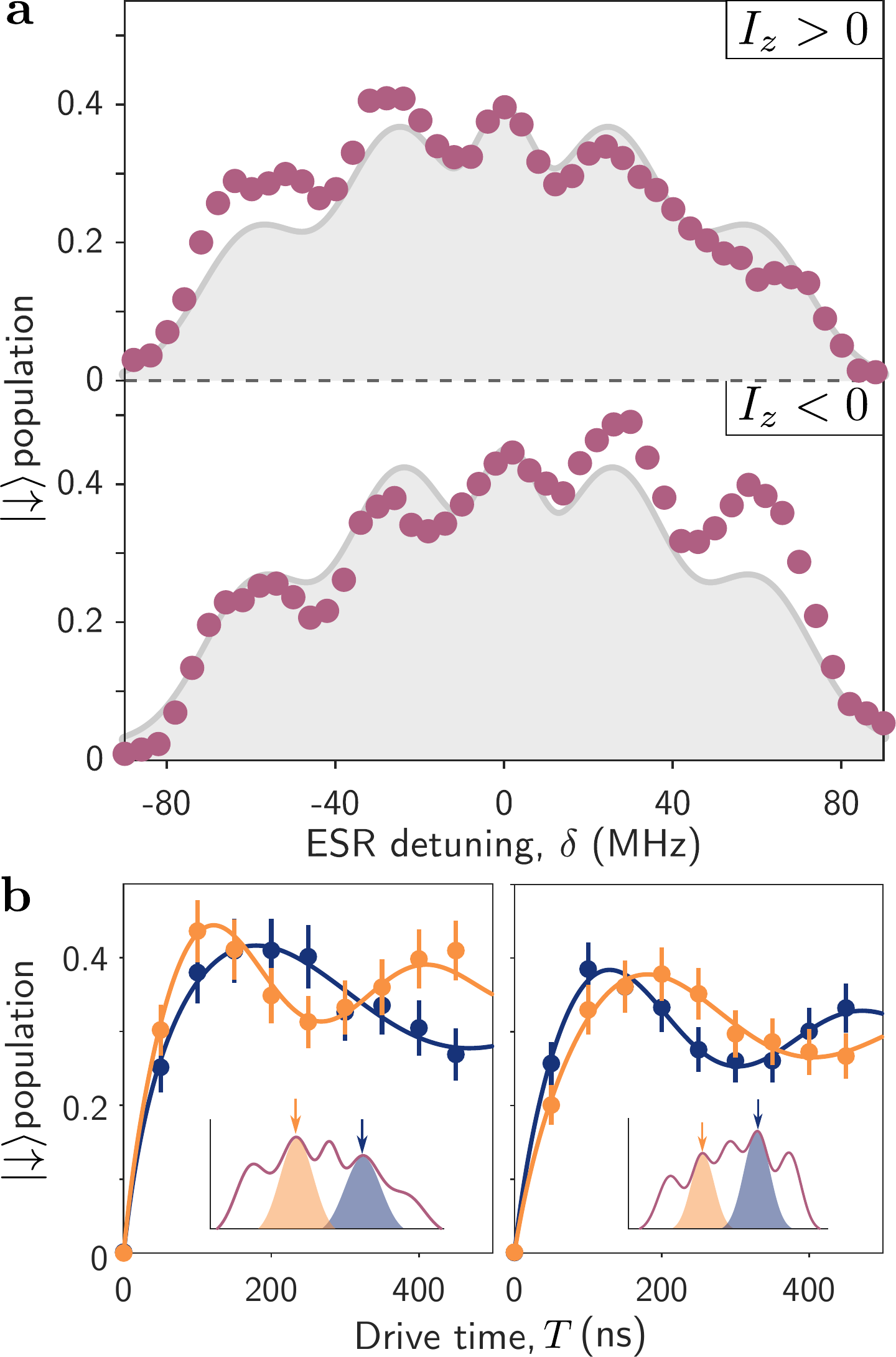}
\caption{\textbf{Magnon sideband asymmetry. a,} Electron $\ket{\downarrow}$ population as a function of ESR detuning $\delta$, following a $1$-$\mu$s drive with Rabi frequency $\Omega_\text{R} = 6.7$\,MHz, for positive (top, $\delta_\text{O}= 7.6$\,GHz) and negative (bottom, $\delta_\text{O}= -4.4$\,GHz) mean-field setpoints. Grey curves are a five-gaussian fit to the spectrum at $\delta_\text{O}= 0$\,GHz. \textbf{b,} Electron $\ket{\downarrow}$ population as a function of drive time $T$, for drive Rabi frequency $\Omega_\text{R} = 23$\,MHz. Left panel: $\delta_\text{O}=6$ GHz; $\delta = -22.3$\,MHz (orange) and $\delta = 17.7$\,MHz (blue). Right panel: $\delta_\text{O}=-2.4$ GHz; $\delta = -25.1$\,MHz (orange) and $\delta = 18.9$\,MHz (blue). Error bars indicate $67\%$ confidence intervals. Solid curves are fits to the phenomenological model $-a(1-T/(b+T))\cos\left(2\pi(\Omega_{\pm1} T+c)\right)+d$, where $\{a,b,c,d,\Omega_{\pm 1}\}$ are free parameters. Left panel: $\Omega_{-1} = 3.6(5)$ MHz (orange), $\Omega_{+1} = 1.8(7)$ MHz (blue). Right panel: $\Omega_{-1} = 2.3(2)$ MHz (orange), $\Omega_{+1} = 3.0(1)$ MHz (blue).}
\label{fig:fig2}
\end{figure}

\section*{Proxy qubit system}

Our proxy qubit interface is shown in Fig.\,1a and consists of an optically controlled electron spin in a semiconductor quantum dot (QD) (see Methods) in contact hyperfine interaction with an ensemble of $N$\,$\sim$\,50,000 nuclear spins distributed among three species: arsenic (total spin $I^{\text{As}}=3/2$), indium ($I^{\text{In}}=9/2$), and gallium ($I^{\text{Ga}}=3/2$). For each species, the average population in each of their Zeeman-split spin states is labelled $\vec{n} = (n_{+3/2}, n_{+1/2}, n_{-1/2}, n_{-3/2})$ (Fig.\,1b) -- for a spin species with $I^{j}>3/2$, such as indium, this description holds qualitatively, while the quantitative treatment is given in S.I. section II.B. Spin-flip transitions between these states can be activated by energy exchange with the electron via a noncollinear hyperfine interaction \cite{Hogele2012} and delivered as collective modes $\Phi_{\pm k}$, i.e. nuclear magnons \cite{Gangloff2019}. From the perspective of a single nucleus this noncollinear hyperfine interaction -- originating as a perturbation on the hyperfine interaction by the nuclear quadrupole interaction of high-spin nuclei with strain-induced electric-field gradients \cite{Urbaszek2013} (S.I. Section II.A) -- allows two types of magnons that change nuclear-spin projections by one ($k=1$) or two ($k=2$) units. When taken over the ensemble, a simple spin-state imbalance, such as thermal polarisation, leads to a corresponding imbalance in the transition strength of magnon excitations. From the proxy qubit's indiscriminate view of the ensemble, the sum effect of all nuclei is the mean-field Overhauser shift,
\begin{equation}
\delta_\text{O}= \frac{3}{2} A^{\text{As}} \mathcal{I}_{z}^{\text{As}} + x\frac{9}{2}  A^{\text{In}} \mathcal{I}_{z}^{\text{In}} + (1-x)\frac{3}{2} A^{\text{Ga}} \mathcal{I}_{z}^{\text{Ga}}\text{,}
\end{equation}
\noindent
on the electron spin resonance (ESR), where $x$ denotes the indium composition. Specific to each species $j$, $A^{j}$ is the hyperfine constant and $\mathcal{I}_z^{j}\in[-1;1]$ is the fraction of the maximal polarisation (i.e. $m = I^j$ for all spins). We denote $I_z$ as the ensemble polarisation over all species. In this picture, the addition of a nuclear magnon of any species changes $I_z$ by one or two units in exchange for an electronic spin flip. This is shown in Fig.\,1c, as an electronic manifold of states, where the polarisation-preserving ESR transition is flanked by polarisation-changing magnon transitions detuned by the nuclear Zeeman energy \cite{Gangloff2019}. Sidestepping the mean-field treatment, the Rabi frequency per magnon mode $\Omega_{\pm k}$ for a nuclear state that contains no coherences \cite{Vandersypen2005, Braun2006,Eble2006,Maletinsky2007,Tartakovskii2007,Latta2009,Hogele2012,Puebla2013,Bluhm2010,Stockill2016}, reveals the underlying single-particle spin states $\vec{n}$ (S.I. section II.B),
\begin{equation}\label{eq:enhance}
\begin{split}
\Omega_{+2}&\propto \sqrt{n_{-3/2}+n_{-1/2}}\\
\Omega_{+1}&\propto \sqrt{n_{-3/2}+n_{+1/2}}\\
\Omega_{-1}&\propto \sqrt{n_{-1/2}+n_{+3/2}}\\
\Omega_{-2}&\propto \sqrt{n_{+1/2}+n_{+3/2}}\text{ .}
\end{split}
\end{equation}

An optical cooling scheme via the electron and the QD's charged exciton state (see Methods) allows us to vary the mean field $\delta_\text{O}$, and thus nuclear polarisation $I_z$ and the populations $\vec{n}$. As depicted in Fig.\,1d, we drive the ESR with a Raman frequency $\omega_{\text{R}}$, generated from microwave-modulated optical fields \cite{Bodey2019}, whilst the optical pumping (repump) beam continuously polarises the electron spin into the $\ket{\uparrow}$ state. As shown in Fig.\,1e and as described in detail in \cite{Gangloff2019}, this continuous drive on a polarised electron generates a feedback loop that locks the mean field $\delta_\text{O}$ precisely to our drive frequency on Raman resonance \cite{Gangloff2019}, $\omega_\text{R}-\omega_\text{e}=\delta_\text{O}$, where $\omega_\text{e} = 25.3$\,GHz is the electron's Zeeman splitting.  When we increase (decrease) the Raman drive frequency $\omega_\text{R}$ slowly relative to relevant relaxation rates in the system, the ensemble polarisation $I_z$ increases (decreases) to maintain resonance via the mean-field Overhauser shift $\delta_\text{O}$. This \emph{dragging} process \cite{Latta2009,Vink2009,Hogele2012} can be exploited until nuclear diffusion processes surpass the Raman locking rates, leading to the disappearance of the dynamical stable point (i.e.\,a bifurcation) and setting a limit on the achievable nuclear polarisation. Over this range, the steady-state fluctuations of the mean-field polarisation $I_z$ are reduced well below their thermal value -- a long-lived, reduced-variance nuclear state (S.I. section III.D), which must exhibit spin-state correlations.

Immediately following the $20$-$\mu$s-long preparation of a nuclear state with a locked mean field $\delta_\text{O}$ (set by $\omega_{\text{R}}$), we reveal the magnon transitions of Fig.~1b,c via the ESR spectrum. Figure 1f shows the electronic $\ket{\downarrow}$ population, read out via resonant optical pumping following $1\,\mu$s of probe Raman drive, as a function of probe Raman detuning $\delta_\text{R}$ from $\omega_{\text{R}}$ and mean field $\delta_\text{O}$ ranging from $-4.4$\,GHz to $13.2$\,GHz. Along the Raman detuning axis, the readout signal reveals one central transition and the expected four satellite transitions at an ESR detuning $\delta$ from the central transition matching $\pm \omega^{j}$ and $\pm 2\omega^{j}$, where $\omega^{j}$ is the nuclear Zeeman energy. Along the mean-field axis, the five resonances have a polarisation-dependent MHz-scale offset, $\delta = \delta_\text{R} - \epsilon$. This is expected from nuclear polarisation, which induces asymmetry in the feedback curve and a small correction $\epsilon$ of the Raman frequency relative to mean field (Fig.\,1e, S.I. section III.E). Crucially, the magnon spectrum is clearly sideband-resolved, allowing us to extract information about the internal state $\vec{n}$ of the nuclear ensemble, at all mean-field setpoints.

\section*{Stokes and Anti-Stokes asymmetry}
Figure 2a highlights magnon spectra taken at two values of the mean field: $\delta_{\text{O}}=7.6$\,GHz (top panel) and $\delta_{\text{O}}=-4.4$\,GHz (bottom panel). For both positive and negative mean-field setpoints, we observe an asymmetry in sideband amplitude at negative ESR detuning ($\delta < 0$) relative to their amplitude at positive ESR detuning ($\delta > 0$). This asymmetry is a function of the mean-field setpoint: a positive $\delta_{\text{O}}$ sets up a positive nuclear polarisation for which polarisation-reducing (negatively detuned) modes are stronger; conversely, a negative $\delta_{\text{O}}$ sets up a negative nuclear polarisation for which polarisation-increasing (positively detuned) modes are stronger. This Stokes vs.\,anti-Stokes process asymmetry is well known from Raman scattering involving states with different occupations, as is the case here with the underlying number of nuclear states available for collective nuclear spin flips in a particular direction.

The amplitude of each magnon sideband has a direct correspondence with the Rabi frequency $\Omega_{\pm k}$ of the activated electron-nuclear exchange responsible for that particular magnon excitation, as per equation (1) (S.I. section II.B). We verify this explicitly with coherent Rabi oscillations induced by a Raman drive set to resonance with a single-magnon mode $\Phi_{\pm1}$, as shown in Fig.\,2b. We observe short-time oscillations whose frequency $\Omega_{\pm 1}$ is a function of whether the excited magnon increases ($+1$, blue data) or decreases ($-1$, orange data) the mean-field nuclear polarisation $I_z$. We confirm that for a positive mean-field shift (left panel), $\Omega_{-1} > \Omega_{+1}$, while for a negative mean-field shift (right panel), $\Omega_{+1} > \Omega_{-1}$.

\section*{Species-resolved spin-state reconstruction}

We wish to identify closely spaced magnon transitions for indium ($\omega^\text{In}=32.7$\,MHz) and arsenic ($\omega^\text{As}=25.3$\,MHz), which requires spectral resolution of a few MHz. A further post-processing step is thus required to extract this species-resolved information from our spectra (Fig.\,1f, Fig.\,2a). Our measurement resolution is limited by the quasi-static mean-field fluctuations that are behind the inhomogeneous linewidth of the ESR, which we measure directly via Ramsey interferometry to be $14$\,MHz (S.I. section III.B). This informs the construction of an optimised Wiener deconvolution filter that removes faithfully this inhomogeneous broadening from our data (S.I. section IV.A). Figure\,3a shows a filtered magnon spectrum and reveals a doublet structure underlying each of the four sideband clusters of Figs.\,1f and 2a; species-specific magnon modes. 

To avoid drive-induced power broadening and optimise signal-to-noise simultaneously, we have set the drive Rabi frequency $\Omega_\text{R}$ to $6.7$\,MHz to be comparable with mean-field fluctuations. In this regime, the activated electron-nuclear exchange is overdamped by the electronic and nuclear dephasing processes (at rate $\Gamma_2$), and at drive times short with respect to $(\Omega_{\pm k}^2/\Gamma_2)^{-1}$, the amplitude of a sideband reflects the electron-nuclear exchange frequency $\Omega_{\pm k}$. We treat each magnon mode as a two-level system with a set dephasing rate (measured independently), defining a lineshape that we fit to all sidebands independently (S.I. section IV), as shown in Fig.\,3a. From these fits, we obtain an exchange frequency $\Omega_{\pm k}$ for each mode, where $k=(1,2)$.

We assemble the dimensionless asymmetry parameter for each magnon type ($k$) from its positive and negative exchange frequencies:
\begin{align}\label{eq:asym}
\nu_{k}&=\frac{\Omega_{-k}^{2}-\Omega_{+k}^{2}}{\Omega_{-k}^{2}+\Omega_{+k}^{2}}\text{.}
\end{align}
\noindent
Indeed, measuring this quantity for both $k=1$ and $k=2$ magnon modes in a spin-3/2 system such as arsenic, we arrive identically at quantities which are specific to each species and do not presume knowledge of the material hyperfine constants (S.I. section II.B), imbalances in spin-state populations, 
\begin{equation}
\begin{split}
n_{+3/2}-n_{-3/2}&=\frac{1}{2}(\nu_{2}+\nu_{1})\\
n_{+1/2}-n_{-1/2}&=\frac{1}{2}(\nu_{2}-\nu_{1})\text{,}
\end{split}
\end{equation}
\noindent
and the asymmetry-commensurate fractional polarisation $\mathcal{I}_{z}^{\star}$:
\begin{equation}\label{eq:imbalance}
\mathcal{I}_{z}^{\star}=\frac{1}{3}(2\nu_{2}+\nu_{1}).
\end{equation}
In the specific case where the nuclei are in a thermal state, this measure of polarisation is equivalent precisely to the previous definition, satisfying $\mathcal{I}_z^{\star}=\mathcal{I}_z$.

\begin{figure*}
\centering
\includegraphics[width=2\columnwidth]{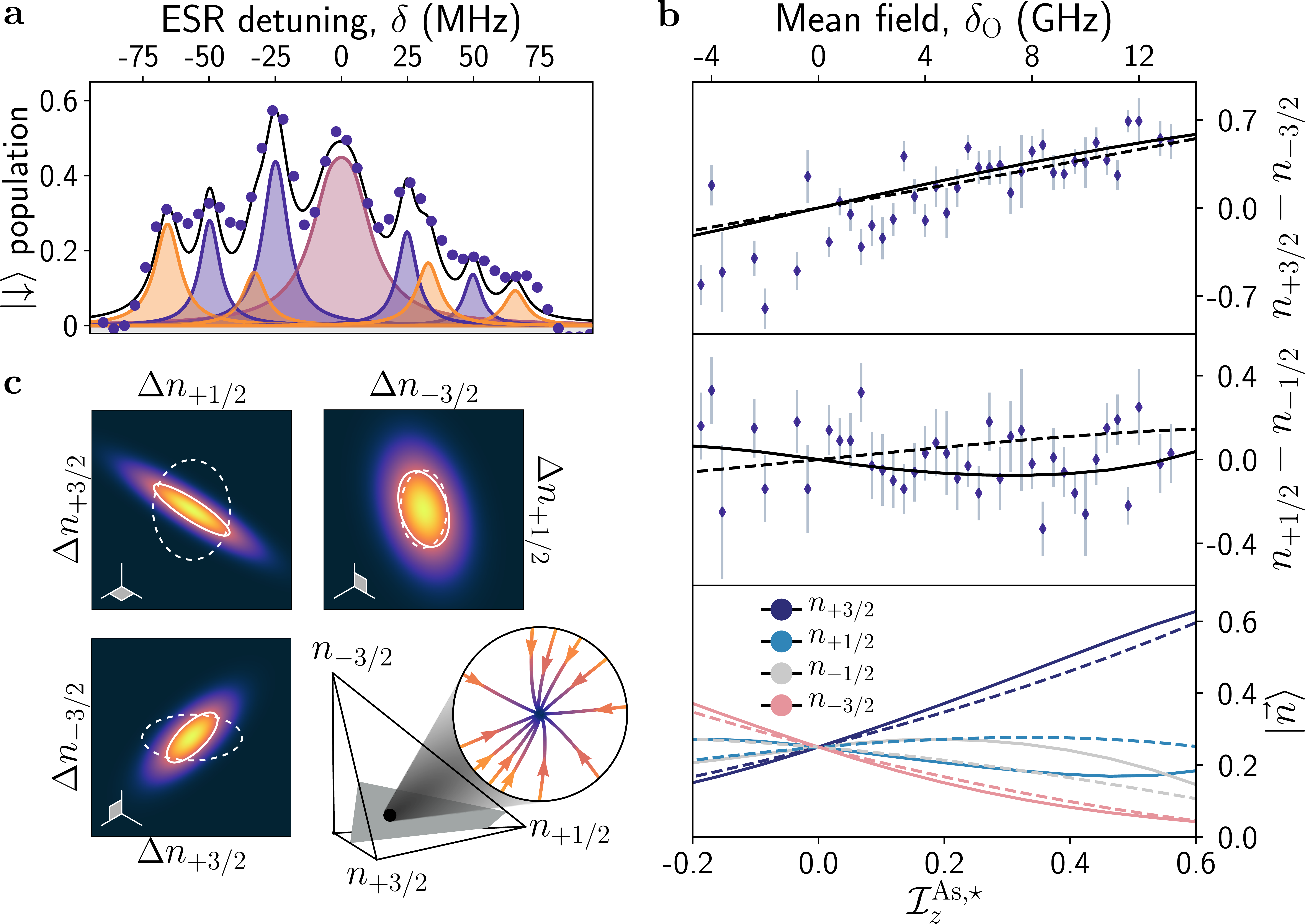}
\caption{\textbf{From sideband asymmetry to species-resolved spin populations. a,} Electron $\ket{\downarrow}$ population at $1$-$\mu$s drive time as a function of ESR detuning $\delta$, deconvolved using a Wiener filter. The black curve is the summed fits of the ESR peak (violet), the indium magnon peaks (orange), and the arsenic magnon peaks (blue). Each resonance lineshape is a driven two-level system (S.I. section IV) with a coherence time $T_2=275(25)$\,ns for arsenic and $T_2=214(14)$\,ns for indium. The fitted Rabi frequencies are $\{\Omega^{\text{In}}_{-2},\Omega^{\text{As}}_{-2},\Omega^{\text{In}}_{-1},\Omega^{\text{As}}_{-1},\Omega^{\text{As}}_{+1},\Omega^{\text{In}}_{+1},\Omega^{\text{As}}_{+2},\Omega^{\text{In}}_{+2}\}$ $=\{0.77(7),0.69(8),0.51(13),1.07(13),0.64(9),0.56(10),\\0.44(9),0.40(9)\}$\,MHz.
\textbf{b,} Population imbalance in the arsenic $3/2$ manifold (top) and $1/2$ manifold (middle) as a function of the mean field. Error bars represent a $67\%$ confidence interval.   Bottom: arsenic spin-state populations as a function of its asymmetry-commensurate fractional polarisation.
All panels: Solid curves are from the 3D Fokker-Planck model. Dashed curves correspond to a thermal distribution.
\textbf{c,} 3D Fokker-Planck modelling of the probability distribution $\vec{n}_{3\text{D}}$ (mean subtracted) for arsenic at the $\mathcal{I}_z^{\text{As},\star}=0.2$ setpoint, projected along three planes. The solid and dashed contours display one standard deviation for the modelled and the thermal (S.I. section V.A) distributions, respectively. Bottom right: 3D space $(n_{+3/2}, n_{+1/2}, n_{-3/2})$ for the simulation. In grey is the plane of constant polarisation passing through the steady state. The inset shows the vector field $d\vec{n}_{3\text{D}}/dt$ in this constant-polarisation plane.}
\label{fig:fig3}
\end{figure*}

The top two panels of Fig.\,3b show the arsenic population imbalances, $n_{+3/2}-n_{-3/2}$ and $n_{+1/2}-n_{-1/2}$, retrieved from filtered spectra taken over the full range of mean-field setpoints. We find that the build up of the mean field is accompanied by a monotonically increasing imbalance in the $3/2$ subspace, whereas the imbalance in the $1/2$ subspace remains close to zero throughout. A comparison with imbalances expected for a thermal state (dashed curves in Fig.\,3b) highlights that while our data for the $3/2$ manifold is consistent with a thermal state, data for the $1/2$ manifold departs from this simplest assumption. To shed light further on the underlying nuclear state, the full spin distribution needs to be reconstructed.

Magnon modes are blind to a polarisation-preserving population transfer from the $3/2$ manifold to the $1/2$ manifold: $(n_{+3/2}+n_{-3/2}) - (n_{+1/2}+n_{-1/2})$ is not measurable in the spectrum. To obtain the populations $\vec{n}$ from imbalances, we require a dynamical model which captures the relative importance of magnon excitation rates in the $1/2$ and $3/2$ manifolds relative to nuclear diffusion rates. To this end, we use the Fokker-Planck formalism with the population normalisation constraint $n_{+3/2} + n_{+1/2} + n_{-1/2} + n_{-3/2} = 1$ and obtain the steady state of the arsenic populations under optical feedback (S.I. section V.A). We visualise the result of this simulation in a three-dimensional space $\vec{n}_{3\text{D}} = (n_{+3/2},\  n_{+1/2},\ n_{-3/2})$. As a generalisation of the one-dimensional feedback function $dI_z/dt$ vs.\,$I_z$ introduced in Fig.\,1e, Fig.\,3c shows the three-dimensional vector field $d\vec{n}_{3\text{D}}/dt$ that flows towards a single stable point $\vec{n}_{3\text{D}}$, with the reconstructed arsenic polarisation $\mathcal{I}_z^{\text{As},\star}$ as setpoint. The model parameters are strongly constrained by the arsenic contribution to the mean-field fluctuations, as quantified independently by the electron inhomogeneous dephasing time $T_2^* = 39$\,ns (S.I. section V.A).

We find good agreement between our imbalance data in Fig.\,3b and our model's predictions, shown as solid curves.  The bottom panel of Fig.\,3b then shows the reconstructed spin populations $\vec{n}$ as a function of the mean field. The thermal state is shown as dashed curves and displays a monotonic behaviour. The non-monotonic nuclear populations prepared by optical cooling are thermal-like only at zero and at the maximum achievable value of polarisation (where the feedback becomes unstable). Strikingly, owing to strong feedback and finite imbalance, the populations deviate consistently from thermal for all intermediate values of polarisation. The dynamics set by the nuclear quadrupolar interaction tune the population imbalance in the $1/2$ manifold: here, when polarising $I_z>0$, a strong first sideband ($\Phi_{+1}$) depletes $+1/2$ in favor of the $+3/2$ and populates the $-1/2$ from $-3/2$ resulting in a $1/2$ manifold imbalance opposing $I_z$. The departure from a thermal distribution is a universal feature of states polarised via \emph{dragging}.

Sub-thermal mean-field fluctuations within the ensemble, $\Delta^2 I_z \ll N$, are a feature of an electron-mediated feedback \cite{Eble2006,Ethier-Majcher2017,Gangloff2019}. To investigate this feature beyond the mean field, we thus reconstruct the steady-state nuclear fluctuations from the mean, $\Delta\vec{n}_{3\text{D}}$, using our Fokker-Planck model. Figure\,3c shows these fluctuations as projections of the probability distribution of $\Delta \vec{n}_{3\text{D}}$, for an arsenic polarisation set to $\mathcal{I}_z^{\text{As},\star}=0.2$.  The grey surface in Fig.\,3c (lower right panel) identifies an $\mathcal{I}_z$-preserving plane, within which the electron-mediated feedback is weak, but towards which it is strong. The $(n_{+1/2}, n_{-3/2})$ plane is almost parallel to this $\mathcal{I}_z$-preserving plane, and consequently, the corresponding projection is nearly thermal. In contrast, the projections onto the orthogonal $(n_{+3/2}, n_{+1/2})$ and $(n_{-3/2}, n_{+3/2})$ planes, experiencing strong feedback, are squeezed and tilted relative to the thermal distribution. This indicates that our non-thermal nuclear state contains spin-state correlations and cannot be described by a factorisable classical distribution. 

\begin{figure}
\includegraphics[width=1\columnwidth]{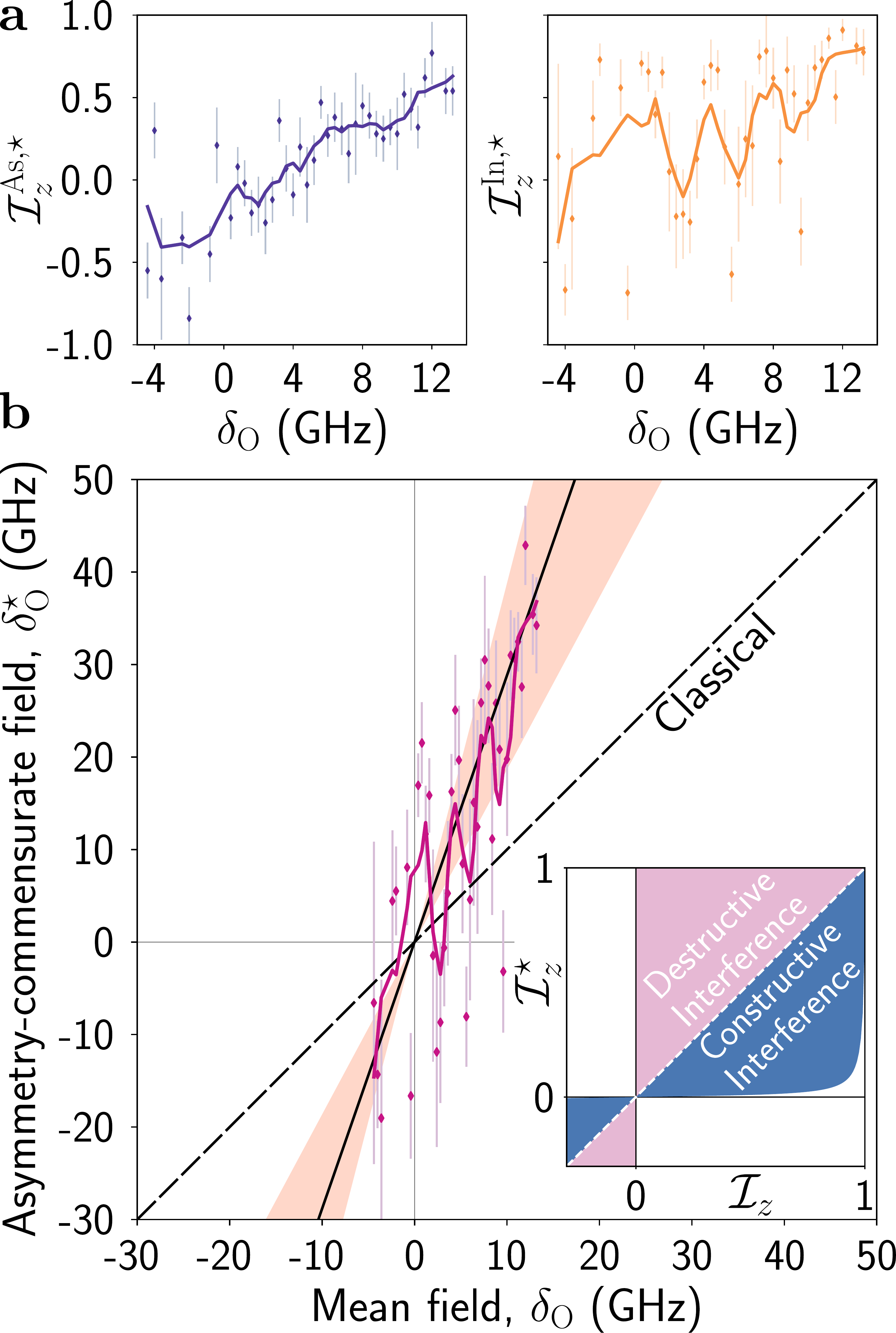}
\caption{\textbf{Fingerprint of a nuclear dark state. a,} Reconstructed polarisation of the arsenic (blue) and indium (orange) nuclear species as a function of the mean field. Accounting for the spin-9/2 character of indium requires applying a scaling function with values in the range $[1;2]$ to Eq.\,\ref{eq:imbalance} (S.I. section II.B.2). \textbf{b,} Derived asymmetry-commensurate mean field versus mean-field setpoint. The solid black curve is a fit to the data with a slope of $2.9(1)$. The shaded coral region indicates the range of indium concentrations $x=0.25$-$0.75$. The dashed line with a slope of $1$ is the mean field that would be reconstructed for a classical nuclear state. Inset: sideband asymmetry parameter as a function of nuclear-spin polarisation in the simpler case of $N$ spin-$1/2$ particles, where there are only two magnon modes characterised by $\Omega_+$ and $\Omega_-$. The asymmetry parameter $(\Omega_-^2 - \Omega_+^2)/(\Omega_-^2 + \Omega_+^2)$ is identically $\mathcal{I}_z$ for a nuclear state with no coherences (dashed line), and will exceed $\mathcal{I}_z$ for nuclear states with sub-radiant character (pink region) and fall short of $\mathcal{I}_z$ for nuclear states with super-radiant character or classical coherences (blue region). In both panels, solid curves are the data passed through a first-order Savitzky-Golay filter with a $1.6$-GHz window. Error bars indicate a 67\% confidence interval.}
\label{fig:fig4}
\end{figure}

\section*{Beyond a classical mean field}

The working assumption for this QD system has been that the $N$-spin nuclear state is classical \cite{Vandersypen2005, Braun2006,Eble2006,Maletinsky2007,Tartakovskii2007,Latta2009,Hogele2012,Puebla2013,Bluhm2010,Stockill2016} -- i.e. it contains no many-body coherences -- which has allowed us to infer spin-state populations straightforwardly from the strength of collective excitations, as per Eqs.\,1-3. However, let us suppose  that in an extreme scenario the total angular momentum of the $N$-spin system sums \emph{coherently} to a single spin -- as for example would manifest as a dark (sub-radiant) state in Dicke's theory \cite{Dicke1954}. In this scenario a mean-field fractional polarisation $\mathcal{I}_z=1/N$ would be the maximum achievable spin projection of the $N$-spin system, where no further upward spin-flip excitations are permissible and where the collective sideband asymmetry attains already its extremum $\nu_1 = 1$ \cite{Taylor2003a}, and $\mathcal{I}_z^{\star}=1$. Generalising this concept, a sideband asymmetry in excess of Eq.\,\ref{eq:asym}, evaluated for a classical state, reveals a nuclear state with reduced total angular momentum. More formally, we can use a classicality bound \cite{Vitagliano2011} to convert the asymmetry parameter of Eq.\,\ref{eq:asym}, and ultimately our derived fractional polarisation of Eq.\,\ref{eq:imbalance}, into an entanglement witness:
\begin{equation}\label{eq:ineq}
    \mathcal{I}_z^\star \leq \frac{\mathcal{I}_z}{1 -  \frac{N}{2}\Delta^2 \mathcal{I}_z + \frac{N}{2}\left(\mathcal{I}_x^2 + \mathcal{I}_y^2\right)}\text{,}
\end{equation}
where $\Delta^2 \mathcal{I}_z \sim (400N)^{-1}$ for our narrowed state \cite{Gangloff2019}, and $\mathcal{I}_x$ and $\mathcal{I}_y$ are transverse coherences, which we do not measure. Violation of this bound necessarily implies some degree of entanglement among nuclei. Because transverse coherences only reduce this classicality bound (right side of Eq.\,\ref{eq:ineq}), we can put a universal condition on our measurements for the manifestation of dark-state many-body coherences (S.I. section VI):
\begin{equation}\label{eq:ineq2}
    \mathcal{I}_z^\star > \mathcal{I}_z \text{.}
\end{equation}

Figure 4a presents the reconstruction of polarisation $\mathcal{I}_z^{j,\star}$ for arsenic and indium as a function of the mean-field setpoint, using Eq.\,\ref{eq:imbalance} with the sideband asymmetry data measured in our magnon spectra. The solid curves are running averages over the data. We infer polarisation $\mathcal{I}_z^{\text{As},\star}$ upwards of $50\%$ for arsenic and $\mathcal{I}_z^{\text{In},\star}$ upwards of $70\%$ for indium. Strikingly,  this is achieved for a mean field $\delta_\text{O}$ of only $13.2$\,GHz; then, what would be the mean field $\delta_\text{O}^\star$ that is commensurate with our asymmetry data? We can construct this asymmetry-commensurate mean field from equation (1) using $\mathcal{I}_z^{j,\star}$ and taking a conservative estimate on gallium polarisation $\mathcal{I}_z^{\text{Ga},\star} \approx (1/2) \mathcal{I}_z^{\text{As},\star}$ (S.I. section V.B). This derived asymmetry-commensurate mean field $\delta_\text{O}^\star$ is shown in Fig.\,4b as a function of the measured mean field $\delta_\text{O}$. The one-to-one (dashed) line, $\delta_\text{O}^\star = \delta_\text{O}$, is the classical prediction in the absence of quantum or classical coherences, i.e. $\mathcal{I}_z^\star = \mathcal{I}_z$. Remarkably, a linear fit to the asymmetry-commensurate mean field (solid curve) exceeds the classical prediction by a factor of $2.9(1)$. While this fit assumes the best estimate for our quantum dot's indium concentration $x=0.5$ \cite{Stockill2016}, the classical prediction is exceeded by a factor of $1.9(1)-3.9(1)$ for $x=0.25-0.75$ (shaded coral area), the complete range of previously reported indium concentrations in different quantum dot systems \cite{Biasiol2011}. This demonstration of $\mathcal{I}_z^\star > \mathcal{I}_z$ constitutes the fingerprint of dark-state coherences in our optically cooled nuclear ensemble, as previously predicted for quantum dots \cite{Imamoglu2003,Taylor2003}, and relayed here as an excess sideband asymmetry. The inset of Fig.\,4b illustrates the two ways in which many-body coherences are manifested for an ensemble of spins (S.I. section VI). On one side of the classical divide (blue shade), a diminished asymmetry may indicate constructive interference in the spin sum, and the presence of a bright (super-radiant) state. On the other side of the classical divide (red shade), an excess asymmetry, as we have observed, is a telltale sign that these coherences are behind a destructive interference in the spin sum, equivalent to a sub-radiant nuclear state. 

 The one-to-all connectivity of our proxy qubit to the dense spin ensemble, unique to this magnon-based approach, renders our experiment sensitive to entanglement among nuclei, and complements tomographic reconstructions that exploit global NMR control and readout of the mean-field Overhauser shift \cite{Chekhovich2017}. An immediate extension of this work is to probe the dynamics of the dark-state coherences from emergence to relaxation. Beyond, our magnon spectroscopy technique could in principle quantify the degree of entanglement among nuclear spins and thus provide the necessary information to extend our classical-equivalent population reconstruction to accommodate the presence of many-body coherences. Moreover, NMR-based access to transverse coherence  \cite{Chekhovich2017} combined with electron-enabled measurements of the total angular momentum of the ensemble and its projection \cite{Jackson2020} would provide a powerful tomographic tool for these entangled many-body states. Honing further control over the nuclear-state coherences shown here offers a route to a quantum memory \cite{Denning2019a} hosted in a low-polarisation decoherence-free subspace \cite{Taylor2003}. Lastly, tuning the interplay between an electron-mediated periodic drive, nuclear interactions and dissipation in the super-radiant regime could provide access to synthetic and emergent many-body phases \cite{Kessler2012a}, including Dicke time crystals \cite{Zhu2019}.
\\\\

\section*{Acknowledgements} 
We thank N.R. Cooper and D.M. Kara for helpful discussions. We acknowledge support from  the US Office of Naval Research Global (N62909-19-1-2115), ERC PHOENICS (617985), EPSRC NQIT (EP/M013243/1), EU H2020 FET-Open project QLUSTER (DLV-862035) and the Royal Society (EA/181068). Samples were grown in the EPSRC National Epitaxy Facility. D.A.G. acknowledges a St John's College Fellowship and C.LG. a Royal Society Dorothy Hodgkin Fellowship.

\section*{Author contributions} 
D.A.G., C.LG., and M.A. conceived the experiments. J.H.B., C.B., D.A.G., G.E.-M., and C.L. carried out the experiments. J.H.B., L.Z., and D.A.G. performed the data analysis. L.Z. performed the theory and simulations with guidance from C.LG. and D.A.G. E.C. and M.H. grew the material. All authors contributed to the discussion of the analysis and the results. All authors participated in preparing the manuscript.

\section*{Competing interests} 
The authors declare no competing interests. 

\section*{References}
\bibliographystyle{naturemagdan.bst}

\section*{Methods}
\label{sec_methods}

\subsection*{Sample structure}
\label{subsec_qddevice}

Self-assembled InGaAs QDs are grown by molecular beam epitaxy and integrated inside a Schottky diode structure to allow charge control \cite{Urbaszek2013}. This comprises a 35-nm tunnel barrier between the n-doped layer and the QDs, and a blocking barrier above the QD layer to prevent charge leakage. The Schottky diode structure is electrically contacted through ohmic AuGeNi contacts to the n-doped layer and a semitransparent Ti gate (6\,nm) is evaporated onto the surface of the sample. A distributed Bragg reflector below and a superhemispherical cubic zirconia solid immersion lens above the QDs maximises photon outcoupling efficiency to 10\% at the first lens for QDs with an emission wavelength around 970\,nm.

Data was taken on two QD devices, made from the same wafer. QD device 1 was used in previous studies \cite{Stockill2016,Ethier-Majcher2017,Bodey2019}, and is the device on which the data presented throughout the main text was taken. QD device 2 was used in previous studies \cite{Gangloff2019,Jackson2020}, and is the device used for some of the data sets presented in the Supplementary Information. 

\subsection*{Experimental setup}
\label{subsec_exp_setup}

The experimental setup is described in detail in S.I. section I. A helium bath cryostat houses the QD devices at 4K. A magnetic field strength $B_z=3.5$\,T is applied transverse to the QD growth axis (Voigt geometry). Two laser beams are combined and sent to the QD: a Raman laser system, which is microwave-modulated, and a resonant readout/repump laser. A cross-polarisation confocal microscope filters resonant laser background, and a grating removes non-resonant background. The filtered signal is sent to a superconducting nanowire single-photon detector.

\subsection*{Cooling and polarisation \emph{dragging} sequence}

We cool the QD nuclear ensemble using an optical Raman scheme detailed in \cite{Gangloff2019}. We optimise the cooling parameters by operating under conditions which maximise the electron $T_{2}^{*}$. The relevant settings during the cooling sequence are: a Raman Rabi frequency $\Omega_\text{R}\approx21$\,MHz and a repump Rabi frequency of $\Omega_{P}\approx0.1\Gamma_{0}/2$ for an excited state linewidth $\Gamma_{0} \approx 150$\,MHz. This gives us an optimal $T_{2}^{*}$ of 39 ns, over an order of magnitude longer than without cooling.

We control the mean-field nuclear polarisation by tuning the frequency of a $22$-GHz LO source whilst running the experimental sequence. We begin a sequence at zero polarisation, where the Raman frequency corresponds to the electron Zeeman energy. From there, we increase the Raman frequency $\omega_\text{R}$ by increasing the LO frequency, thereby increasing the electron spin splitting at a rate of $0.04\ \mathrm{GHz\ s}^{-1}$. In this way, we are able to drag the electron spin resonance over a range exceeding $17\,\mathrm{GHz}$. We compensate the unwanted shift in transition frequency to the optical excited state, due to the Overhauser shift of the ground and excited states, by applying a compensating linear ramp to the QD device gate voltage. This tunes the trion transition frequency via the DC Stark shift.

\subsection*{Experimental spectroscopy sequence}

Throughout our experiments, we keep the time ratio of cooling to spectroscopy around 4, meaning that we spend 80\% of the time cooling the nuclear ensemble. The cooling sequence lasts for $20\,\mu \mathrm{s}$ and the spectroscopy sequence for $5\,\mu \mathrm{s}$. Using these timescales ensures that the steady state of the nuclear ensemble is primarily determined by the cooling sequence, and thereby maintains Raman resonance. By rapidly tuning the frequency and amplitude of an Arbitrary Waveform Generator output we generate experimental control pulses of a chosen detuning $\delta_\text{R}$ and Rabi frequency $\Omega_\text{R}$. When taking a drive-time dependence, such as for Fig.\,2b, we pair two pulses of increasing and decreasing lengths such that total Raman power is conserved, allowing us to stabilise the time-averaged laser power to within 1\%. 

\subsection*{Data processing and normalisation}

\emph{Background subtraction} -- Our raw measurements of the magnon spectrum include a constant background count rate due to optically induced electron spin relaxation, via a mechanism external to the QD \cite{Bodey2019}. We subtract this constant background, which is equal to the count rate at large Raman detuning $\delta_\text{R} \sim 100$\,MHz, from our data. When integrated over a typical spectrum, the background count rate constitutes $22\%$ of the uncorrected signal.

\emph{Polarisation-dependent count rate} -- We observe a systematic decrease in count rate as the nuclear ensemble is polarised so as to increase the mean-field shift on the ESR. This is because the collected readout fluorescence passes through a narrowband ($\sim 20$\,GHz) optical grating in order to remove the laser background from the far-detuned Raman beams. When we polarise, spin-flipping photon emission will change in frequency due to the change in ESR splitting. The collection efficiency of these photons therefore decreases as nuclear polarisation increases from zero. In order to correct for this effect, we normalise data taken at each polarisation by its average count rate.

\emph{From count rate to electron-spin population} -- Throughout the main text, read-out fluorescence counts have been converted to electron $\ket{\downarrow}$ population. In Figs.\,1, 2a, and 3 we use the fit to a time evolution of the magnon spectrum (S.I. section III.C) to fix the count rate which corresponds to 50\% $\ket{\downarrow}$ population. In Fig. 2b, we assume that the system, when on resonance with the ESR, will reach a steady state of 50\% $\ket{\downarrow}$ population at long drive time.

\section*{Data availability} 
All data that support the plots within this paper and other findings of this study are available from the corresponding authors upon reasonable request.

\section*{Code availability} 
All code that was used in the analysis of data shown in this paper is available from the corresponding authors upon reasonable request.

\section*{Additional information}

\noindent
\textbf{Supplementary information} is available for this paper.

\noindent
\textbf{Correspondence and requests for materials} should be addressed to C.LG. or M.A.

\end{document}



\title{Supplementary Materials for: Witnessing quantum correlations in a nuclear ensemble via an electron spin qubit}

\author{Dorian A.\,Gangloff\textsuperscript{1,*}}
\author{Leon Zaporski\textsuperscript{1,*}}
\author{Jonathan H.\,Bodey\textsuperscript{1,*}}
\author{Clara Bachorz\textsuperscript{1}}
\author{Daniel M.\,Jackson\textsuperscript{1}}
\author{Gabriel \'Ethier-Majcher\textsuperscript{1}}
\author{Constantin Lang\textsuperscript{1}}
\author{Edmund Clarke\textsuperscript{2}}
\author{Maxime Hugues\textsuperscript{3}}
\author{Claire Le Gall\textsuperscript{1,$\dagger$}}
\author{Mete Atat\"ure\textsuperscript{1,$\dagger$}}

\noaffiliation

\affiliation{Cavendish Laboratory, University of Cambridge, JJ Thomson Avenue, Cambridge, CB3 0HE, UK}
\affiliation{EPSRC National Epitaxy Facility, University of Sheffield, Broad Lane, Sheffield, S3 7HQ, UK}
\affiliation{Universit\'e C\^ote d'Azur, CNRS, CRHEA, rue Bernard Gregory, 06560 Valbonne, France
\\ \ \\
\textsuperscript{*}\,These authors contributed equally to this work.
\\
\textsuperscript{$\dagger$}\,Correspondence should be addressed to: cl538@cam.ac.uk; ma424@cam.ac.uk.
\\ \ \\
}


\maketitle

\tableofcontents


\section{Experimental techniques}


\subsection{Quantum dot devices}\label{sec:device}



Data was taken on two QD devices, made from the same wafer. QD device 1 was used in previous studies \cite{Bodey2019,Stockill2016,Ethier-Majcher2017,Huthmacher2018}, and is the device on which the data was taken throughout the main text and this supplement, unless otherwise stated. QD device 2 was used in previous studies \cite{Gangloff2019}, and is the device used for some of the data sets presented in this supplement. 

Self-assembled InGaAs QDs are grown by molecular beam epitaxy and integrated inside a Schottky diode structure to allow charge control \cite{Urbaszek2013}. This comprises a 35nm tunnel barrier between the n-doped layer and the QDs, and a blocking barrier above the QD layer to prevent charge leakage. The Schottky diode structure is electrically contacted through ohmic AuGeNi contacts to the n-doped layer and a semitransparent Ti gate (6 nm) is evaporated onto the surface of the sample. A distributed Bragg reflector below the QDs maximises photon outcoupling efficiency. The photon collection is further enhanced with a superhemispherical cubic zirconia solid immersion lens on the top Schottky contact of the device. We estimate a photon-outcoupling efficiency of 10\% at the first lens for QDs with an emission wavelength around 970 nm. 

\subsection{Experimental schematic}

\begin{figure*}
\includegraphics[width=1.8\columnwidth]{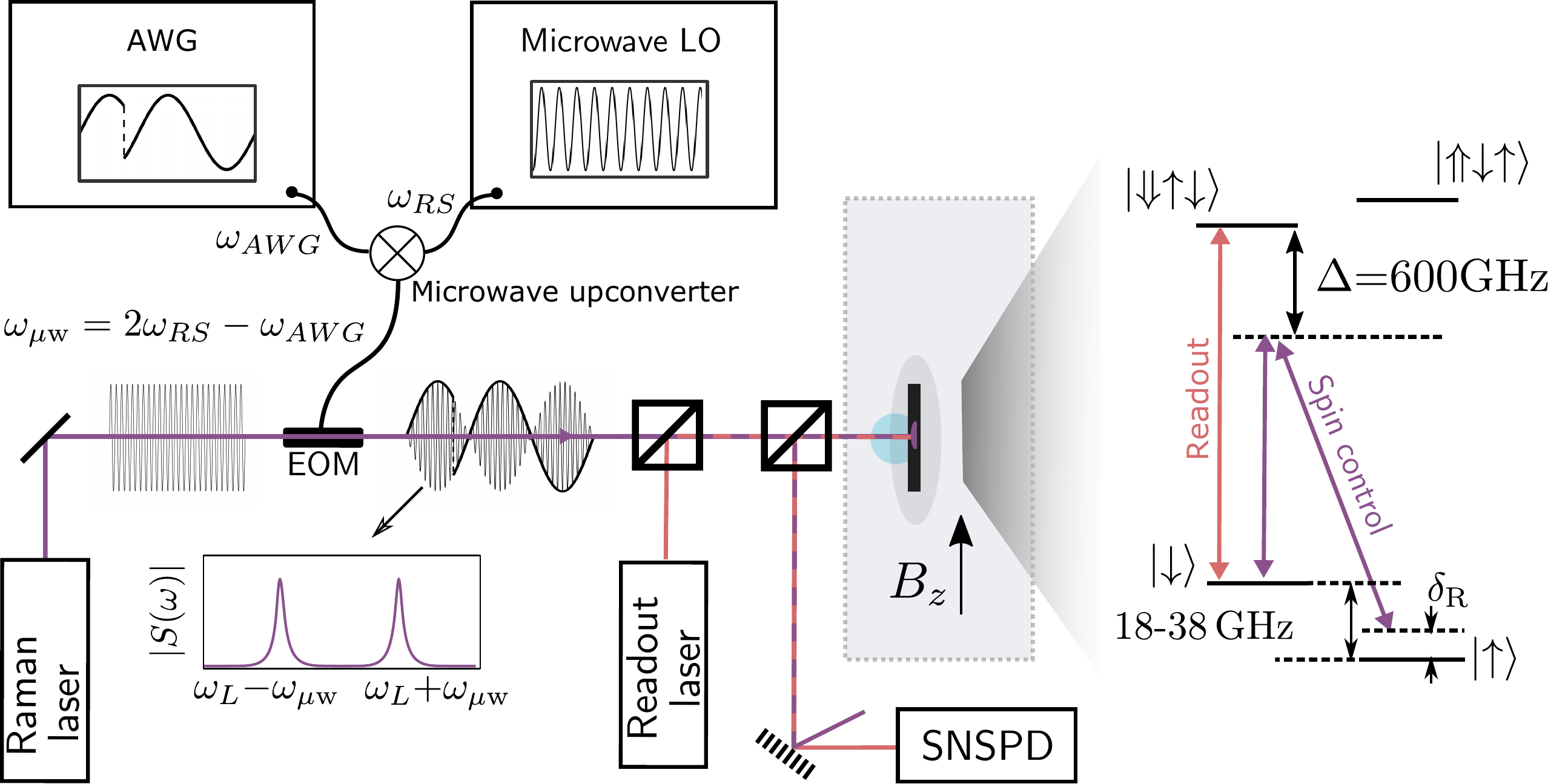}
\caption{\textbf{Experimental setup schematic:} A continuous wave Raman laser at frequency $\omega_{L}$ travels through an electro-optic modulator (EOM), where it is mixed with a microwave signal derived from the upconversion of an arbitrary waveform generator (AWG) signal by a microwave-frequency local oscillator (LO). This (to first-order) splits the Raman laser spectrally into two coherent sidebands, separated by twice the microwave frequency $\omega_{\mu \mathrm{w}}$. These are combined with a resonant readout laser, and sent to the QD. The QD is under an in-plane magnetic field, inside a bath cryostat. The Raman laser beams are far-detuned from the trion excited states, and drive a two-photon process between the electron spin states. The readout laser is resonant with the excited states. Collected light is passed through a grating, which removes background originating from the non-resonant Raman laser. We filter the resonant laser background using a crossed-polariser scheme. Detection makes use of a superconducting nanowire single-photon detector (SNSPD).}
\label{fig:expsetup}
\end{figure*}




A schematic of the experiment is shown in Fig. \ref{fig:expsetup}. A helium bath cryostat houses the QD device at 4K. A magnetic field $B_z=3.5$\,T is applied transverse to the QD growth axis (Voigt geometry). Two laser beams are combined and sent to the QD: a Raman laser system, which is microwave-modulated, and a resonant readout/repump laser (MogLabs diode laser). A cross-polarisation confocal microscope filters resonant laser background, and a grating removes non-resonant background. The filtered signal is sent to a superconducting nanowire single photon detector (SNSPD, Quantum Opus One).

\subsection{Laser system}

The Raman laser system is based on a Toptica BoosTA tapered amplifier seeded by a Toptica DL Pro diode laser. This is far-detuned from the optically excited states by $\Delta=600$\,GHz. This travels through a fibre-based EOSPACE electro-optic modulator (EOM) which is driven with a microwave waveform. The microwave waveform is generated by mixing a Rohde\&Schwarz  LO microwave source (operated at $\omega_\text{RS}\in [5,10]$\,GHz) with the signal from a Tektronix arbitrary waveform generator (AWG70001A, 25\,GS/s) (operated around $\omega_\text{AWG}=300$\,MHz). An Analog Devices wideband microwave IQ-mixer frequency-doubles the Rohde\&Schwarz signal, and downshifts it using the AWG signal. 

The first-order sidebands of the optical field after the EOM are two coherent laser fields, separated in frequency by twice the microwave drive frequency $\omega_{\mu\text{w}}$. These Raman beams pass through a quarter-wave plate, arriving at the QD with a circular polarisation. They address the electron spin states with a two-photon detuning (i.e. Raman frequency) of $\delta_\text{R}$.

\subsection{Polarisation \emph{dragging} sequence}

\begin{figure}
\includegraphics[width=\columnwidth]{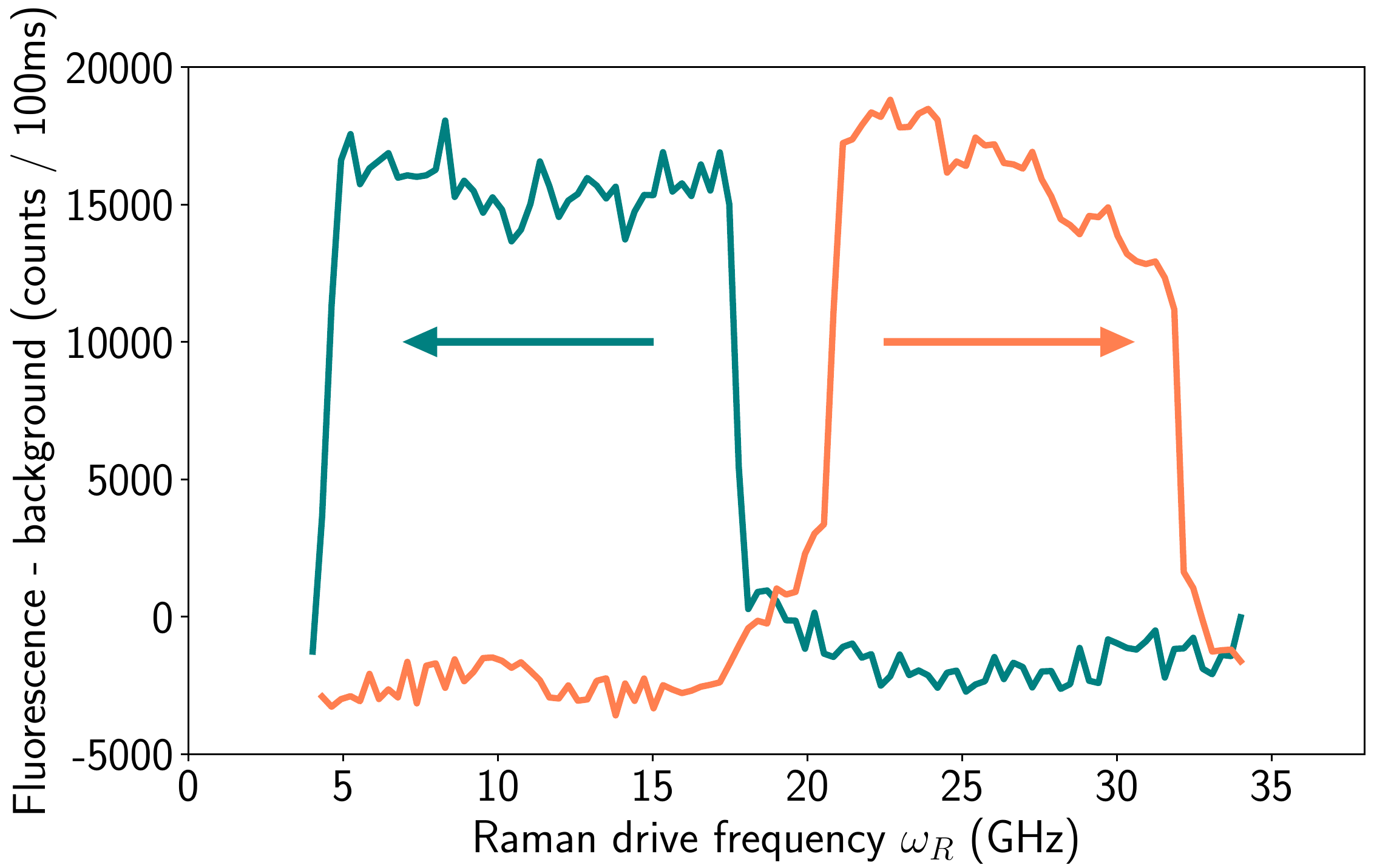}
\caption{\textbf{Raman dragging:} resonance fluorescence signal collected during the polarisation sequence shown as a function of the electron spin splitting, as measured by $\omega_\text{R} = 2\omega_{\mu \text{w}}$, for increasing (decreasing) spin splitting in orange (green). This data was taken on QD device 2, at a magnetic field $B_z = 3$\,T.}
\label{fig:dragging}
\end{figure}

We control the mean-field nuclear polarisation by tuning the frequency of the Rohde\&Schwarz LO source $\omega_\text{RS}$ whilst running the experimental sequence, which consists mostly of cooling. Tuning the LO in turn tunes $\omega_{\mu\mathrm{w}}$, altering the stable lockpoint of the feedback function (see Fig. 1 of the main text). We begin a sequence at zero polarisation, where the Raman sideband splitting corresponds to the electron Zeeman energy. From there, we increase the sideband splitting (two-photon detuning/Raman frequency $\omega_\text{R}$) by increasing the Rohde\&Schwarz frequency. The linear range of the feedback (Fig. 1e, main text) is set by the nuclear Zeeman energy ($\approx 40\,\mathrm{MHz}$). During the preparation of a polarised state, the frequency is increased in steps of 1 MHz, to stay well within this linear regime of feedback.
Also, the electron spin resonance (ESR) is increased at a rate of $0.04\ \mathrm{GHz\ s}^{-1}$. In this way, we are able to drag the electron spin resonance over a $>17$\,GHz range for QD device 1, and over a $>25$\,GHz range for QD device 2 (as in Fig.\,\ref{fig:dragging}). We compensate the unwanted shift in transition frequency to the optical excited state, due to the Overhauser shift of the ground and excited states, by applying a compensating linear ramp to the QD device gate voltage. This tunes the trion transition frequency via the DC Stark shift. Left unchecked, the generated Overhauser shift would cause our repump laser to lose resonance. A typical resonance fluorescence signal for this \emph{dragging} process is shown in Fig.\,\ref{fig:dragging}, where the falling edges indicate a loss of feedback stability on the mean-field shift, as described in the main text. When taking magnon spectra (e.g. Fig. 1 of the main text), the polarisation sequence is stopped before this loss of stability and of resonance fluorescence.

\subsection{Experimental spectroscopy sequence}

In Figs. 1f and 2a of the manuscript, we use a time-averaged measurement, with the following elementary sequence:
\begin{enumerate}
\item We cool the nuclei optically for $20\,\mu\mathrm{s}$. This step uses an optical Raman scheme detailed in \cite{Gangloff2019}. We optimise the cooling parameters by operating under conditions which maximise the electron $T_{2}^{*}$ - an inhomogeneous dephasing time, inversely proportional to the magnitude of nuclear polarisation fluctuations. The relevant settings during the cooling sequence are: a Raman Rabi frequency $\Omega_\text{R}\approx21$\,MHz and a repump Rabi frequency of $\Omega_{P}\approx0.1\Gamma_{0}/2$ for an excited state linewidth $\Gamma_{0} \approx 150$\,MHz. This gives us an optimal $T_{2}^{*}$ of $39\,\mathrm{ns}$, over an order of magnitude longer than without cooling.
\item We probe the nuclei for $5\,\mu\mathrm{s}$. In this step, we first initialise the electron spin. We then measure the electron spin-state population after a single ESR probe pulse of duration $1\,\mu\mathrm{s}$ , of electron Rabi frequency $\Omega_\text{R}\approx6.7$\,MHz  and at a given detuning $\delta_\text{R} = \omega_\text{R}^{\text{(probe)}} - \omega_\text{R}$ (i.e. relative to the preparation). The electron spin-initialisation and read-out pulses are both 100ns: a pulse from the resonant readout/repump laser excites population from $\ket{\downarrow}$ to one of the trion excited states; from there, it decays radiatively into the ground state manifold (Fig. \ref{fig:expsetup}); the pulse length is much longer than the excited state lifetime. The collected fluorescence is proportional to the population of $\ket{\downarrow}$; and at the end of this readout/repump pulse, the electron is into $\ket{\uparrow}$. The remainder in this step is “dead-time”, where no control pulses act on the electron.
\end{enumerate}
This elementary sequence (prep+single probe of the spectrum at a fixed $\delta_R$) is repeated $\sim$ 2,400,000 times per data point (60s of integration). Then we step $\delta_{R}$ by 4MHz, and perform another time-averaged measurement at this new detuning. 

Figure 2b presents a measurement which also alternates a cooling and a probe step, but where we instead scan the length of the ESR probe pulse at a fixed $\delta_{\text{R}}$. During this measurement, we pair two pulses of increasing and decreasing lengths such that total Raman power is conserved, allowing us to stabilise the time-averaged laser power to within 1\%.

We keep the time ratio of cooling-to-spectroscopy around 20-to-5, meaning that we spend 80\% of the time cooling the nuclear ensemble. This ensures that the steady-state of the nuclear ensemble is primarily determined by the cooling sequence \cite{Jackson2020}.

\subsection{Data processing and normalisation}

\subsubsection{Background subtraction}\label{t1bgd}

Our raw measurements of the magnon spectrum include a constant background count rate, as determined by the readout signal we obtain at large Raman detuning $\delta_\text{R}$. This is due to optically induced electron spin relaxation, via a mechanism external to the QD \cite{Bodey2019}. We subtract this constant background, which is equal to the count rate at large Raman detuning $\delta_\text{R}$, from our data. When integrated over a typical spectrum, the background count rate constitutes $22\%$ of the uncorrected signal.

\subsubsection{Polarisation-dependent count rate}

We observe a systematic decrease in count rate as the nuclear ensemble is polarised so as to increase the mean-field shift on the ESR. This is because the collected readout fluorescence passes through a narrowband ($\sim 20$\,GHz) optical grating in order to remove the laser background from the far-detuned Raman beams. When we polarise, Raman photon emission will change in frequency due to the change in ESR splitting. The collection efficiency of Raman photons therefore decreases as nuclear polarisation increases from zero. In order to correct for this effect, we normalise data taken at each polarisation by its average count rate.

\subsubsection{From count rate to electron-spin population}

Throughout the main text, read-out fluorescence counts have been converted to electron $\ket{\downarrow}$ population. In Figs.\,1, 2(a), and 3 we use the fit to the time evolution of the magnon spectrum (later presented in Fig. \ref{fig:2d_decon}) to fix the count rate which corresponds to 50\% $\ket{\downarrow}$ population. In Fig. 2(b), we assume that the system, when on resonance with the ESR, will reach a steady-state of 50\% $\ket{\downarrow}$ population at long drive time.



\section{Model}\label{sec:model}
\subsection{System Hamiltonian}
The effective Hamiltonian of the system for a single spin species is given by \cite{Gangloff2019}:
\begin{equation}\label{full_hamiltonian_conclusion}
\begin{split}
\hat{H}=&\overbrace{\delta \hat{S}_z + \Omega \hat{S}_x}^{\mathrm{ESR~drive}} +\overbrace{\omega_\text{n} \sum_j  \hat{I}_z^j}^{\mathrm{Nuclear~Zeeman~splitting}}\\& - \overbrace{\sum_j a^j \hat{I}_z^j \hat{S}_z}^{\mathrm{Hyperfine~interaction}} +\overbrace{\hat{H}^0_Q+\hat{V}_Q^{\prime\prime}}^{\mathrm{Quadrupolar~interaction}}
\end{split}
\end{equation}
The first two terms model the ESR drive, in the frame rotating with the frequency of the drive. The third term introduces a nuclear Zeeman interaction. The fourth term stands for the hyperfine interaction between the central electron and the nuclei, in the limit of a high external magnetic field. The hyperfine constant \textit{per nucleus} $a^j$ is taken constant for a given species (a box approximation: $a=A/N$). The sums run over the nuclei of that species.

Moreover:
\begin{equation}
\begin{split}
\hat{H}^0_Q&=\sum_j \frac{B_Q^j}{2}(2\sin^2 \theta-\cos^2 \theta)\hat{I}_z^{j2}\\
\hat{V}_Q^{\prime \prime}&= -\Omega \hat{S}_y [(\hat{\Phi}_{+1}+\hat{\Phi}_{-1})+(\hat{\Phi}_{+2}+\hat{\Phi}_{- 2})]
\end{split}
\end{equation}
are the terms originating from the strain-induced quadrupolar interaction, that offset the quantisation axis, thus introducing a correction to the hyperfine interaction.

The term $\hat{V}_Q^{\prime \prime}$, derived perturbatively via the Schrieffer-Wolff transformation, enables the coherent electron-nuclear exchange. In particular, the magnon injection is effectuated by the action of the spin-wave operators:
\begin{equation}\label{full_h_creation_annihilation}
\begin{split}
\hat{\Phi}_{\pm 1} &\equiv \sum_j\mp i\underbrace{\sin 2 \theta \Big(\frac{aB^j_Q}{2\omega_\text{n}^2}\Big)}_{\equiv \alpha_1}[\hat{I}^j_\pm \hat{I}^j_z+ \hat{I}^j_z\hat{I}^j_\pm] \equiv \sum_j \hat{\Phi}^j_{\pm 1} \\
\hat{\Phi}_{\pm 2} & \equiv \sum_j \mp i \underbrace{\frac{1}{2}\cos^2\theta\Big(\frac{aB^j_Q}{2\omega_\text{n}^2}\Big)}_{\equiv\alpha_2} \hat{I}^{2j}_\pm \equiv \sum_j \hat{\Phi}^j_{\pm 2}\\
\end{split}
\end{equation}
Where we distinguished collective and single-spin operators $\hat{{\Phi}}_{\pm k}$ and $\hat{{\Phi}}^j_{ \pm k}$, respectively, for processes changing the nuclear polarisation by one ($k=1$) or two ($k=2$) units. 

Within a single-spin manifold transition between $\frac{1}{2}$ and $-\frac{1}{2}$
is not allowed due to the form of $\hat{{\Phi}}^j_{\pm 1}$.

\subsection{Relating Exchange frequencies to spin-state populations and polarisation}
Action of the $\hat{{\Phi}}_{\pm k}$ operator on the initial (pure) state $\ket{M^{(0)}}$ will bring it to
\begin{equation}
\ket{M^{(1)}}=\frac{\hat{{\Phi}}_{\pm k}\ket{M^{(0)}}}{\sqrt{\bra{M^{(0)}}\hat{{\Phi}}_{\mp k}\hat{{\Phi}}_{\pm k}\ket{M^{(0)}}}}\,.
\end{equation}
At times when dynamics are restricted to the $\{\ket{M^{(0)}},\ket{M^{(1)}}\}$ manifold, the electron-nuclear exchange frequency $\Omega_{\pm k}$ is proportional to the matrix element of the $\hat{\Phi}_{\pm k}$ operator
\begin{equation}
\Omega_{\pm k}=\Omega \bra{M^{(1)}} \hat{{\Phi}}_{\pm k} \ket{M^{(0)}}\,,
\end{equation}
where $\Omega$ is the Rabi frequency of the ESR drive.
Assuming that the system is initialised in a pure product state:
\begin{equation}\label{product_state_considered}
\ket{M^{(0)}}=\bigotimes_j \ket{m^j}
\end{equation}
the exchange frequency is found as
\begin{equation}
\Omega_{\pm k}^2=\Omega^2\sum_j |\bra{m^j \pm k} \hat{{\Phi}}^j_{ \pm k} \ket{m^j}|^2
\end{equation}
Across a single spin-$I^{j}$ species, the above sum can be reduced to a sum over single spin states weighed by their fractional populations $n_{I^{j}}$, $n_{I^{j}-1}$, ...,  $n_{-I^{j}}$, such that
\begin{equation}\label{exc_freq8}
\Omega_{\pm k}^2=N\alpha_k^2 \Omega^2 \sum^I_{m=-I} n_m |P^{(k)}_{\pm}(I^{j},m)|^2 \equiv \eta_{\pm k}^2 \Omega^2\,,
\end{equation}
where constants $\alpha_k$ were defined in Eq.\,\ref{full_h_creation_annihilation}, and $P^{(k)}_{\pm}(I^{j},m^{j})=\bra{m^j \pm k} \hat{{\Phi}}^j_{ \pm k} \ket{m^j}/(N\alpha_k)$ are numerical pre-factors resulting from the form of the single-spin magnon laddering operators, given explicitly by
\begin{equation}
\begin{split}
P^{(1)}_{\pm}(I^{j},m^j)=&(2m^j\pm 1)\sqrt{I^{j}(I^{j}+1)-m^j(m^j\pm 1)} \\
P^{(2)}_{\pm}(I^{j},m^j)=&\sqrt{I^{j}(I^{j}+1)-(m^j\pm 2)(m^j\pm 1)}\\&\cross\sqrt{I^{j}(I^{j}+1)-m^j(m^j\pm 1)}
\end{split}
\end{equation}
It should be noted that away from zero polarisation, the enhancement factors $\eta_{\pm k}$ differ among the $\pm k$ sideband transitions, and directly reflect the populations of single spin states.

One can relate the exchange frequencies for all four sideband processes to the polarisation of partaking spin owing to the identity:
\begin{equation}
\begin{split}
&2(|P^{(2)}_{+}(I^{j},m^{j})|^2-|P^{(2)}_{-}(I^{j},m^{j})|^2)\\&+(|P^{(1)}_{+}(I^{j},m^{j})|^2-|P^{(1)}_{-}(I^{j},m^{j})|^2)=\\
&-2m^{j}(4I^{j2}+4I^{j}-3)
\end{split}
\end{equation}
which translates to: 
\begin{equation}
2\frac{\Omega_{+2}^2-\Omega_{-2}^2}{N\alpha_2^2\Omega^2}+\frac{\Omega_{+1}^2-\Omega_{-1}^2}{N\alpha_1^2\Omega^2}=-2I^{j}(4I^{j2}+4I^{j}-3)I_z/I_z^{\mathrm{max}}
\end{equation}

\subsubsection{Exact treatment of $I^{j}=3/2$}

For the particular case of $I^{j}=\frac{3}{2}$, Eq.\,\ref{exc_freq8} gives:
\begin{equation}
\begin{split}
\Omega_{+2}&=\sqrt{12N}\Omega \cos^2 \theta \frac{a B_Q}{4\omega_\text{n}^2}\sqrt{n_{-3/2}+n_{-1/2}}\\
\Omega_{+1}&=\sqrt{12N}\Omega \sin 2\theta \frac{a B_Q}{2\omega_\text{n}^2}\sqrt{n_{-3/2}+n_{1/2}}\\
\Omega_{-1}&=\sqrt{12N}\Omega \sin 2\theta \frac{a B_Q}{2\omega_\text{n}^2}\sqrt{n_{-1/2}+n_{3/2}}\\
\Omega_{-2}&=\sqrt{12N}\Omega \cos^2 \theta \frac{a B_Q}{4\omega_\text{n}^2}\sqrt{n_{1/2}+n_{3/2}}\\
\end{split}
\end{equation}
where we substituted back in the exact expressions for $\alpha_k$ from Eq.\,\ref{full_h_creation_annihilation}. Since $\sum_m n_m =1$, the assumption of an initial product state leads to the emergence of two constants of motion:
\begin{equation}\label{constsofmotion}
\begin{split}
\Omega^2_{+2}+\Omega^2_{-2}&=12 N \Omega^2 \cos^4 \theta \Big (\frac{a B_Q}{4\omega_\text{n}^2}\Big)^2\\
\Omega^2_{+1}+\Omega^2_{-1}&=12 N \Omega^2 \sin^2 2\theta \Big (\frac{a B_Q}{2\omega_\text{n}^2}\Big)^2
\end{split}
\end{equation}
This motivates the definition of a self-referenced dimensionless parameter - the magnon asymmetry - as in Eq.\,3 of the main text. 

Magnon asymmetries for both first and second sideband processes are linked to the single spin state populations via:
\begin{equation}
\begin{split}
\nu_1&= n_{3/2}+n_{-1/2}-n_{1/2}-n_{-3/2}\\\nu_2&=n_{3/2}+n_{1/2}-n_{-1/2}-n_{-3/2}
\end{split}
\end{equation}
which is used to arrive at the expressions in Eqs.\,4 and \,5 of the main text.

The constants of motion (Eq.\,13) are crucial to go beyond Eq.\,11 and establish a self-referenced measure of the polarisation of a spin $I^{j}=3/2$, for any distribution of population. For a spin $I^{j}>3/2$, the quantity $(\Omega^2_{+k}+\Omega^2_{-k})$ is no longer constant in general, and we must assume a type of population distribution in order to link the magnon sideband asymmetry to polarisation.

\subsubsection{Treatment of $I^{j}>3/2$ with thermal-state approximation}

Equation 5 of the main text, which equates the measured quantity $\frac{1}{3}(2\nu_2+\nu_1)$ to the single-species polarisation $\mathcal{I}_{z,\frac{3}{2}}^\star$, is a definition that holds true only for $I^j=\frac{3}{2}$. For $I^j>\frac{3}{2}$ we still measure the same quantity ($\frac{1}{3}(2\nu_2+\nu_1)$), but as we will show below, in this case it slightly underestimates the polarisation. 

Our definition imposes $\mathcal{I}_{z,I^j}^\star$ to match $\mathcal{I}_{z,I^j}$ for a thermal state of any spin $I^j$. We use this to generalise $\mathcal{I}_{z,\frac{3}{2}}^\star$ to higher-spin species. For a thermal state we can straightforwardly
calculate $\mathcal{I}_{z,I^j}$ (which is our definition of $\mathcal{I}_{z,I^j}^\star$) and $\frac{1}{3}(2\nu_2+\nu_1)$ (which is the experimentally measured parameter). Plotting the former against the latter yields Fig. S3, and gives us the high-spin correction to Eq. 5 of the main text. Figure S3 shows that $\mathcal{I}_{z,\frac{3}{2}}^\star$ is an underestimate of the polarisation $\mathcal{I}_{z,\frac{9}{2}}^\star$.

We have used this function to reconstruct the polarisation of indium ($I^{j}=9/2$) shown in Fig.\,4a of the main text, from measurements of its magnon mode asymmetry.
The feedback-induced deviations of the spin-state populations from those of a thermal state, already small for spin-$\frac{3}{2}$ species (see Fig. 3b, main text), should be even smaller for a spin-$\frac{9}{2}$ species such as indium. This scaling operation (up to a factor 2) thus presents a more realistic estimate of the indium contribution to the Overhauser field. Treating indium as a spin-$\frac{3}{2}$ species instead (no scaling), the asymmetry-commensurate Overhauser field would exceed the classical bound by 1.9(1).

\begin{figure}
\includegraphics[width=0.9\columnwidth]{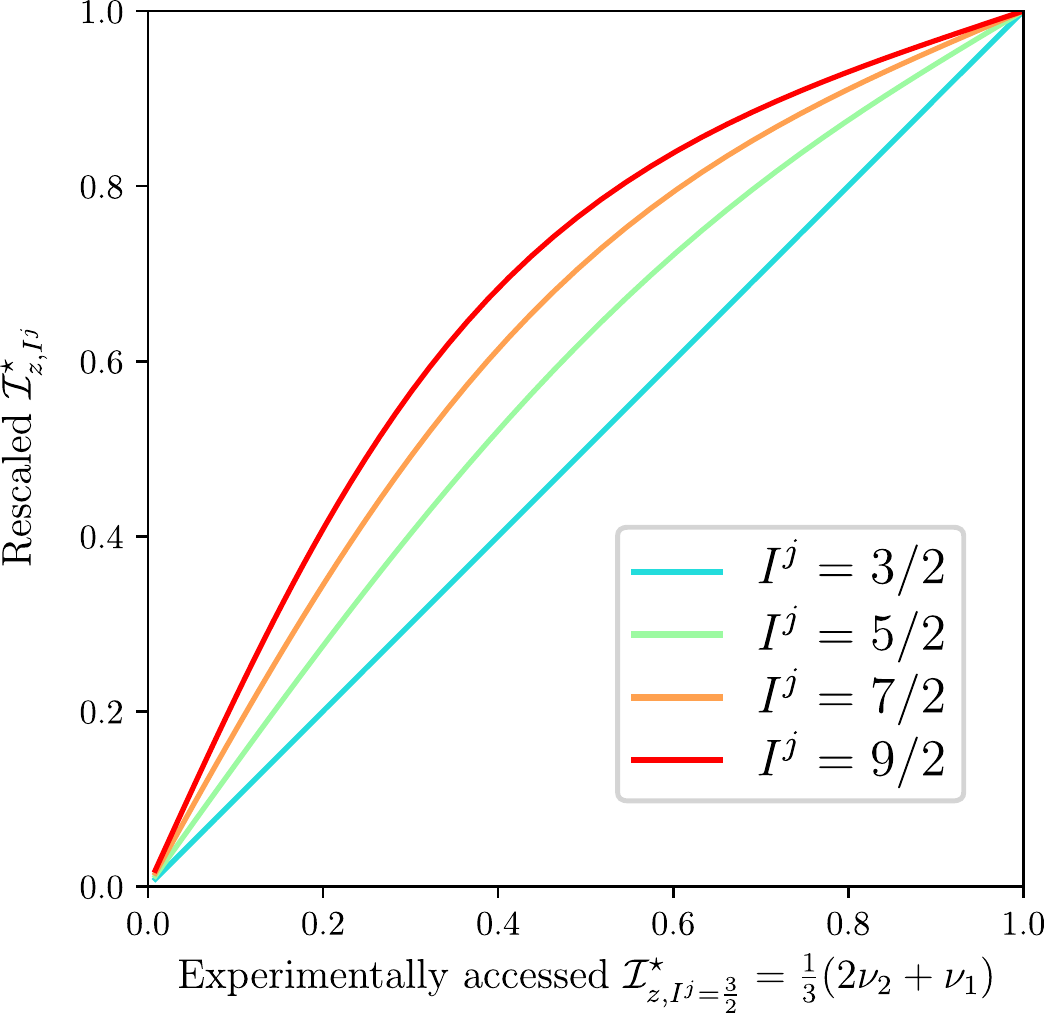}
\caption{\textbf{High-spin corrections to the spin-$\frac{3}{2}$ model:} Rescaling of the experimentally accessed asymmetry-commensurate polarisation for a high-spin nucleus.}
\label{fig:Ipol}
\end{figure}


\subsubsection{Mixed state}

The above considerations are generalised easily to arbitrary states (including mixed states), through:
\begin{equation}
\Omega^2_{\pm k}= \Omega^2\mathrm{Tr}\hat{\rho}_n\hat{{\Phi}}_{\mp k}\hat{{\Phi}}_{\pm k}
\end{equation}
where $\hat{\rho}_n$ stands for a density operator, representative of the nuclear state of the system. In the case of a mixture of product states considered in Eq. \ref{product_state_considered}, the statements involve the expectation values of polarisation, and manifold imbalances. The consequence of non-vanishing steady-state coherences is discussed in section \ref{steady_state_coherences}.

\section{Complementary Measurements}
\subsection{Homogeneous dephasing}\label{sec:hahnecho}


\begin{figure}
\includegraphics[width=1\columnwidth]{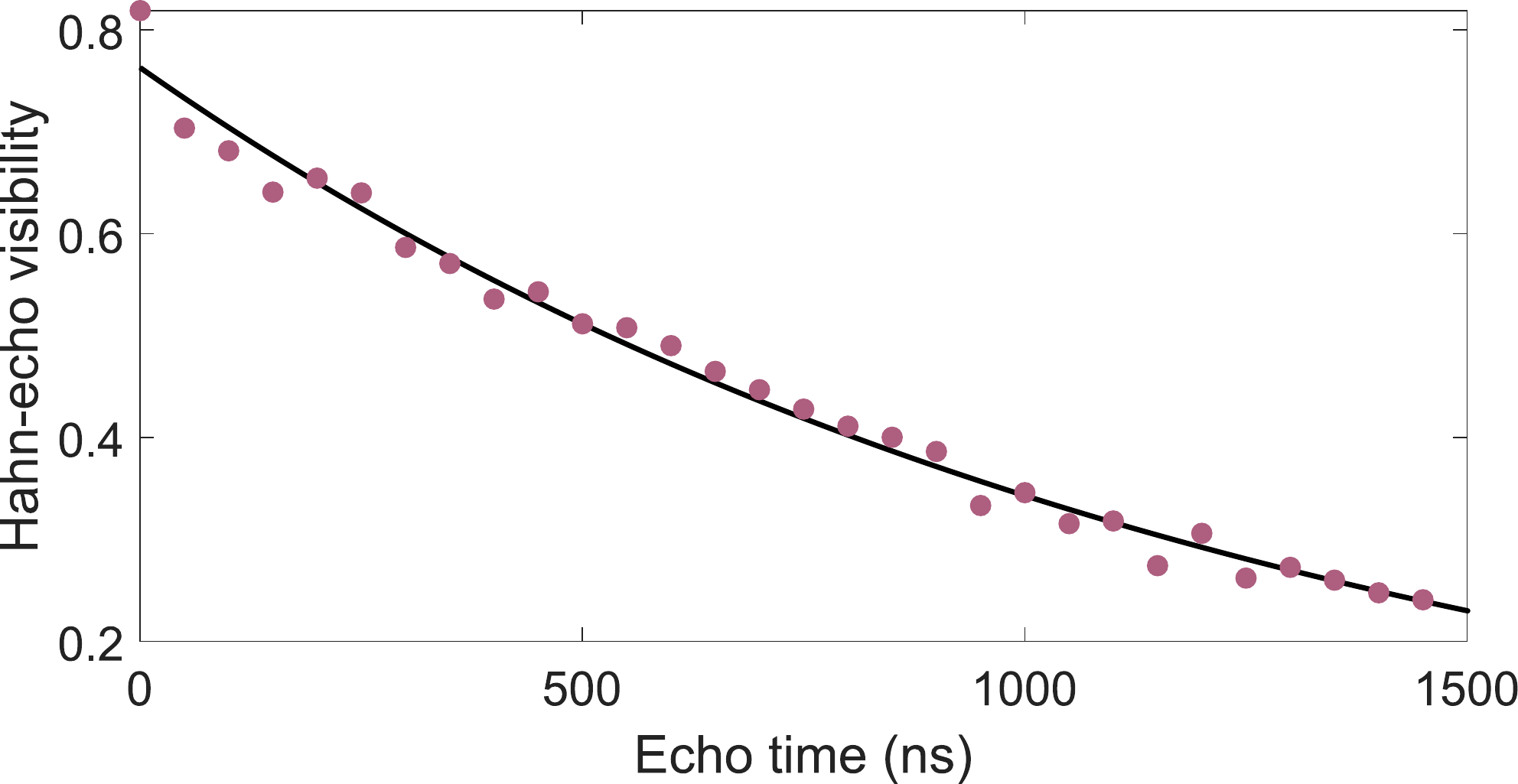}
\caption{\textbf{Hahn echo measurement:} Extracted visibility as a function of echo time (pink circles). Fitting to an exponential decay (black curve) yields a dephasing time of $T_{\mathrm{HE}}=1251\pm62\ \mathrm{ns}$.}
\label{fig:hahnecho}
\end{figure}


We measure the electron spin dephasing time $T_{\mathrm{HE}}$ using the Hahn echo sequence, which consists of two $\frac{\pi}{2}$ pulses separated by an echo time, and a refocusing $\pi$ pulse placed at half the echo time. The $\pi$ pulse filters noise which is static over the duration of the pulse sequence, and tuning the phase of the final $\frac{\pi}{2}$ pulse allows the remaining coherence of the electron spin to be measured. Figure \ref{fig:hahnecho} depicts the dependence of the Hahn echo visibility on echo time yielding a Hahn-Echo time of $1.2$\,$\mu$s.


\subsection{Inhomogeneous dephasing}\label{t2starmeasurement}

\begin{figure}
\includegraphics[width=0.8\columnwidth]{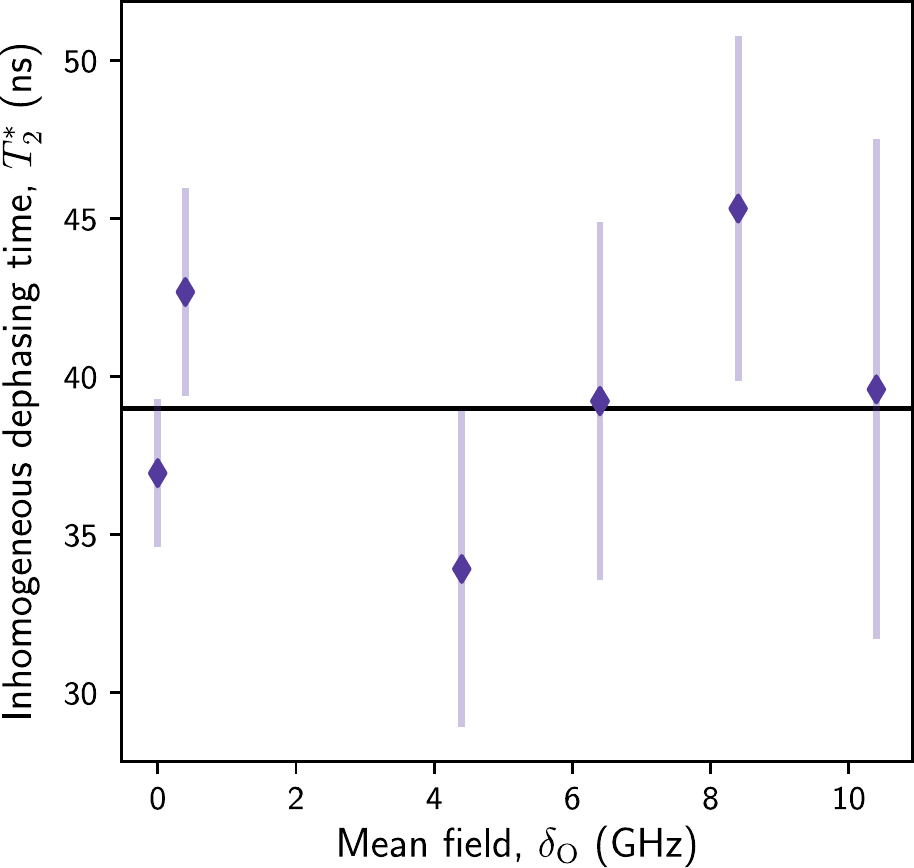}
\caption{\textbf{Ramsey measurement:} Fitted free induction decay time of the electron spin, measured using Ramsey interferometry (purple circles). This is plotted as a function of Overhauser shift. Error bars indicate $\pm1\sigma$. The black curve is a constant fit at $T_{2}^{*}=39\pm1.5~[\mathrm{ns}]$.}
\label{fig:ramsey}
\end{figure}

We measure $T_2^*$ using Ramsey interferometry \cite{Stockill2016}. In the time domain, we observe a Gaussian decay profile of the electron spin coherence which we fit with $\propto e^{-(t/T_2^*)^2}$. 
Figure \ref{fig:ramsey} presents the values measured with increasing polarisation. Our $T_{2}^{*}$ data is polarisation-independent within error bars and equal to $T_{2}^{*}=39\pm1.5~[\mathrm{ns}]$ (solid line in Fig. S4).

The inhomogeneous dephasing time, $T_{2}^{*}$, appears throughout the experiment as a limit on spectral resolution: the effect of the inhomogeneous dephasing on the magnon spectrum is equivalent to that of a Gaussian convolution filter, with width:
\begin{equation}\label{widtht2*}
\sigma_\delta =\frac{\sqrt 2 }{2\pi T_2^*}
\end{equation}
$T_{2}^{*}$ is set by the performance of the nuclear preparation - see section \ref{sec:Phase_space_flow} for a complete discussion of mechanisms that influence it. 

\subsection{Hyperspectral map of magnon modes}
We have measured the magnon spectrum as a function of the Raman drive time from $0$ to $1.5$\,$\mu$s, at a near-zero mean-field shift. The resulting two-dimensional map, shown in the left-most panel of Fig.\,\ref{fig:2d_decon}, shows the electron-spin population as a function of ESR detuning $\delta$ and Raman drive time. This data is used to constrain the dephasing of magnon modes in the magnon spectrum modelling, as analysed in section
\ref{sec:magnon_spec_t}.

\subsection{Nuclear spin decay}\label{sec:nucleart1}
\begin{figure}
\includegraphics[width=\columnwidth]{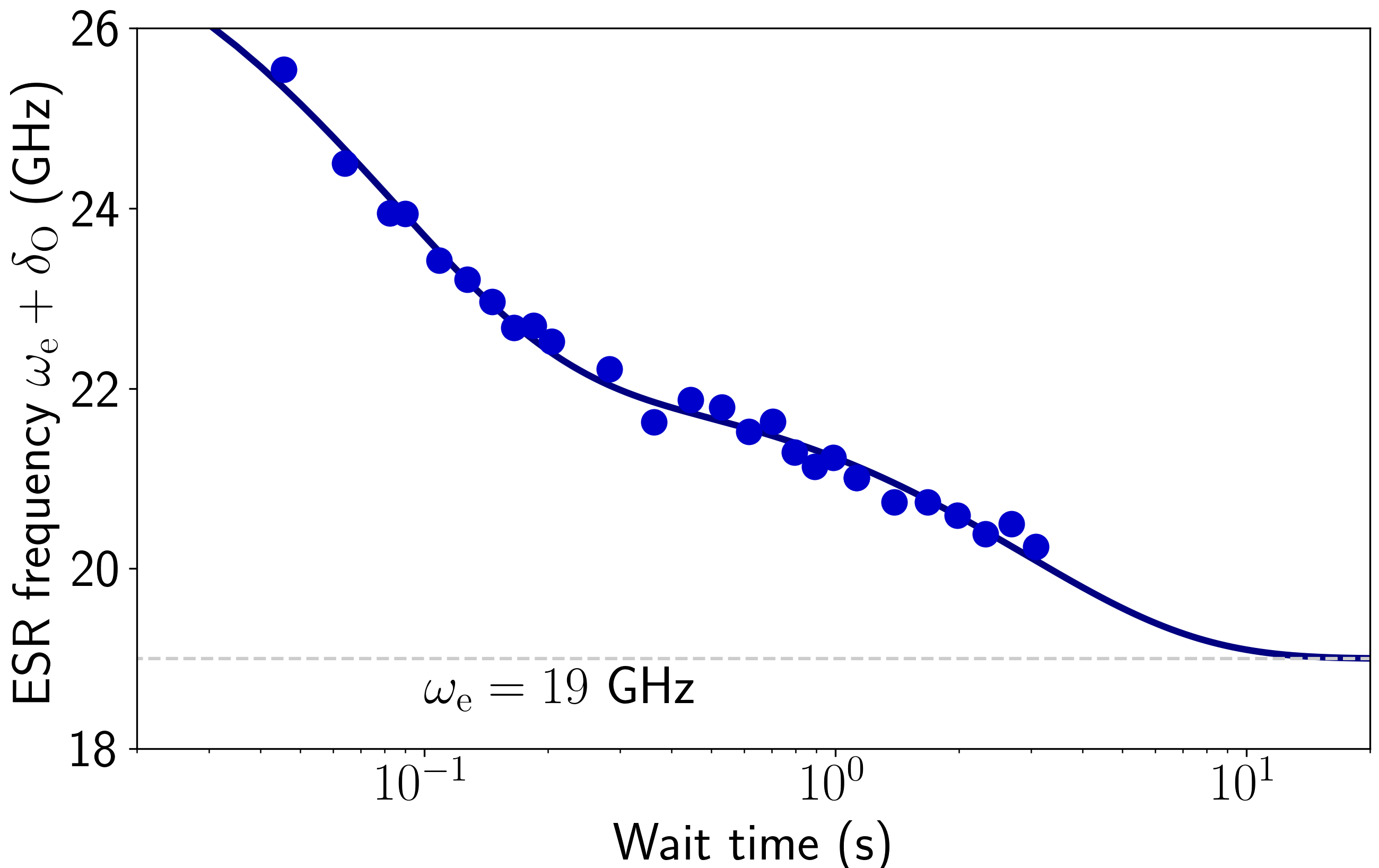}
\caption{\textbf{Mean-field relaxation measurement:} Electron spin splitting as a function of a waiting time $t$, during which the mean nuclear field relaxes. This data was taken on QD device 2, at a magnetic field $B_z = 3$\,T. The solid curve is a fit to the bi-exponential function $\omega_\text{R}(t) = A\exp(-t/\tau_1) + B\exp(-t/\tau_2) + 19$\,GHz.}
\label{fig:nucleart1}
\end{figure}

Using QD device 2 at a field $B_z = 3$\,T, we measure the relaxation of the mean-field shift following its preparation at a $\sim 25.5$\, GHz electron-spin splitting (the electron-spin splitting at zero nuclear polarisation is $19$\,GHz). The cooling beams (Raman and resonant) are turned off and a waiting time from $0.05$ to $3$\,s follows. The resonant beam is then turned on again as the gate is swept across all four trion state resonances, allowing us to infer the electron spin splitting spectrally. The result is shown in Fig.\,\ref{fig:nucleart1}, where the data clearly exhibits two relaxation timescales. Using a biexponential fit, we find that the initial fast decay of $\sim 50\%$ of the mean-field polarisation occurs over a characteristic time $\tau_1 = 80\pm 7$\,ms, while the slower decay occurs over a characteristic time $\tau_2 = 2.9 \pm 0.2$\,s.

This fitted timescale $\tau_1$ is relevant to our modelling of the cooled nuclear spin populations in section \ref{fokker_planc_section}, where it sets a constraint on the nuclear spin diffusion rate $\Gamma_\text{nuc} \sim 1/\tau_1$ that limits the cooling.

The fitted timescale $\tau_2$ exceeds by a factor of $10^6$ our electron-spin relaxation time of a few tens of $\mu$s; this process is dominated by coupling to the back-contact (tunnel barrier of 35\,nm in our devices, see section\,\ref{sec:device}). For identical magnetic field of $3.5$\,T, this is in close quantitative agreement with recent reports on the ratio of electron and nuclear spin relaxation times for tunnel barriers below 42\,nm \cite{Gillard2021}. The parameter $\tau_2$ is not used further in our analysis.

\begin{figure}
\includegraphics[width=\columnwidth]{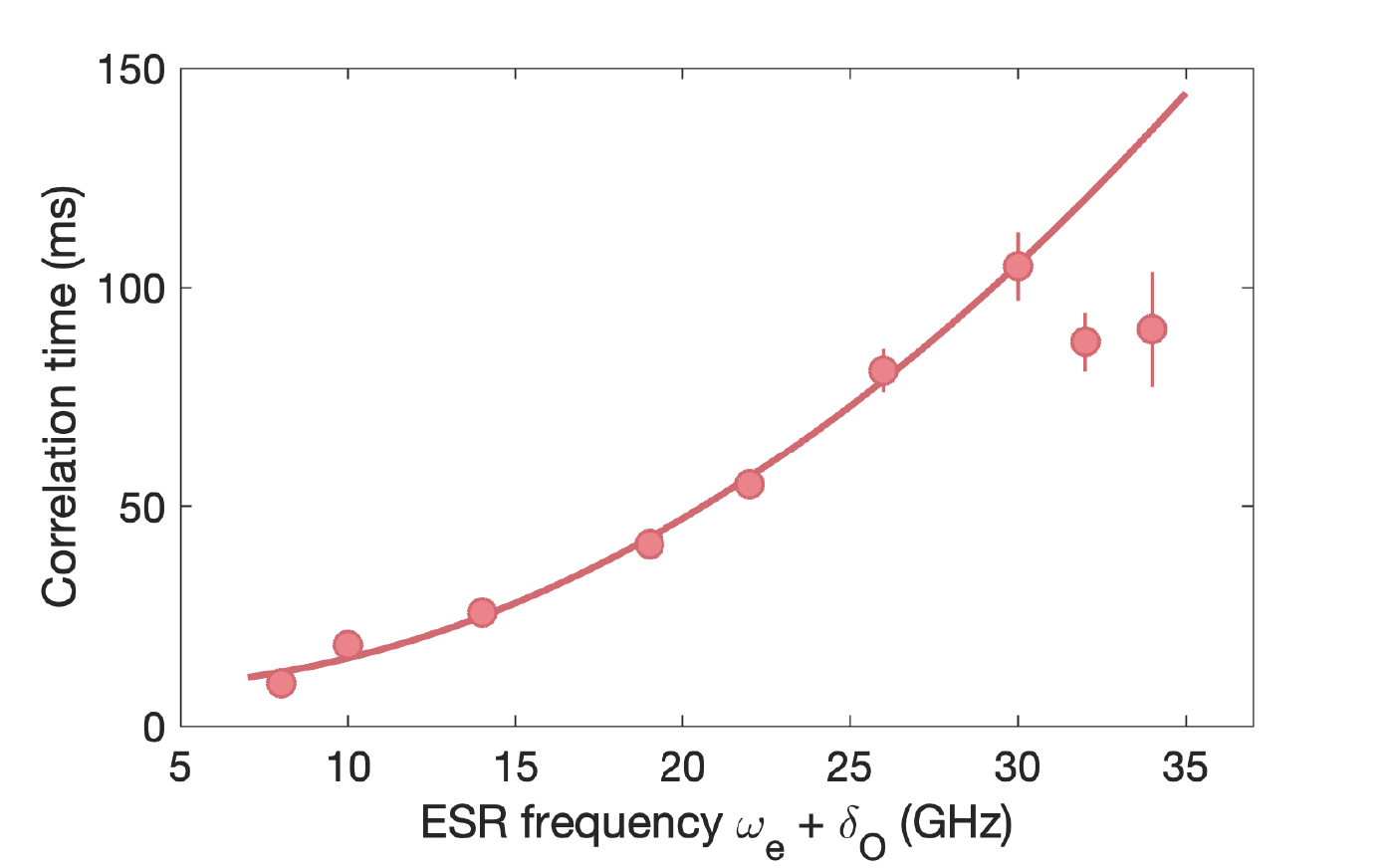}
\caption{\textbf{Nuclear correlation time:} Characteristic correlation time of the mean-field fluctuations, $\sim 1/T_2^*$, as a function of the electron-spin splitting. This data was taken on QD device 2, at a magnetic field $B_z = 3$\,T. The solid curve is a quadratic function $a\omega_\text{R}^2 + b\omega_\text{R} + c$, $a=0.13(3)$\,ms/GHz$^2$, $b=-1(1)$\,ms/GHz, $c=10(9)$\,ms. Error bars indicate a $67\%$ confidence interval.}
\label{fig:t2starrelax}
\end{figure}

We also measure the characteristic relaxation time of the mean-field fluctuations, which are reduced from thermal equilibrium by our optical cooling, as a function of the mean field. Following the preparation of a variable electron spin splitting using our polarisation sequence, we turn the cooling beams off, introduce a wait time up to 200\,ms, and measure the electronic $T_2^*$ using a Ramsey interferometry sequence. We find that the variance of the mean-field, $\sim (1/T_2^*)^2$, relaxes exponentially with a characteristic \emph{correlation} time \cite{Ethier-Majcher2017}. This correlation time is shown as a function of the prepared electron spin splitting in Fig.\,\ref{fig:t2starrelax}. We find that the correlation time for mean-field values leading to electron-spin splittings of $22-25$\,GHz is the same as the $\tau_1 \sim 80$\,ms measured as a fast decay of the mean field itself in Fig.\,\ref{fig:nucleart1}. We also see that, with the exception of extremal points close to regions of instability, the correlation grows quadratically with the electron spin splitting (solid curve in Fig.\,\ref{fig:t2starrelax}). Taken together, these observations point to a fast nuclear relaxation mechanism for the mean-field statistics (both mean and variance) governed by electronic degrees of freedom; a plausible candidate is the three-body RKKY-type interaction that arises in the second-order of the hyperfine interaction \cite{Wust2016}.

\subsection{ESR offset, $\epsilon$}\label{sec:epsilon}

\begin{figure}
\includegraphics[width=1\columnwidth]{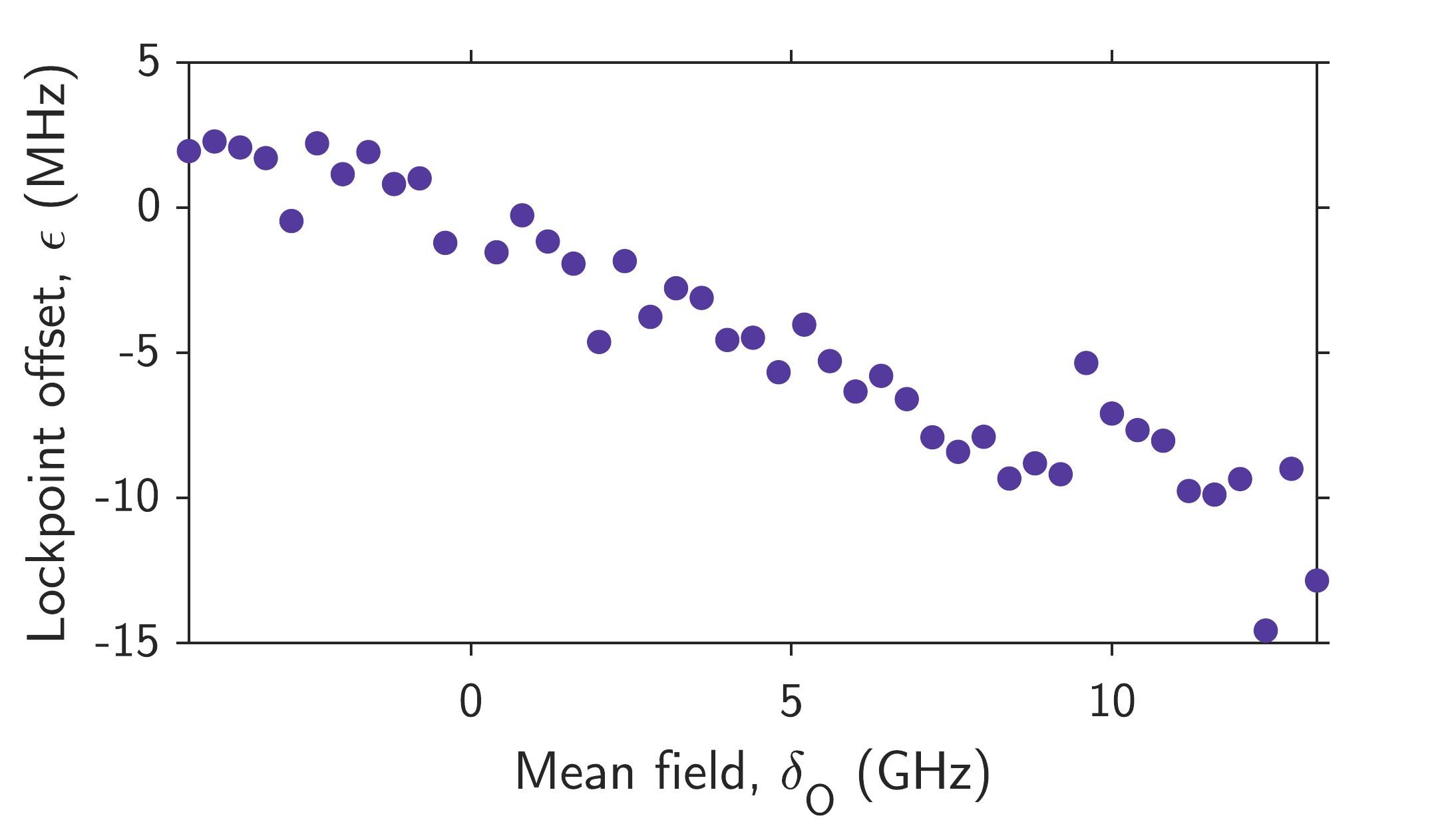}
\caption{\textbf{ESR offset:} Spectral offset, $\epsilon$, fitted as common-mode offset on all sideband and ESR features on unfiltered spectra from hyperspectral map (Fig.\,1f main text), as a function of mean-field shift.}
\label{fig:epsilon}
\end{figure}

The hyperspectral magnon map shown in Fig.\,1f of the main text, as a function of Raman detuning $\delta_\text{R}$ and mean-field shift $\delta_{O}$, exhibits a MHz-scale offset that scales linearly with the GHz-scale mean-field shift -- owing to a nuclear polarisation-induced shift in the nuclear feedback curve (main text Fig.\,1e). In more detail, we extract $\epsilon$ from a fit to the unfiltered spectra of Fig.\,1f, as an overall offset common to all sideband and ESR features. We show $\epsilon$ as a function of mean-field shift in Fig.\,\ref{fig:epsilon}.

The \emph{ESR detuning} is then defined as a corrected Raman detuning: $\delta = \delta_{R} - \epsilon$.


\section{Supplementary notes on data analysis}

\subsection{Wiener deconvolution}\label{sec:wiener}

\begin{figure}
	\includegraphics[width=0.8\columnwidth]{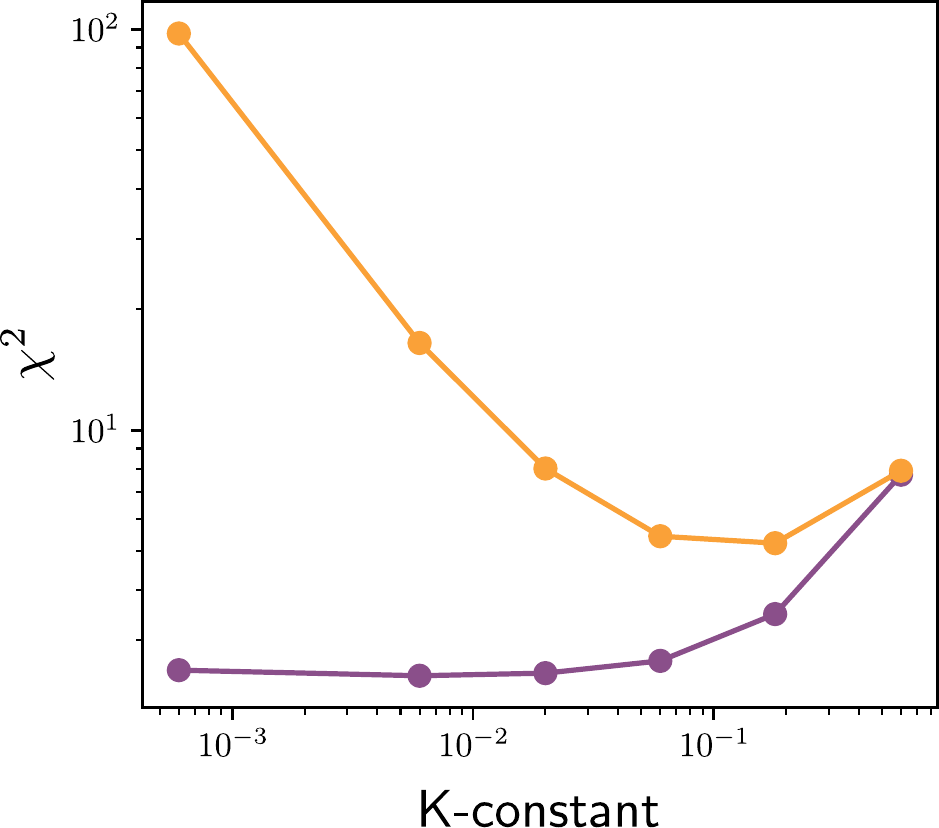}
	\caption{\textbf{Assessment of the Wiener filtering performance}. The orange curve illustrates the dependence of the sum of the squared residuals of the model fitted to the deconvolved signal (Eq.\,\ref{chisq}), as a function of the $K$-value used in the Wiener filter to perform this deconvolution. The purple curve indicates the same quantity obtained from the unprocessed data and the convolved model.}
	\label{fig:optWiener}
\end{figure}

Fitting the magnon spectrum with a model that accounts for an electron $T_2^*$ process -- i.e. a convolution with a Gaussian profile of width given by Eq.\,\ref{widtht2*} -- is computationally intensive. Since $T_2^*$ is known through an independent measurement (\hyperref[t2starmeasurement]{section \ref{t2starmeasurement}}), its effect can be factored out by deconvolving prior to fitting. This significantly reduces the computational cost of the fitting procedure. In the face of noise inherent to any real measurement, deconvolution has to be done with great care. Here we outline how this can be optimally achieved.

The Fourier transform of the data - $\tilde{D}(q)=\int\,d\delta e^{-i2\pi q \delta}D(\delta)$ - is expected to be of the form:
\begin{equation}
\tilde{D}(q)=\tilde{G}(q)\tilde{S}(q)+\tilde{N}(q)
\end{equation}  
where $\tilde{G}(q)$ and $\tilde{S}(q)$ are the Fourier transforms of the Gaussian with width $\sigma_\delta$ and of the signal that we wish to extract, respectively. 

Multiplying $\tilde{D}(q)$ directly by the inverse Gaussian filter: 
\begin{equation}
\tilde{G}^{-1}(q)=e^{2\pi^2 \sigma_\delta^2q^2}
\end{equation}
where $\sigma_\delta$ is taken from Eq.\,\ref{widtht2*}, would amplify the high-frequency part of noise $\tilde{N}(q)$, and would distort the filtered signal $\tilde{S}(q)$. Instead, we use the Wiener filter \cite{wiener2013extrapolation}:
\begin{equation}
\tilde{W}(q)=\frac{\tilde{G}^*(q)}{|\tilde{G}(q)|^2+K}
\end{equation}
where the constant $K$ should be set close to the ratio of noise-to-signal spectral densities ($|\tilde{N}|^2/|\tilde{S}|^2$). Without \textit{a priori} knowledge of the spectral noise density, we find $K$ by optimizing the filtering performance. 

The main contributions to $\tilde{N}$ come from drifts that we take as uncorrelated with signal. This guarantees that for the optimal $K$, the mean-square error of the extracted signal and the true signal, namely:
\begin{equation}\label{chisq}
\chi^2 \equiv \int_{-\infty}^\infty \, dq |\tilde{W}\tilde{D}-\tilde{S}|^2
\end{equation}
is minimised. For each value of $K$ taken from a broad range ($10^{-3}$ - $10^{-1}$), we fit the deconvolved spectra with our model (section \ref{sec:tls_model}) and evaluate the sum of the squared residuals. This sum (orange curve in Fig. \ref{fig:optWiener}) is minimised for $K=0.06$, which we conclude to be the optimal filtering parameter.

\begin{figure*}
	\includegraphics[width=2\columnwidth]{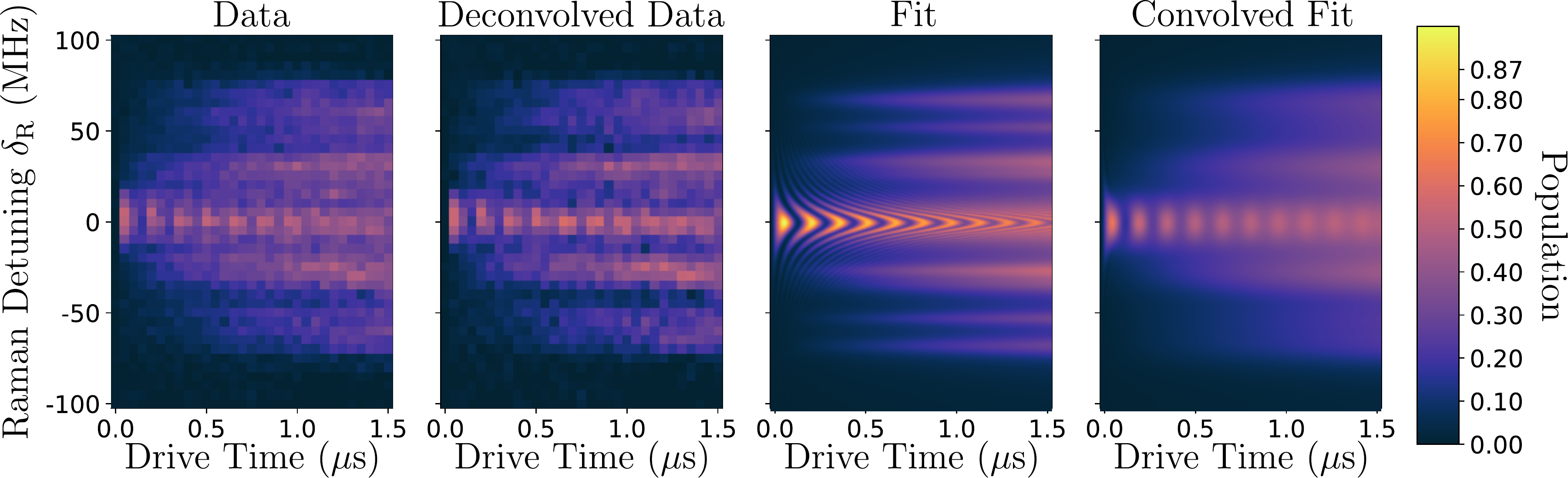}
	\caption{\textbf{Constraining the damping parameters:} Electron $\ket{\downarrow}$ population is measured as a function of drive time and ESR detuning $\delta_{\mathrm{R}}$, at mean field $\delta_\text{O}=0$. This reveals Rabi oscillations on the central ESR transition, alongside the emergence of four magnon transitions for each of two nuclear species (left). We deconvolve this data to factor out the $T_{2}^{*}$ effects (middle left), and fit the result with a model incorporating nine independent two-level systems (middle right). This fit is then convolved with a Gaussian profile corresponding to the respective $T_{2}^{*}$ (right).}
	\label{fig:2d_decon}
\end{figure*}

We further assess the goodness of the fit against the raw data. We convolve the fitted model and evaluate the sum of the squared residuals (purple curve in Fig. \ref{fig:optWiener}). This sum reaches a minimum below $K\sim 10^{-1}$, and increases for $K$ below $\sim10^{-3}$ where excessive amplification of high frequencies of the noise occurs. The Wiener filter at $K=0.06$ operates close to this minimum, and has the benefit of simultaneously minimising the $\chi^2$ from Eq.\,\ref{chisq}.

\subsection{Model function used to fit the data}\label{sec:tls_model}
When modelling the ESR drive, or any of the sideband processes, we approximate the time evolution of their lineshapes as that of independent two-level systems (TLSs), each governed by the following optical Bloch equations:
\begin{equation}
\begin{split}
\dot{\hat{\rho}}=&-\frac{i}{2}[\Omega \hat{\sigma}_x+\delta_{\mathrm{R}} \hat{\sigma}_z, \hat{\rho} ]+\frac{1}{2T_2}L(\hat{\sigma}_z) \hat{\rho}\\
&+\Gamma_+L(\hat{\sigma}_+)\hat{\rho}+\Gamma_-L(\hat{\sigma}_-)\hat{\rho}
\end{split}
\end{equation}
Here the dissipative dynamics are modelled by Lindblad operators  $L(\hat{\sigma}_\alpha)\hat{\rho}=\hat{\sigma}_\alpha\hat{\rho}\hat{\sigma}^\dagger_\alpha-\frac{1}{2}\{\hat{\sigma}^\dagger_\alpha \hat{\sigma}_\alpha,\hat{\rho}\}$. In the Bloch vector representation, i.e. $\hat{\rho}=\frac{1}{2}(1+\vec{s}\cdot \hat{\vec{\sigma}})$, the optical Bloch equations are equivalent to:
\begin{equation}
\begin{pmatrix}
\dot{s}_x \\ \dot{s}_y \\ \dot{s}_z
\end{pmatrix}=\underbrace{\begin{pmatrix}
	-\Gamma_2 & - \delta_{\mathrm{R}} & 0 \\ 
	\delta_{\mathrm{R}} & -\Gamma_2 & -\Omega \\
	0 & \Omega & -(\Gamma_-+\Gamma_+)
	\end{pmatrix}}_M \begin{pmatrix}
s_x \\ s_y \\ s_z
\end{pmatrix} +\begin{pmatrix}
0\\ 0 \\ \Gamma_+-\Gamma_-
\end{pmatrix}
\end{equation}
Where $\Gamma_2=(\Gamma_-+\Gamma_+)/2+1/T_2$. The time evolution of the Bloch vector is then found as:
\begin{equation}
\vec{s}(t)=e^{Mt}(\vec{s}(0)-\vec{s}(\infty))+\vec{s}(\infty)
\end{equation}
where the steady state Bloch vector components are:
\begin{equation}
\begin{split}
s_x(\infty)&=\frac{\delta_{\mathrm{R}} \Omega(\Gamma_+-\Gamma_-)}{\Omega^2 \Gamma_2 +(\Gamma_-+\Gamma_+)(\delta_{\mathrm{R}}^2+\Gamma_2^2)}\\
s_y(\infty)&=-\frac{\Gamma_2 \Omega(\Gamma_+-\Gamma_-)}{\Omega^2 \Gamma_2 +(\Gamma_-+\Gamma_+)(\delta_{\mathrm{R}}^2+\Gamma_2^2)}\\
s_z(\infty)&=\frac{(\delta_{\mathrm{R}}^2+\Gamma_2^2)(\Gamma_+-\Gamma_-)}{\Omega^2 \Gamma_2 +(\Gamma_-+\Gamma_+)(\delta_{\mathrm{R}}^2+\Gamma_2^2)}\\
\end{split}
\end{equation}

The magnon spectrum and its time dependence (in the absence of optical pumping) are broadened and damped by three types of processes: \begin{itemize}
	\item Inhomogeneous dephasing - a $T_2^*$ process,
	\item Homogeneous pure dephasing - a $T_2$ process,
	\item Slow, optically induced $T_1$ process, modelled by $\Gamma_+=\Gamma_- \ne 0$.
\end{itemize}
The optically induced $T_1$ process is subtracted from the data, as described in section \ref{t1bgd}. The $T_2^*$ effects are factored out through a Wiener deconvolution, as described in section \ref{sec:wiener}. Since, under the above conditions, the TLS saturates to $\vec{s}(\infty)=\vec{0}$, the lineshapes undergo expansion from their short-time limit to infinitely wide, allowing the rate of homogeneous pure dephasing to be constrained by a fit to the time dependence of the spectrum.



\subsection{Setting dephasing times from the magnon spectrum time-dependence}\label{sec:magnon_spec_t}

\begin{figure*}
	\includegraphics[width=2\columnwidth]{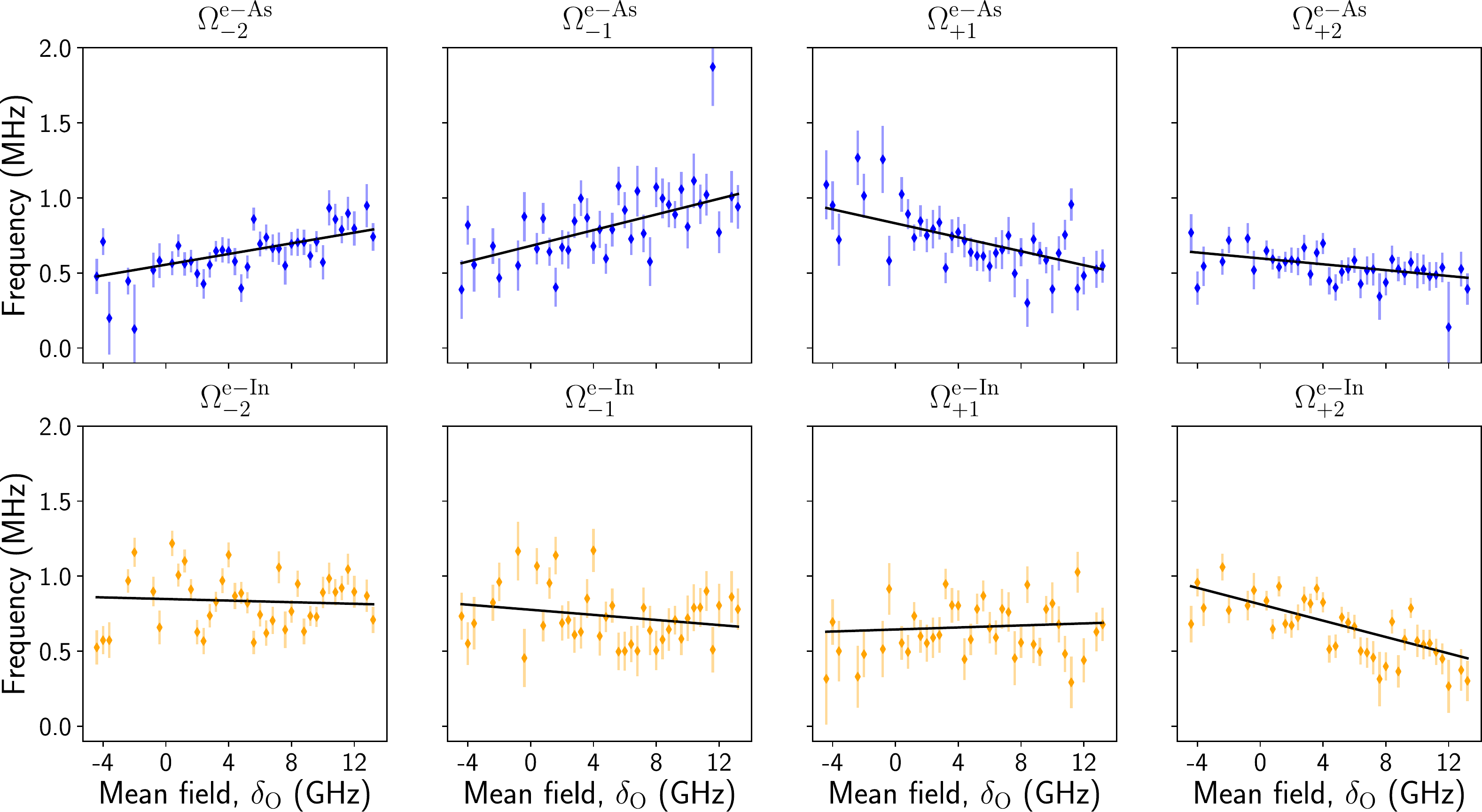}
	\caption{\textbf{Extracted electron-nuclear exchange frequencies:} The upper panels illustrate the exchange frequencies of arsenic nuclei used to reconstruct their polarisation (as per Eq.\,4 of the main text), as a function of Overhauser shift from the zero polarisation lock-point (see also panel (a) of Fig. 4 of the main text). The lower panels illustrate the same parameters, but for indium nuclei.}
	\label{fig:excfreqs}
\end{figure*}

Following the application of a Wiener deconvolution to the data (section \ref{sec:wiener}), nine independent TLS models (section \ref{sec:tls_model}) are fitted to the time dependence of the magnon spectrum, as shown in Fig.\,\ref{fig:2d_decon}. The lineshape at $\delta_{\mathrm{R}}=0 $\,MHz corresponds to the central ESR. The lineshapes at $\delta_{\mathrm{R}}=\pm \omega^{\mathrm{As}}, \pm 2 \omega^{\mathrm{As}} $, for $ \omega^{\mathrm{As}} = 25.4 $\,MHz correspond to the first and second sideband transitions of the arsenic species, and those at
$\delta_{\mathrm{R}}=\pm \omega^{\mathrm{In}}, \pm 2 \omega^{\mathrm{In}} $, for $ \omega^{\mathrm{In}} = 32.7 $\,MHz are the same transitions of the indium species. The ESR Rabi frequency is fitted as $\Omega=6.70\pm 0.01 [\mathrm{MHz}]$, whereas the exchange frequencies $\Omega_{\pm k}^j$ for the sideband transitions are all of the order $\sim 10^2 ~[\mathrm{kHz}]$.

From the fit shown in Fig.\,\ref{fig:2d_decon}, we obtain the pure dephasing times for the ESR, arsenic magnons, and indium magnons:
\begin{itemize}
	\item Electron homogeneous dephasing time, fitted as $T_2=4.55\pm 0.19 ~[\mu\mathrm{s}]$. We note that this $T_2 > T_{\mathrm{HE}}$ (section \ref{sec:hahnecho}) as the continuous Rabi drive dynamically decouples the electron spin beyond the simple Hahn-Echo time.
	\item Pure dephasing time for electron-arsenic interaction, $T_2^\mathrm{e-As}=275.4\pm 24.2 ~[\mathrm{ns}]$
	\item Pure dephasing time for electron-indium interaction, $T_2^\mathrm{e-In}=213.8\pm 14.3 ~[\mathrm{ns}]$
\end{itemize}


We do not resolve features arising from interactions with gallium nuclei, and therefore do not include them in our fitting. This is consistent with our previous reports of magnons in these systems \cite{Gangloff2019,Jackson2020}, and is likely to stem from gallium's weaker quadrupolar coupling \cite{Stockill2016}. Taking typical literature values for its quadrupolar constant $B_{Q}$ and quadrupolar angle $\theta$ \cite{Bulutay2012}, alongside its Zeeman energy, we expect gallium's magnon modes to be a factor of 15 (4) weaker than those for arsenic, for processes changing $I_{z}$ by 1 (2) unit(s) \cite{Jackson2020}. Given that arsenic and indium interactions are already weak compared to their dephasing rates, we can safely neglect all modes relating to gallium.

\subsection{Fitting the species-resolved spectra (data presented in the main text)}\label{spectrafits}

\begin{figure}
	\includegraphics[width=0.9\columnwidth]{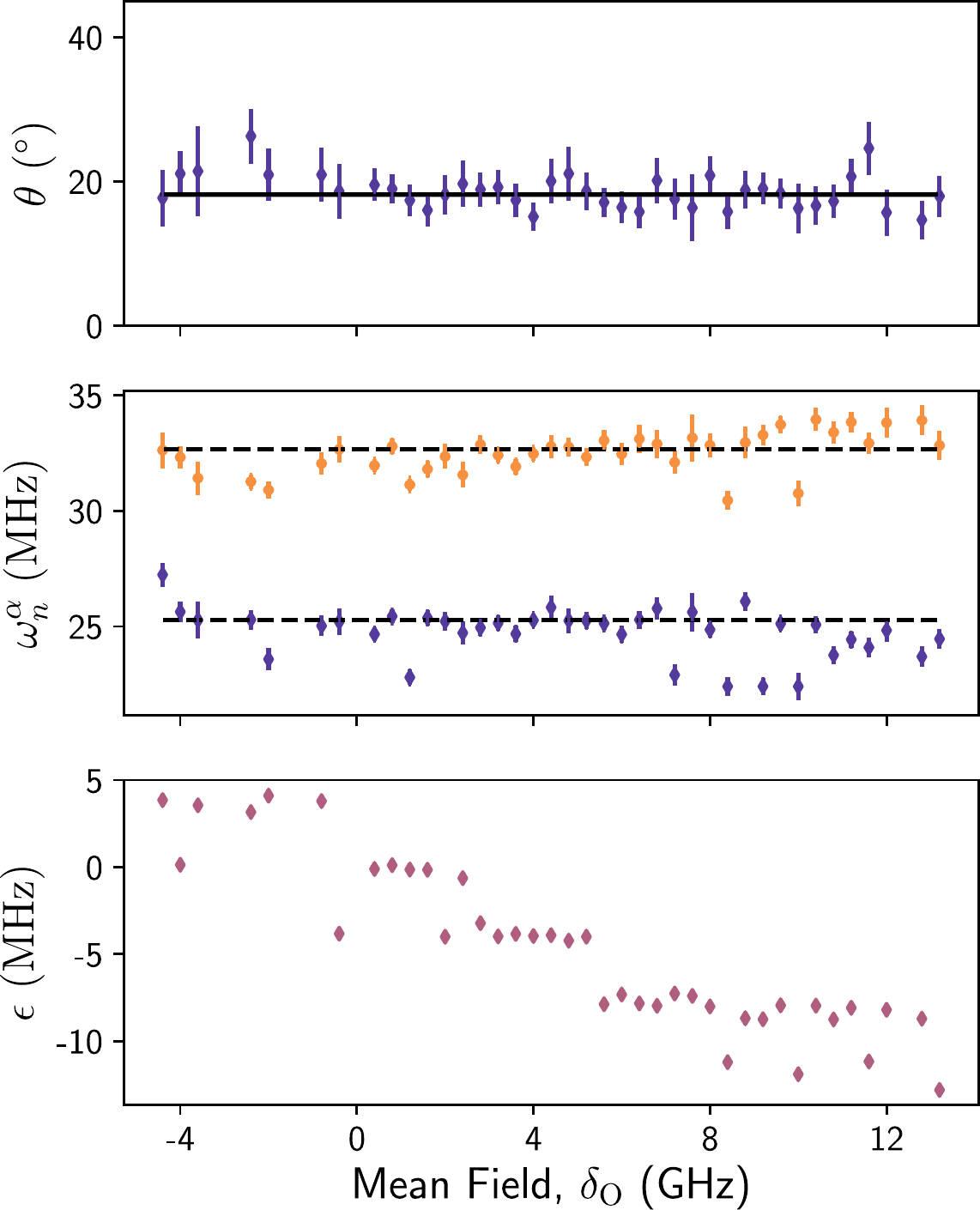}
	\caption{\textbf{Fit parameters and identified quadrupolar angle:} The top panel illustrates fitted values of the arsenic quadrupolar angle (purple circles) with a constant fit of $\theta=18.5^\circ$ (black curve). The middle panel shows fitted arsenic (purple) and indium (orange) Zeeman frequencies. The dashed lines are literature values \cite{Stockill2016}. The data in the bottom panel are the fitted ESR offsets (section \ref{sec:epsilon}).}
	\label{fig:rem_par}
\end{figure}

With the dephasing parameters fixed from the fit in the previous section (\ref{sec:magnon_spec_t}), we now focus on individual spectra taken at a $1$\,$\mu$s drive time, for a range of mean-field shifts: $-4.4$ to $13.2$\,GHz in steps of $0.4$\,GHz. We let the fit identify the relative offset of the centre of the ESR peak ($\epsilon$, see also section \ref{sec:epsilon}), and leave the Zeeman frequencies for indium and arsenic as free parameters. On top of these three parameters, the exchange frequencies for each of the sideband transitions ($\Omega^{j}_{-2}$, $\Omega^{j}_{-1}$, $\Omega^{j}_{+1}$, $\Omega^{j}_{+2}$ for $j=\mathrm{As},\mathrm{In}$) are fitted from the TLS lineshape at $t=1~[\mu \mathrm{s}]$ drive time (see section \ref{sec:tls_model}).

A summary of fitted exchange frequencies is shown in Fig. \ref{fig:excfreqs}. Figure\,3a from the main text illustrates the sum of the nine lineshapes that constitutes a fit to the deconvolved spectrum.

The fitted Zeeman frequencies and central peak offset are plotted in Fig.\,\ref{fig:rem_par}, together with $1\sigma$ error bars.  Orange data points are associated to indium, purple to arsenic. Dashed lines stand for the Zeeman frequencies of indium and arsenic under an external magnetic field of $3.5$\,T, converted from their nuclear $g$-factors found in the literature \cite{Stockill2016}: $\omega^{\mathrm{As}}=25.4$\,MHz, $\omega^{\mathrm{In}}=32.7$\,MHz.

The piecewise constant structure in the fitted central peak offset $\epsilon$ is an artefact of the fit's initial guess on the offset being identified with the position of the maximum point within the central peak. Since changing $\epsilon$ by a small amount, while holding other parameters constant, has a strong effect on the goodness of fit, this parameter stays close to its initial value. The step-to-step distance is consistent with the sampling of the magnon spectra at detunings spaced by $4$\,MHz.

Besides the reconstruction of population imbalances (as in Fig.\,3 of the main text), we use the exchange frequencies to reconstruct the quadrupolar angle for arsenic, via an expression easily derived from the identities of Eq.\,\ref{constsofmotion}:
\begin{equation}
\tan \theta =\frac{1}{4}\sqrt{\frac{\Omega^2_{+1}+\Omega^2_{-1}}{\Omega^2_{+2}+\Omega^2_{-2}}}
\end{equation}
The arsenic quadrupolar angles are evaluated for each data point, and displayed in the top panel of Fig.\,\ref{fig:rem_par}, together with a constant fit of $\theta=18.45^\circ$, which we use in the Fokker-Planck modelling of the phase-space flow (section \ref{sec:Phase_space_flow}).

\subsection{Robustness of the fitting results}
We verify that our data analysis is extremely robust against reasonable changes in the model parameters:

\emph{Wiener deconvolution $K$-constant} --
For values of $K$ from the range of $0.02$ - $0.1$ (see Fig. \ref{fig:optWiener}), the fitted exchange frequencies and damping parameters remain within their respective $1\sigma$ confidence intervals.

\emph{Inhomogeneous dephasing time $T_2^*$} --
A change of $T_2^*$ used in modelling by $\pm 10$\,ns has no visible effect on the observed trends of the extracted exchange frequencies, plotted in Fig. \ref{fig:excfreqs}. 

\emph{Homogeneous dephasing timescales} --
For a two-fold increase in dephasing rate ($T_2^{\mathrm{e-x}}\to T_2^{\mathrm{e-x}}/2$ for $x=\mathrm{As,In}$, as well as $T_2\to T_2/2$), we recover the same fitted parameters, albeit with higher uncertainties. For dephasing rates reduced two-fold ($T_2^{\mathrm{e-x}}\to 2T_2^{\mathrm{e-x}}$ for $x=\mathrm{As,In}$, as well as $T_2\to 2T_2$) we observe changes in our fitted parameters, and higher uncertainties. As an illustration, the asymmetry-commensurate arsenic polarisation is $\mathcal{I}_z^{\mathrm{As},\star}\approx0.056\delta_{\text{O}}/\text{GHz}$ (within $1\sigma$) for dephasing times smaller or equal to those used in the manuscript, and is $\mathcal{I}_z^{\mathrm{As},\star}\approx0.084\delta_{\text{O}}/\text{GHz}$ when dephasing times are doubled. The excess sideband asymmetry we report is thus robust against a reasonable choice of dephasing. 




\section{Supplementary notes on the Fokker-Planck modelling of nuclear spin distributions}\label{sec:Phase_space_flow}
\subsection{Modelling of the Arsenic Ensemble (3D F-P)}\label{3dfp}

\subsubsection{Master Equation}

\begin{figure}
	\includegraphics[width=0.65\columnwidth]{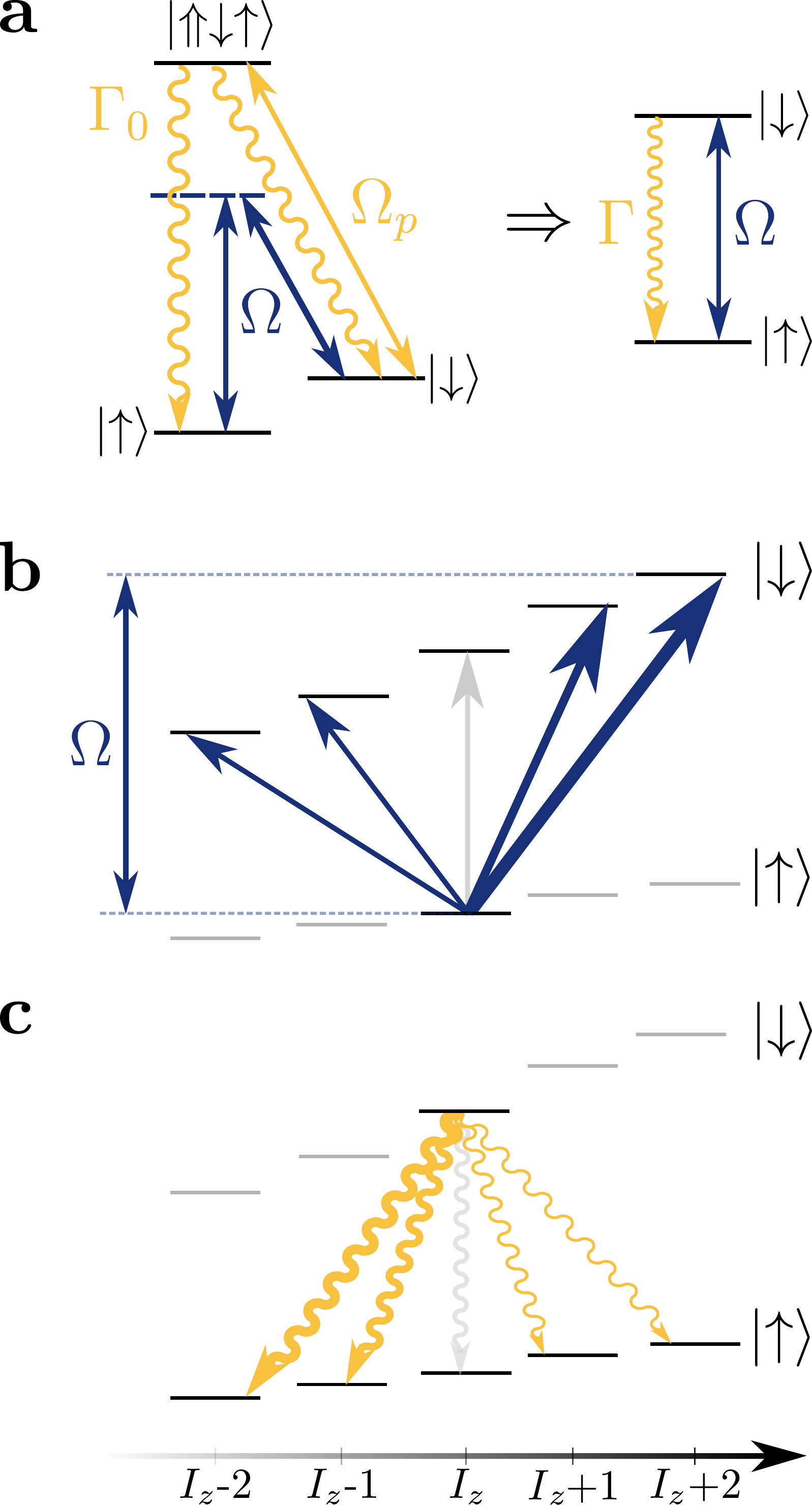}
	\caption{\textbf{Dynamics of Raman cooling:} \textbf{a} Central spin control is effectuated through a two-photon Raman drive of effective Rabi frequency $\Omega$. Second laser resonant with $\ket{\downarrow} \to \ket{\Uparrow \downarrow \uparrow}$ transition drives the system to a short-lived ($1/\Gamma_0$) trion state, that decays to the electron $\ket{\uparrow}$ and $\ket{\downarrow}$ with equal probabilities. Eliminating the trion state adiabatically, one arrives at the effective two-level system with a Rabi drive $\Omega$ and decay $\Gamma$. \textbf{b} For a detuned ESR drive, the electron-nuclear spin flips induce a directional drift in the nuclear phase-space, that brings the ensemble into the state satisfying the $\delta=0$ condition. \textbf{c} Spontaneous sideband processes introduce diffusion into the phase-space dynamics, and are represented here for $I_z>0$. }
	\label{fig:phasflow}
\end{figure}

In order to reconstruct spin-state populations (and their distribution), we adapt the master equation treatment of feedback control of a nuclear spin ensemble devised by W. Yang and L. J. Sham \cite{Yang2013}. Our model constitutes an extension of their work, that captures the electron-nuclear spin-flip processes enabled by the strain-induced quadrupolar interaction \cite{Urbaszek2013} : under our Raman drive, the electron spin polarisation is modulated at the Rabi frequency; this results in a time-dependent Knight field, which when driving at the nuclear Zeeman energy, enables nuclear spin-flips. The model assumes vanishing nuclear coherences and takes the approximation of Markovian dynamics. This is a reasonable approximation given that the electronic quasi-steady state in a given Overhauser field is to first order governed by the optical spin pumping ($\Gamma$), which happens on a faster timescale than the nuclear back-action \cite{Yang2013}. 

The master equation governing the cooling dynamics for the nuclear density operator $\hat{\rho}_n$ is: 
\begin{equation}\label{YS_starting_point}
\begin{split}
\frac{d}{dt}\hat{\rho}_n=-\sum_{j}\sum_{k=1,2}[\hat{\Phi}^j_{-k},\hat{\Phi}^j_{+k}W^k_+(I_z)\hat{\rho}_n]\\
-\sum_j\sum_{k=1,2}[\hat{\Phi}^j_{+k},\hat{\Phi}^j_{-k}W^k_-(I_z)\hat{\rho}_n]
\end{split}
\end{equation}

Polarisation-dependent Raman scattering rates $W^k_\pm$ are approximated as those of a two-level system constituted by $\ket{I_z}$ and $\ket{I_z\pm k}$ states, as done in \cite{Hogele2012,Gangloff2019}. The rate of electronic optical pumping $\Gamma$ is related to the natural linewidth $\Gamma_0$ of the exciton state $\ket{\Uparrow \downarrow \uparrow}$ in a lambda system, by:
\begin{equation}
\Gamma=\frac{\Gamma_0}{4}\frac{2(\Omega_p/\Gamma_0)^2}{1+2(\Omega_p/\Gamma_0)^2}
\end{equation}
where $\Omega_p$ is the Rabi frequency of the optical pumping field (Fig. \ref{fig:phasflow}a). We take:
\begin{equation}\label{ramanscattering}
W_{\pm k} (\delta_{\text{e}},I_z)=\frac{\Gamma}{2}\frac{\Omega^2/\Gamma\Gamma_2}{1+\Omega^2/\Gamma\Gamma_2+(\Delta_{\pm k}(\delta_{\text{e}},I_z)/\Gamma_2)^2}
\end{equation} 
with:
\begin{equation}
\Delta_{\pm k}(\delta_{\text{e}})=\delta_{\text{e}}-\frac{1}{2}a(2I_z\pm k) \mp k\omega_\text{n}
\end{equation}
being a detuning of the effective two-level drive from the $\ket{I_z} \to \ket{I_z\pm k}$ transition (Fig. \ref{fig:phasflow}b). We have introduced $\delta_{\text{e}} = \omega_\text{R} - \omega_\text{e}$ as a laser detuning from the electron Zeeman splitting. The strength of pure dephasing is:
\begin{equation}
\Gamma_2=\frac{\Gamma}{2}+\frac{1}{T_2}
\end{equation}
The decrease of polarisation fluctuations is a result of strong phase-space flow towards the state $I_z$ defined by the mean field $\delta_\text{O}=\sum_j a^j I_z^j$ where $\delta_{\text{e}} = \delta_\text{O}$. The cooling performance is dependent on the competition of Raman scattering rates with spontaneous sideband processes and nuclear spin diffusion (Fig.\,\ref{fig:phasflow}c). 

Since the system comprises multiple spin-species, we partition the mean field into the respective spin-species components:
\begin{equation}
\delta_\text{O}=\delta_\text{O}^{\mathrm{In}}+\delta_\text{O}^{\mathrm{Ga}}+\delta_\text{O}^{\mathrm{As}}
\end{equation}
each defining a species-specific nuclear polarisation. 

\subsubsection{Steady state solution}\label{fokker_planc_section}
We now model the phase-space flow of the arsenic nuclear state. Here the lockpoint is set by $\delta_\text{O}^{\mathrm{As},\star}$ - a component of the total mean field reconstructed from the asymmetry measurement. 

Following Ref.\,\cite{Yang2013}, we seek a steady-state probability distribution for the system to be in state $\ket{\vec{N}}\equiv \ket{N_{+\frac{3}{2}},N_{+\frac{1}{2}},N_{-\frac{1}{2}},N_{-\frac{3}{2}}}$, where $N_m$ is the number of spins in projection $m$. This is given by:
\begin{equation}
p(\vec{N},t)=\text{Tr}\hat{\rho}_n(t)\delta_{\vec{N}^\prime, \vec{N}}
\end{equation}
which, using Eq.\,\ref{YS_starting_point}, is shown to evolve according to a rate equation:

\begin{equation}\label{big_rate_eqn}
\begin{split}
&\frac{\partial}{\partial t}p(\vec{N},t)=\\
&-\sum_m N_m \eta_m^2 W_{+1}[I_z(\vec{N})]p(\vec{N},t)\\
&+\sum_m (N_m+1) \eta_m^2 W_{+1}[I_z(\vec{N}^{(m,m+1)})]p(\vec{N}^{(m,m+1)},t)\\
&-\sum_m N_{m+1} \eta_m^2 W_{-1}[I_z(\vec{N})]p(\vec{N},t)\\
&+\sum_m (N_{m+1}+1) \eta_m^2 W_{-1}[I_z(\vec{N}^{(m+1,m)})]p(\vec{N}^{(m+1,m)},t)\\
&-\sum_m N_m \epsilon_m^2 W_{+2}[I_z(\vec{N})]p(\vec{N},t)\\
&+\sum_m (N_m+1) \epsilon_m^2 W_{+2}[I_z(\vec{N}^{(m,m+2)})]p(\vec{N}^{(m,m+2)},t)\\
&-\sum_m N_{m+2}  \epsilon_m^2 W_{-2}[I_z(\vec{N})]p(\vec{N},t)\\
&+\sum_m (N_{m+2}+1)  \epsilon_m^2 W_{-2}[I_z(\vec{N}^{(m+2,m)})]p(\vec{N}^{(m+2,m)},t)\\
\end{split}
\end{equation}
with single-spin magnon-operator matrix elements:
\begin{equation}
\begin{split}
\eta_m&=|\bra{m+1}\hat{\Phi}^j_{+1}\ket{m}| \\
\epsilon_m&= |\bra{m+2}\hat{\Phi}^j_{+2}\ket{m}|
\end{split}
\end{equation}
and:
\begin{equation}
\vec{N}^{(k,l)}=[..,N_k+1,..,N_l-1,..]
\end{equation}
being a concise notation for a microstate accessible from $\vec{N}$ through a spin-flip that changes projection of a single spin from $m=l$ to $m=k$. In addition to the stimulated Raman scattering rates induced by the ESR drive ($\propto W_{\pm k}$), spontaneous Raman scattering rates ($\propto \Gamma_{\mathrm{nc}}$) induced by optically pumping the electron ($\Gamma$) have to be incorporated \cite{Yang2013,Hogele2012}, as per Fig. \ref{fig:phasflow}c. At last, we account for a slow nuclear spin decay ($\Gamma_{\mathrm{nuc}}$) - a $\Delta I_z =\pm 1$ thermalisation mechanism - as seen experimentally in section \ref{sec:nucleart1}. Embedding these processes in the master equation is done through the following extension:
\begin{equation}\label{added_diffusion}
\begin{split}
&\eta_m^2 W_{\pm 1}[I_z(\vec{N})]\to\eta_m^2 \Bigg(W_{\pm 1}[I_z(\vec{N})]+\\
&\underbrace{\frac{\Gamma}{4}\frac{\Omega^2/\Gamma\Gamma_2}{1+\Omega^2/\Gamma\Gamma_2+((\delta_{\text{e}}-aI_z(\vec{N}))/\Gamma_2)^2}}_{\Gamma_{\mathrm{nc}}[I_z(\vec{N})]/2}\Bigg)+\frac{\Gamma_{\mathrm{nuc}}}{2} \\
&\epsilon_m^2 W_{\pm 2}[I_z(\vec{N})]\to\epsilon_m^2 \Bigg(W_{\pm 2}[I_z(\vec{N})]+\\
&\underbrace{\frac{\Gamma}{4}\frac{\Omega^2/\Gamma\Gamma_2}{1+\Omega^2/\Gamma\Gamma_2+((\delta_{\text{e}}-aI_z(\vec{N}))/\Gamma_2)^2}}_{\Gamma_{\mathrm{nc}}[I_z(\vec{N})]/2}\Bigg)
\end{split}
\end{equation}
The nuclear spin decay is the only process included in the master equation that allows $\frac{1}{2} \leftrightarrow -\frac{1}{2}$ transitions within the single-spin manifold.

Rate equation \ref{big_rate_eqn} has a unique steady-state solution, which is hard to find exactly in the $N \gg 1$ limit. However, in this limit, treating $n_q=N_q/N$ as continuous variables is well-justified, allowing the equation to be expanded in the small parameter $N^{-1}$ up to second order, which turns it into a Fokker-Planck equation \cite{Risken1984}.

In doing this, it will be convenient to define functions $g_{i,\pm k}(\vec{n})$, $g^n_{i,\pm 1}(\vec{n})$ and a linear differential operator $\mathcal{D}_{k,l}$:
\begin{align}
g_{i,\pm k}(\vec{n}) \equiv &12 \alpha_k^2 N n_i (W_{\pm k}(\vec{n})+\Gamma_{\mathrm{nc}}(\vec{n})/2)\\
g^{n}_{i,\pm 1}(\vec{n}) \equiv &Nn_i\Gamma_{\mathrm{nuc}}/2 \\
\mathcal{D}_{k,l} \equiv &N^{-1}\Big[\frac{\partial}{\partial n_k}-\frac{\partial}{\partial n_l}\Big]+\\
&\frac{N^{-2}}{2}\Big[\frac{\partial^2}{\partial n_k^2}+\frac{\partial^2}{\partial n_l^2}-2\frac{\partial^2}{\partial n_k \partial n_l}\Big]
\end{align}
with $\alpha_1=\sin{2\theta}\Big(\frac{aB^{j}_{Q}}{2\omega_{n}^{2}}\Big)$ and $\alpha_2=\frac{1}{2}\cos^2{\theta}\Big(\frac{aB_{Q}^{j}}{2\omega_{n}^{2}}\Big)$. The steady-state probability distribution $p(\vec{n})$ of the master equation, with $\vec{n}$ as in the main text, is then a solution to the partial differential equation:

\begin{equation}\label{fullofops}
\begin{split}
0=&\mathcal{D}_{\frac{1}{2},\frac{3}{2}}((g_{\frac{1}{2},+1}+g^n_{\frac{1}{2},+1})p)+\mathcal{D}_{-\frac{1}{2},\frac{1}{2}}(g^n_{-\frac{1}{2},+1}p)\\
+&\mathcal{D}_{-\frac{3}{2},-\frac{1}{2}}((g_{-\frac{3}{2},+1}+g^n_{-\frac{3}{2},+1})p)\\
+&\mathcal{D}_{\frac{3}{2},\frac{1}{2}}((g_{\frac{3}{2},-1}+g^n_{\frac{3}{2},-1})p)+\mathcal{D}_{\frac{1}{2},-\frac{1}{2}}(g^n_{\frac{1}{2},-1}p)\\
+&\mathcal{D}_{-\frac{1}{2},-\frac{3}{2}}((g_{-\frac{1}{2},-1}+g^n_{-\frac{1}{2},-1})p)\\
+&\mathcal{D}_{-\frac{1}{2},\frac{3}{2}}(g_{-\frac{1}{2},+2}p)+\mathcal{D}_{-\frac{3}{2},\frac{1}{2}}(g_{-\frac{3}{2},+2}p)\\
+&\mathcal{D}_{\frac{3}{2},-\frac{1}{2}}(g_{\frac{3}{2},-2}p)+\mathcal{D}_{\frac{1}{2},-\frac{3}{2}}(g_{\frac{1}{2},-2}p)+\mathcal{O}(N^{-3})
\end{split}
\end{equation}
Here, all functions and their derivatives are evaluated at $\vec{n}$. To recast Eq.\,\ref{fullofops} in the more familiar form of a steady-state Fokker-Planck equation, we proceed with defining a drift vector and a diffusion tensor, with $\partial_i \equiv \frac{\partial}{\partial n_i}$:

\begin{equation}\label{fokkerplanck_usualform}
\begin{split}
&0=-\begin{pmatrix} \partial_{\frac{3}{2}} \\ \partial_{\frac{1}{2}} \\ \partial_{-\frac{1}{2}} \\ \partial_{-\frac{3}{2}}
\end{pmatrix}^T \cdot
\Bigg\{ N^{-1} \begin{pmatrix}
v_{\frac{3}{2}} \\ v_\frac{1}{2} \\ v_{-\frac{1}{2}} \\ v_{-\frac{3}{2}}
\end{pmatrix} - \frac{N^{-2}}{2} \\
 &\Bigg[ 
\begin{pmatrix} \partial_{\frac{3}{2}} \\ \partial_{\frac{1}{2}} \\ \partial_{-\frac{1}{2}} \\ \partial_{-\frac{3}{2}}
\end{pmatrix}^T
\begin{pmatrix}
D_{\frac{3}{2},\frac{3}{2}} & D_{\frac{3}{2},\frac{1}{2}} & D_{\frac{3}{2},-\frac{1}{2}} & 0 \\
D_{\frac{1}{2},\frac{3}{2}} & D_{\frac{1}{2},\frac{1}{2}} & D_{\frac{1}{2},-\frac{1}{2}} & D_{\frac{1}{2},-\frac{3}{2}} \\
D_{-\frac{1}{2},\frac{3}{2}} & D_{-\frac{1}{2},\frac{1}{2}} & D_{-\frac{1}{2},-\frac{1}{2}} & D_{-\frac{1}{2},-\frac{3}{2}}\\
0 & D_{-\frac{3}{2},\frac{1}{2}} & D_{-\frac{3}{2},-\frac{1}{2}} & D_{-\frac{3}{2},-\frac{3}{2}}
\end{pmatrix}
\Bigg]^T\Bigg\} p(\vec{n})
\end{split}
\end{equation}
The drift terms are given by:
\begin{equation}
\begin{split}
v_{\frac{3}{2}} =& (g_{\frac{1}{2},+1}+g^n_{\frac{1}{2},+1})-(g_{\frac{3}{2},-1}+g^n_{\frac{3}{2},-1})\\&+g_{-\frac{1}{2},+2}-g_{\frac{3}{2},-2}\\
v_{\frac{1}{2}} =& -(g_{\frac{1}{2},+1}+g^n_{\frac{1}{2},+1}) +(g_{\frac{3}{2},-1}+g^n_{\frac{3}{2},-1})\\&+(g^n_{-\frac{1}{2},+1}-g^n_{\frac{1}{2},-1})+g_{-\frac{3}{2},+2}-g_{\frac{1}{2},-2}\\
v_{-\frac{1}{2}}=&(g_{-\frac{3}{2},+1}+g^n_{-\frac{3}{2},+1})-(g_{-\frac{1}{2},-1}+g^n_{-\frac{1}{2},-1})\\&+(g^n_{\frac{1}{2},-1}-g^n_{-\frac{1}{2},+1})-g_{-\frac{1}{2},+2}+g_{\frac{3}{2},-2}\\
v_{-\frac{3}{2}}=&-(g_{-\frac{3}{2},+1}+g^n_{-\frac{3}{2},+1})+(g_{-\frac{1}{2},-1}+g^n_{-\frac{1}{2},-1})\\&-g_{-\frac{3}{2},+2}+g_{\frac{1}{2},-2}
\end{split}
\end{equation}
And the elements of the diffusion tensor are:
\begin{equation}
\begin{split}
D_{\frac{3}{2},\frac{3}{2}}=& (g_{\frac{1}{2},+1}+g^n_{\frac{1}{2},+1})+(g_{\frac{3}{2},-1}+g^n_{\frac{3}{2},-1})\\&+g_{-\frac{1}{2},+2}+g_{\frac{3}{2},-2}\\
D_{\frac{1}{2},\frac{1}{2}}=& (g_{\frac{1}{2},+1}+g^n_{\frac{1}{2},+1})+(g_{\frac{3}{2},-1}+g^n_{\frac{3}{2},-1})\\&+(g_{\frac{1}{2},-1}^n+g_{-\frac{1}{2},+1}^n)+g_{-\frac{3}{2},+2}+g_{\frac{1}{2},-2}\\
D_{-\frac{1}{2},-\frac{1}{2}}=&(g_{-\frac{3}{2},+1}+g^n_{-\frac{3}{2},+1})+(g_{-\frac{1}{2},-1}+g_{-\frac{1}{2},-1}^n)\\&+(g_{\frac{1}{2},-1}^n+g_{-\frac{1}{2},+1}^n)+g_{-\frac{1}{2},+2}+g_{\frac{3}{2},-2}\\
D_{-\frac{3}{2},-\frac{3}{2}}=&(g_{-\frac{3}{2},+1}+g^n_{-\frac{3}{2},+1})+(g_{-\frac{1}{2},-1}+g_{-\frac{1}{2},-1}^n)\\&+g_{-\frac{3}{2},+2}+g_{\frac{1}{2},-2}\\
D_{\frac{1}{2},\frac{3}{2}}=& -(g_{\frac{1}{2},+1}+g_{\frac{1}{2},+1}^n+g_{\frac{3}{2},-1}+g_{\frac{3}{2},-1}^n)\\
D_{\frac{1}{2},-\frac{1}{2}}=& -(g_{\frac{1}{2},-1}^n+g_{-\frac{1}{2},+1}^n)\\
D_{-\frac{3}{2},-\frac{1}{2}}=&-(g_{-\frac{3}{2},+1}+g_{-\frac{3}{2},+1}^n+g_{-\frac{1}{2},-1}+g_{-\frac{1}{2},-1}^n)\\
D_{-\frac{1}{2},\frac{3}{2}}=&-(g_{\frac{3}{2},-2}+g_{-\frac{1}{2},+2})\\
D_{-\frac{3}{2},\frac{1}{2}}=&-(g_{\frac{1}{2},-2}+g_{-\frac{3}{2},+2})\\
D_{i,j}=&D_{j,i}
\end{split}
\end{equation}

Finally, we use the physical constraint on populations, $\sum_i n_i=1$, to reduce the dimensionality of the Fokker-Planck equation by one. This is a straightforward exercise in multivariate calculus. In the following, we will assume that drift vector and diffusion tensor are defined in a three-dimensional space of variables $n_{\frac{3}{2}}$, $n_{\frac{1}{2}}$, $n_{-\frac{3}{2}}$, satisfying $n_i\geq0$ and $\sum_i n_i\leq1$.

\subsubsection{Numerical approach to solving the problem}
%
\emph{Setup} -- In this section we continue working with a single spin-species: arsenic. The lock-points of the flow are again set by $ \delta_\text{O}^{\mathrm{As},\star}$ - arsenic components of total mean field reconstructed from the asymmetry measurements. The objective of the following is to find a steady-state probability distribution of the state in the phase space of single-spin state populations $\vec{n}$.

As in the one-dimensional case (Fig.\,1e of the main text), the drift terms in Eq.\,\ref{fokkerplanck_usualform} lead to a sizeable feedback in the range $\Delta I_z \sim \omega_n/a$, corresponding to a volume small compared to that of the entire phase space. This makes domain discretisation ill-suited for finding the steady-state probability distribution.

Instead, we develop a method relying on knowing the position of the centre of steady state probability distribution - $\vec{n}_0$. This point is found numerically by following the flow lines in the phase-space, towards a stable point that satisfies $\delta_{\text{e}}=\delta_\text{O}^{\mathrm{As},\star}$ within $\epsilon$ ($I_z$ locked to the Raman drive). The technique developed in this section allows to reconstruct the steady-state distribution in the vicinity of $\vec{n}_0$, which we \textit{a priori} expect to be a sufficiently good approximation, since the strong feedback is confined to a small phase-space volume.  




\emph{General method} -- The diffusion component of Eq.\,\ref{fokkerplanck_usualform} scales like $N^{-1}$ as compared to the drift term. This makes our case an excellent candidate for the  WKB approximation \cite{Risken1984}. We use the ansatz:
\begin{equation}
p(\vec{n})\propto\text{exp}\Big(-N w(\vec{n})\Big)
\end{equation}
which brings Eq.\,\ref{fokkerplanck_usualform} into the form:
\begin{equation}\label{equation_wkb}
0=\vec{v}\cdot \grad w +\frac{1}{2}(\grad w)^TD\grad w +\mathcal{O}(N^{-1})
\end{equation}
Since $N\sim10^5$ this is an excellent approximation, and it reduces greatly the complexity of the problem.

Then, since $D$ is positive-definite everywhere except at the boundaries, where at least one of its components is zero, we have that at $\vec{v}=0$ we also have $\grad w=0$. This means that the probability distribution has a local maximum at the zero of the flow field $\vec{v}$ which we denoted as $\vec{n}_0$. 

We then expand $\vec{v}$ and $\grad w$ around $\vec{n}_0$ up to first order in $\delta \vec{n}=\vec{n}-\vec{n}_0$:
\begin{equation}
\begin{split}
\grad w = S \delta \vec{n} +\mathcal{O}(\delta \vec{n}^2)\\
\vec{v}= J \delta \vec{n} +\mathcal{O}(\delta \vec{n}^2)
\end{split}
\end{equation}
Here, $J$ stands for Jacobian ($J_{ij}=\frac{\partial v_i}{\partial n_j}|_{\vec{n}=\vec{n}_0}$) and:
\begin{equation}
S_{ij}=\frac{\partial^2 w}{\partial n_i\partial n_j} \Bigg\lvert_{\vec{n}=\vec{n}_0}
\end{equation}
is a symmetric, Hessian matrix, which we will show to be positive-definite if $\vec{n}_0$ is a stable point. In the vicinity of a stable lock-point, matrix $S$ defines a Gaussian probability distribution:
\begin{equation}\label{fokkerplanck_solution_gaussian}
p(\vec{n}_0+\delta \vec{n})\propto\text{exp}\Big(-\frac{N}{2} \delta \vec{n}^TS \delta \vec{n} +\mathcal{O}(\delta \vec{n}^2)\Big)
\end{equation}
Neglecting the $\mathcal{O}(N^{-1})$ terms, we further reduce Eq.\,\ref{equation_wkb} to:
\begin{equation}\label{almostfinal}
0=\delta  \vec{n}^T (J^TS)\delta  \vec{n}+\delta  \vec{n}^T (\frac{1}{2}SDS)\delta  \vec{n}+\mathcal{O}(\delta \vec{n}^2)
\end{equation}
which is a sum of two quadratic forms. We now use the fact that for a general vector $\vec{x}$ and matrix $M$:
\begin{equation}
\begin{split}
\vec{x}^TM\vec{x}&=M_{ij}x_ix_j\\&=\frac{1}{2}(M_{ij}+M_{ji})x_ix_j+
\frac{1}{2}\underbrace{(M_{ij}-M_{ji})}_{\text{anti-symmetric}}\overbrace{x_ix_j}^{\text{symmetric}}\\&=\frac{1}{2}\vec{x}^T(M+M^T)\vec{x}
\end{split}
\end{equation}
so for Eq.\,\ref{almostfinal} to hold for arbitrary $\delta\vec{n}$, the sum of symmetric matrix components of $J^TS$ and $SDS$ has to be zero:
\begin{equation}\label{controltheoryeqn}
J^TS+SJ=-SDS
\end{equation}
Provided that the zero of $\vec{v}$ we identified is stable (i.e. all eigenvalues of $J$ are negative), $S$ comes out positive-definite, as desired. Note that this construction does not require $J$ to be symmetric, so it is suited for flows with non-zero curl as well (which is numerically verified to be the case in our problem). 

We now outline a way to solve Eq.\,\ref{controltheoryeqn} for $S$. 
We begin by multiplying by $S^{-1}$ from both sides: 
\begin{equation}\label{controltheoryeqn2}
S^{-1}J^T+JS^{-1}=-D
\end{equation}
and then diagonalize $J$ and $J^T$ such that:
\begin{equation}
\begin{split}
J^T&=U^{-1}\Lambda U \\
J&=V^{-1} \Lambda V \\
\Lambda &= \text{diag}(\lambda_1,\lambda_2,\lambda_3)
\end{split}
\end{equation}
We then define $\tilde{S}^{-1}_{ij}=-D_{ij}/(\lambda_i+\lambda_j)$.
Finally:
\begin{equation}
S^{-1}=V^{-1}\tilde{S}^{-1}U
\end{equation}
We have now found the covariance matrix - $S^{-1}$ - of the multivariate Gaussian distribution $p(\vec{n})$, which together with $\vec{n}_0$ fully defines the steady-state probability distribution in the vicinity of a lock-point. 


The computational complexity of arriving at this solution using our method is low. The core of the problem lies in finding $\vec{n}_0$ numerically, which can be done by following the flow lines in the basin of attraction of the stable point. The computational cost of this procedure is comparable to that of the gradient descent algorithm. In the present case, the vector field is not a gradient of a scalar field, but its analytical form is known, and is expected to have stable zeros.  


Our feedback (Eq.\,\ref{big_rate_eqn}) is non-linear, and bistable, so in order to avoid convergence to the trivial solution $\ket{\vec{n}}=\ket{0.25,0.25,0.25,0.25}$ we pick an initial guess $\ket{\vec{n}_{\mathrm{ini}}}=\ket{0.85,0.05,0.05,0.05}$, which corresponds to a polarisation much higher than the ones we lock the system to, and which is predicted to flow to the point $\ket{\vec{n}_0}$ by one-dimensional mean-field theory. 


Entries of $D$ and $J$ are evaluated symbolically, so it is only the diagonalisation and matrix inversion that are the possible sources of numerical errors. Since the matrices involved are $3\cross3$ and real, the errors are negligible.

\emph{Solution} -- The covariance matrices found numerically are non-diagonal, which implies the existence of classical correlations in the optically prepared nuclear state (of a single species). A spectral analysis of these matrices shows that the eigenvector associated with the smallest eigenvalue is also a normal to the plane of constant polarisation $I_z$ - this is expected of the feedback mechanism designed to narrow down the fluctuations in $I_z$, as highlighted in the main text. 

Under the assumption of a single nuclear spin species, the electron's inhomogeneous dephasing time predicted by the model is:
\begin{equation}
T_{2,\mathrm{Model}}^*=\frac{\sqrt 2 (2\pi)^{-1}}{ \sqrt{ \langle \Delta^2(\frac{3}{2}A \mathcal{I}_z) \rangle}}
\end{equation}
Thanks to the convenient form of the identified solution, fluctuations in $I^{\mathrm{As}}_z$ can be calculated easily:
\begin{equation}
\langle \Delta^2 I^{\mathrm{As}}_z \rangle=N^{\mathrm{As}}\vec{b}^TS^{-1}\vec{b}
\end{equation}
where we define $\vec{b}$:
\begin{equation}
I^{\mathrm{As}}_z / N^{\mathrm{As}}=\underbrace{(2\hat{x}+\hat{y}-\hat{z})}_{\vec{b}} \cdot \vec{n}-1 / 2
\end{equation}
as a vector decomposed in a basis of unit vectors $\{\hat{x},\hat{y},\hat{z}\}$ along $n_{3/2}$, $n_{1/2}$ and $n_{-3/2}$, respectively. The fluctuations of the fractional polarisation are: $\langle \Delta^2 \mathcal{I}^{\mathrm{As}}_z \rangle=\langle \Delta^2 I^{\mathrm{As}}_z \rangle/(\frac{3}{2}N^\mathrm{As})^2$.

\emph{Relating the $T_{2,\mathrm{Model}}^*$ in our single-species model to the measured $T_2^*$} -- 
How do we relate $T_{2,\mathrm{Model}}^*$ to the experimentally measured $T_2^*$ from section \ref{t2starmeasurement}?
Our measured $T_2^*$ is the sum effect of all three species:

\begin{multline}\label{multispeciest2star}
T_2^*=\\\frac{\sqrt 2 (2\pi)^{-1}}{\sqrt{\langle\Delta^2(\frac{3}{2}A^{\mathrm{As}}\mathcal{I}^{\mathrm{As}}_z+\frac{9}{2}xA^{\mathrm{In}}\mathcal{I}^{\mathrm{In}}_z+\frac{3}{2}(1-x)A^{\mathrm{Ga}}\mathcal{I}^{\mathrm{Ga}}_z)\rangle}}
\end{multline}
where we labelled the concentration of indium as $x$ .

We need to make some simple assumptions in order to connect our model value to our experimental value. Here we make two: no correlations between polarisations of species-characteristic sub-ensembles, and the same degree of fractional polarisation narrowing for each species ($ \langle \Delta^2 \mathcal{I}^{\mathrm{As}}_z \rangle= \langle \Delta^2  \mathcal{I}^{\mathrm{In}}_z \rangle= \langle \Delta^2  \mathcal{I}^{\mathrm{Ga}}_z \rangle$). Under these assumptions the model inhomogeneous dephasing time is rescaled by a factor $\sim$2 compared to the measured $T_2^*$:

\begin{equation}
T_{2}^*\approx \underbrace{\frac{\sqrt{(\frac{3}{2}A^{\mathrm{As}})^2}}{\sqrt{( \frac{3}{2}A^{\mathrm{As}})^2+(\frac{9}{2}x A^{\mathrm{In}})^2+(\frac{3}{2}(1-x) A^{\mathrm{Ga}})^2 }}}_{\sim 1/2}T_{2,\mathrm{Model}}^*
\end{equation}


We note that if instead we had made the opposite assumption of correlated sub-ensembles, our scaling factor would be only $\sqrt{2}$ different.
Indeed, for sub-ensembles fully correlated by the feedback, the $T_2^*$ scales as $N^{-1/2}$ for $N$ spins. Thus, our $T_{2,\mathrm{Model}}^*$, which contains $N/2$ spins, would be of order $\sqrt 2$ $T_2^*$. Most importantly, our reconstruction is insensitive to $T_{2}^{*}$ on this scale.

\emph{Fitting procedure} -- In order to match the model to the  measured $T_2^*$ and population imbalances (Fig 3b of the main text), we leave $\Gamma_{\mathrm{nuc}}$ as a free parameter in a fit, and fix the remaining parameters to the values strongly informed by previous studies on dots from the same wafer (all summarised in the Table III, at the end of the supplementary materials). 
Calculated values of $T_2^*$ fall in the $30-40~(\mathrm{ns})$ interval.

\subsubsection{Thermal state}

We now calculate the covariance matrix that the thermal state of corresponding polarisation would feature, as per Fig.\,3 of the main text. 

The single-spin partition function, expressed as a function of the thermodynamic $\beta$, is given by:
\begin{equation}
Z_1=2\cosh\frac{3}{2}\beta + 2 \cosh\frac{1}{2} \beta
\end{equation}
A known polarisation $I_z$ results in a numerically solvable equation for $\beta$:
\begin{equation}
\langle I_z \rangle =-\frac{\partial}{\partial \beta} \ln Z_1^N= -N \frac{3\sinh\frac{3}{2}\beta +\sinh\frac{1}{2}\beta}{2\cosh\frac{3}{2}\beta + 2 \cosh\frac{1}{2}}
\end{equation}
For a given $\beta$, the probability for the single spin to have the projection $m$ is:
\begin{equation}
p_1(m)=e^{-\beta m}/Z_1
\end{equation}
which in turn is used to calculate the thermal probability distribution following:
\begin{equation}
\begin{split}
&P_{\mathrm{th}}(N_{\frac{3}{2}},N_{\frac{1}{2}},N_{-\frac{1}{2}},N_{-\frac{3}{2}})=\\
&\underbrace{p_1(3/2)^{N_{\frac{3}{2}}}p_1(1/2)^{N_{\frac{1}{2}}}p_1(-1/2)^{N_{-\frac{1}{2}}}p_1(-3/2)^{N_{-\frac{3}{2}}}}_{\text{microstate probability}}\\&\times\underbrace{\frac{N!}{N_{\frac{3}{2}}!N_{\frac{1}{2}}!N_{-\frac{1}{2}}!N_{-\frac{3}{2}}!}}_{\text{microstate degeneracy}}
\end{split}
\end{equation}
For large $N$, we can use the Stirling approximation:
\begin{equation}
N! \approx \frac{\sqrt{2\pi N}}{e^N} N^N=\frac{\sqrt{2\pi N}}{e^N}N^{N_{\frac{3}{2}}}N^{N_{\frac{1}{2}}}N^{N_{-\frac{1}{2}}}N^{N_{-\frac{3}{2}}}
\end{equation}
Then:
\begin{equation}
P_{\mathrm{th}}(N_{\frac{3}{2}},N_{\frac{1}{2}},N_{-\frac{1}{2}},N_{-\frac{3}{2}})\propto\prod_{m\in\{\pm\frac{3}{2},\pm\frac{1}{2}\}}\frac{(Np_1(m))^{N_{m}}}{N_{m}!}
\end{equation}
so the statistics is Poissonian, with $\langle N_{m} \rangle =\langle \Delta^2 N_{m} \rangle = N p_1(m)$.

In the limit of $N_{m} \sim N$, the Stirling approximation applied to each factorial turns the Poissonian distribution into a multivariate Gaussian distribution:
\begin{equation}
\begin{split}
P_{\mathrm{th}}(N_{\frac{3}{2}},&N_{\frac{1}{2}},N_{-\frac{1}{2}},N_{-\frac{3}{2}})\\ &\propto\prod_{m\in\{\pm\frac{3}{2},\pm\frac{1}{2}\}}\exp{-\frac{(N_{m}-\langle  N_{m} \rangle)^2}{2\langle  N_{m} \rangle}}
\end{split}
\end{equation}
The corresponding thermal covariance matrix $S^{-1}_{\mathrm{th}}$ is given by:
\begin{equation}
S^{-1}_{\mathrm{th}}= \mathrm{diag}(\langle n_{\frac{3}{2}} \rangle,\langle n_{\frac{1}{2}} \rangle,\langle n_{-\frac{3}{2}} \rangle)
\end{equation}
It features polarisation fluctuations $\sqrt{\langle \Delta^2 I_z \rangle}$ an order of magnitude higher than those of the non-equilibrium steady state we prepare optically.  






\subsection{Modelling of Gallium Polarisation (2D F-P)}

While information about arsenic and indium was extracted directly from our data, we cannot draw any analogous statements on the gallium sub-ensemble due to its weaker coupling to the electron and thus its apparent absence from our magnon spectra (see Section \ref{sec:magnon_spec_t}).
However, we can intuit that the relative degree of polarisation of the two species is a function of the corresponding strength of the feedback $A^{j}/\omega^{j}$ each species experiences \cite{Yang2013}:
\begin{equation}\label{galliumpolarization_expectedratio}
\frac{\mathcal{I}_z^{\mathrm{As}}}{\mathcal{I}_z^{\mathrm{Ga}}}\approx\frac{A^{\mathrm{As}}/\omega^{\mathrm{As}}}{A^{\mathrm{Ga}}/\omega^{\mathrm{Ga}}} \sim 2
\end{equation}

To derive formally this relative degree of polarisation we use a two-species Fokker-Planck model.
Without loss of generality, we treat both gallium and arsenic as spin-$\frac{1}{2}$ species. Assuming vanishing coherences, this reduces the description of the flow to the two-dimensional space of fractional polarisations spanned by $\mathcal{I}_z^{\mathrm{As}}$ and $\mathcal{I}_z^{\mathrm{Ga}}$.

To estimate the co-dependence of $\mathcal{I}_z^{\mathrm{As}}$ and $\mathcal{I}_z^{\mathrm{Ga}}$, 
we look for stable points of the flow with a lock-point set as:
\begin{equation}\label{2d_lockpoint}
\delta= \delta_\text{O}^\mathrm{Ga} + \delta_\text{O}^\mathrm{As} 
\end{equation}

This 2D Fokker-Planck model is analogous to the 3D case, from section \ref{3dfp}. The only difference lies in the expressions for the phenomenological Raman scattering rates, and the rates of spontaneous sideband processes, which both need to now account for the presence of the second species. 
The Raman scattering rates for the first and second species - $W^{\mathrm{As}/\mathrm{Ga}}_{\pm1}(\delta_\text{e},I_z^{\mathrm{As}},I_z^{\mathrm{Ga}})$, respectively - are the counterparts of the expression from Eq.\,\ref{ramanscattering} with the only difference lying in:
\begin{equation}
\begin{split}
\Delta^{\mathrm{As}/\mathrm{Ga}}_{\pm 1}(\delta_\text{e})=&\delta_\text{e}-\frac{1}{2}a^{\mathrm{As}/\mathrm{Ga}}(2I^{\mathrm{As}/\mathrm{Ga}}_z\pm 1) \mp \omega^{\mathrm{As}/\mathrm{Ga}} \\&-a^{\mathrm{Ga}/\mathrm{As}}I^{\mathrm{Ga}/\mathrm{As}}_z
\end{split}
\end{equation}
A similar modification enters the spontaneous sideband processes (as included in Eq.\,\ref{added_diffusion}), where now their rate is given by:
\begin{equation}
\begin{split}
&\Gamma_{\mathrm{nc}}(\delta_\text{e},I_z^{\mathrm{As}},I_z^{\mathrm{Ga}})=\\ &=\frac{\Gamma}{2}\frac{\Omega^2/\Gamma\Gamma_2}{1+\Omega^2/\Gamma\Gamma_2+((\delta_\text{e}-a^{\mathrm{As}} I^{\mathrm{As}}_z-a^{\mathrm{Ga}} I^{\mathrm{Ga}}_z)/\Gamma_2)^2}
\end{split}
\end{equation}

Solving this 2D Fokker-Planck model uses the method already outlined in the section \ref{3dfp} for our 3D model.

Neglecting slow nuclear-spin diffusion, competition between species-characteristic sideband strengths has an effect on the covariance matrix of the steady-state probability distribution, but not its centre.

\begin{figure}
	\includegraphics[width=0.8\columnwidth]{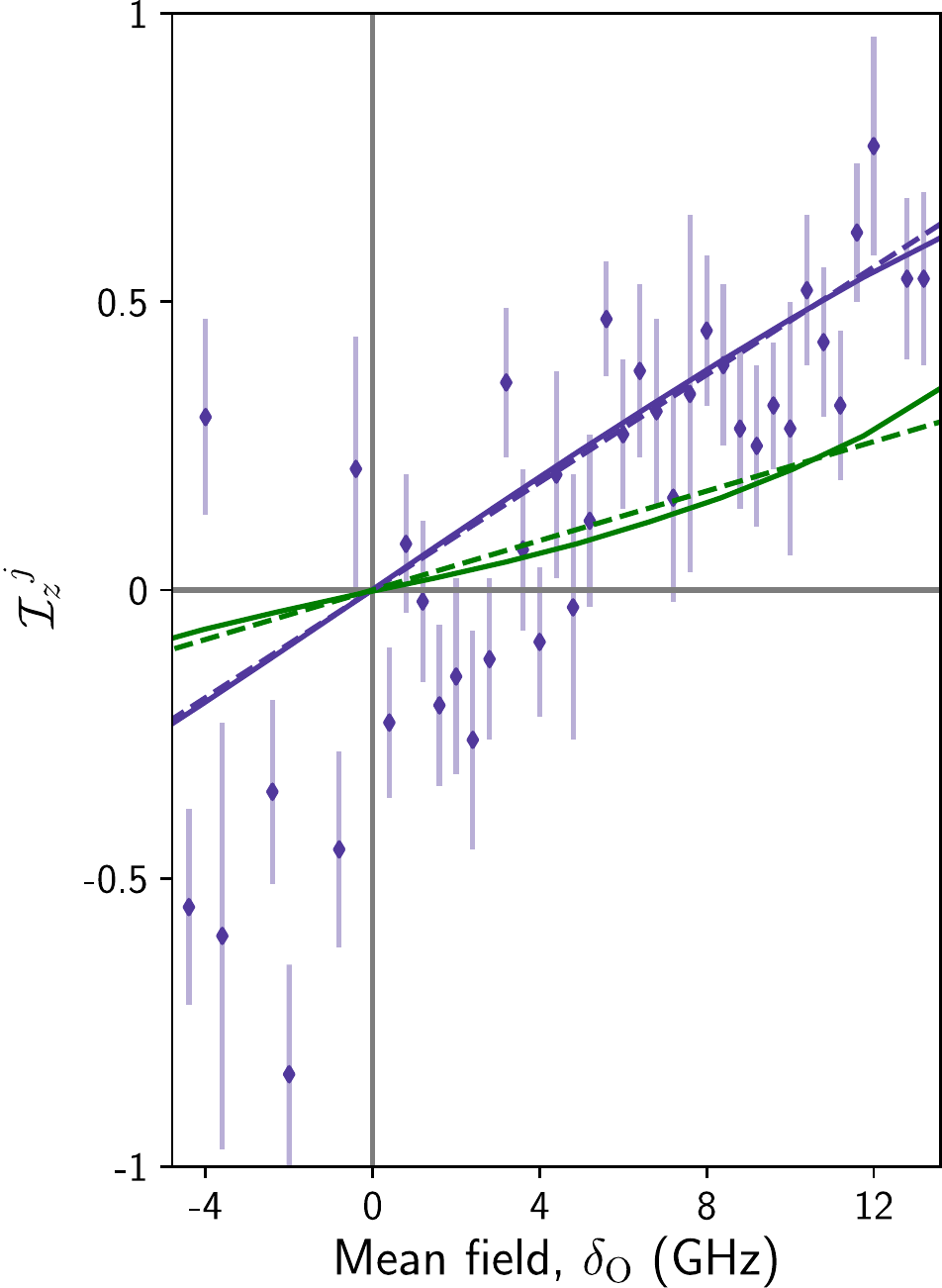}
	\caption{\textbf{Reconstruction of gallium polarisation:} Theoretical polarisation of arsenic (solid purple curve) and gallium (solid green curve), as a function of mean field, alongside linear fits to both (dashed curves). The mean field input to the model has been scaled such that we consider the range of arsenic polarisation which we access experimentally (purple diamonds).}
	\label{fig:2species}
\end{figure}

The 2-D Fokker-Planck model returns the stable points $(\mathcal{I}^{\mathrm{As}}_z,\mathcal{I}^{\mathrm{Ga}}_z)$ - a pair for each total Overhauser field $\delta$. We connect our simulation to experimental observation by considering our simulation output over the range of experimentally observed arsenic polarisation (Fig. 4a), i.e. we simply rescale our simulation setpoint $\delta$ such that experimental and simulated arsenic polarisations are in the same range. In Fig. \ref{fig:2species} we plot these stable points (solid curves), alongside the experimental data for $\mathcal{I}^{\mathrm{As}}_z$ (purple diamonds).

Over this range, the ratio of gallium and arsenic polarisation is approximately fixed. As a result, we perform linear fits of the simulations (dashed lines), and obtain $\mathcal{I}_z^{\mathrm{Ga}} = 0.46 \mathcal{I}_z^{\mathrm{As}}$ (in agreement with our expectation from Eq. \ref{galliumpolarization_expectedratio}). This is the ratio mentioned in the main text.

\section{Supplementary notes on steady-state coherences}\label{steady_state_coherences}

Here we outline the physical mechanisms that can lead to an enhanced asymmetry of the magnon-injection processes, and show that our measurements (Fig. 4) can only be explained by a degree of entanglement among nuclei. To convey the general concept with simplicity, we will consider an ensemble of spin-$\frac{1}{2}$ nuclei ($\mathrm{SU(2)}$ symmetry group), in which the magnon creation/annihilation operators are simply the total angular momentum laddering operators:

\begin{equation}
\begin{split}
    \hat{I}_\pm &= \hat{I}_x \pm i\hat{I}_y\\
    &= \sum_j \hat{I}^j_\pm
\end{split}
\end{equation}

The concepts illustrated by the $\mathrm{SU(2)}$ model generalise straightforwardly to a high-spin ensemble interfaced with a proxy qubit. Assuming a uniform electron-nuclear coupling, the lowest type of symmetry exhibited by the system is permutation invariance. In the case of spin-$\frac{1}{2}$ systems, this leads to the conservation of total spin $I$ (a quantum number associated to the operator $\vec{I}=\sum_j \vec{I}^j$), thereby partitioning the Hilbert space into sub-radiant and super-radiant sub-spaces of well defined $I$, or - in other words - well-defined exchange symmetries \cite{Dicke1954}. The proper mathematical treatment of the ensemble of permutation invariant $m$-level systems requires working with SU($m$) symmetry groups. Nonetheless, for permutation-invariant systems violating the conservation of momentum (like the ensemble considered in the manuscript), the conservation of exchange symmetry still holds. As a result, the physical mechanism of many-body interference, destructive or constructive interference, leads to analogous quantum phases of the ensemble: sub-radiant and super-radiant.

\subsubsection{General entanglement witness}

Taking an ensemble of $N$ spin-$1/2$ particles, the following inequality holds for any separable state:

\begin{equation}\label{eq:witness1}
\langle \Delta \hat{I}_x^2 \rangle + \langle \Delta \hat{I}_y^2 \rangle + \langle \Delta \hat{I}_z^2 \rangle \geq \frac{N}{2} \text{,}
\end{equation}
\noindent
where $\langle \Delta \hat{I}_\mu^2 \rangle$ is the variance of collective operator $\hat{I}_\mu$; violation of this bound implies entanglement \cite{Vitagliano2011}.

\subsubsection{Asymmetry parameter as an entanglement witness}\label{subsubsec:asymmetry}

In this general quantum picture, the magnon exchange frequencies are correlators of collective spin operators $\hat{I}_\pm$:
\begin{equation}
\Omega_\pm^2 \propto \langle \hat{I}_\mp \hat{I}_\pm \rangle
\end{equation}

This expectation value is correct for any many-body state, quantum or classical. The asymmetry parameter, as we measure it, is simply a ratio of these correlators, where we use $[\bullet,\bullet]$ and $\{\bullet,\bullet\}$ to denote the commutator and anti-commutator respectively:

\begin{equation}
    \begin{split}
        \nu &= \frac{\Omega^2_--\Omega_+^2}{\Omega^2_-+\Omega_+^2}\\
    &=\frac{\langle [\hat{I}_+,\hat{I}_-]\rangle}{\langle \{\hat{I}_+,\hat{I}_-\}\rangle}
    \end{split}
\end{equation}

This is then expressed as expectation values of the collective spin operators $\hat{I}_x,\hat{I}_y,\hat{I}_z$:

\begin{equation}\label{correlator_relations}
\begin{split}
\langle [\hat{I}_+,\hat{I}_-]\rangle&=2\langle \hat{I}_z \rangle\\
\langle \{\hat{I}_+,\hat{I}_-\}\rangle&=2\langle \hat{I}_x^2 + \hat{I}_y^2 \rangle\\
\end{split}
\end{equation} 

The asymmetry parameter is then simply a ratio of longitudinal $z$ first moment of polarisation to transverse $x,y$ second moments of polarisation:

\begin{equation}\label{eq:asymmetry}
    \begin{split}
        \nu &= \frac{\langle \hat{I}_z \rangle}{\langle \hat{I}_x^2 \rangle + \langle \hat{I}_y^2 \rangle}
    \end{split}
\end{equation}

We then use the relation between variance and second moments $\langle \hat{I}_x^2 \rangle = \langle \Delta \hat{I}_x^2 \rangle + \langle \hat{I}_x \rangle^2$ to solve for the noise terms as a function of asymmetry:

\begin{equation}
    \begin{split}
        \langle \Delta \hat{I}_x^2 \rangle + \langle \Delta \hat{I}_y^2 \rangle &= \frac{\langle \hat{I}_z \rangle}{\nu} - \langle \hat{I}_x \rangle^2 - \langle \hat{I}_y \rangle^2
    \end{split}
\end{equation}

\noindent
Now using the general entanglement witness, Eq.\,\ref{eq:witness1}, we can express the asymmetry parameter as a formal entanglement witness:

\begin{equation}
        \nu \leq \frac{\langle \hat{I}_z \rangle}{\frac{N}{2} - \langle \Delta \hat{I}_z^2 \rangle + \langle \hat{I}_x \rangle^2 + \langle \hat{I}_y \rangle^2}
\end{equation}

This expression can be recast in terms of mean fractional polarisations $\mathcal{I}_\mu = \langle \hat{I}_\mu\rangle /I_z^{\text{max}}$, where $I_z^{\text{max}} = N/2$ for spin-$1/2$, and the asymmetry-commensurate longitudinal fractional polarisation $\mathcal{I}_z^\star = \nu$:

\begin{equation}\label{eq:asympolarbound}
    \begin{split}
        \mathcal{I}_z^\star &\leq \frac{\mathcal{I}_z}{1 -  \frac{N}{2}\Delta^2 \mathcal{I}_z + \frac{N}{2} \left(\mathcal{I}_x^2 + \mathcal{I}_y^2\right)}
    \end{split}
\end{equation}
\noindent
Violation of this general inequality then necessarily implies the presence of entanglement in the spin ensemble. 

\subsubsection{Universal features}

While Eq.\,\ref{eq:asympolarbound} was derived for a spin-$1/2$ ensemble, certain universal features can be extracted that only depend on fractional polarisations and are independent of the spin character of the ensemble: 

\begin{itemize}
    \item In any system, the entanglement condition is inversely proportional to transverse coherences $\mathcal{I}_x^2 + \mathcal{I}_y^2$; i.e. entanglement is present at lower values of  $\mathcal{I}_z^\star$ for a state with transverse coherences.
    \item In a system where the longitudinal fluctuations are highly sub-thermal, $\Delta^2 \mathcal{I}_z \sim 0$, Eq.\,\ref{eq:asympolarbound} becomes: 
    
\begin{equation}\label{eq:asympolarbound2}
    \begin{split}
        \mathcal{I}_z^\star &\leq \frac{\mathcal{I}_z}{1 + \frac{N}{2} \left(\mathcal{I}_x^2 + \mathcal{I}_y^2\right)}\text{,}
    \end{split}
\end{equation}

    which necessarily implies that the following observation is sufficient to prove the presence of entanglement:

\begin{equation}\label{eq:asympolarbound3}
    \begin{split}
        \mathcal{I}_z^\star > \mathcal{I}_z \text{.}
    \end{split}
\end{equation}

\end{itemize}

\subsubsection{Experimental observations}

Our experimental observations are that the nuclear ensemble following optical preparation has longitudinal fluctuations (measured in section \ref{t2starmeasurement} as the electronic $T_2^*$) reduced by a factor of $\sim400$ relative to their thermal value \cite{Ethier-Majcher2017,Gangloff2019}, i.e. $\Delta^2 \mathcal{I}_z \ll 1/N$ . We have no direct measurement of transverse coherences $\mathcal{I}_x^2 + \mathcal{I}_y^2$, however we have shown that such coherences would reduce the asymmetry for which the state remains classical. Our measurements shown in Fig.\,4 reveal that for our nuclear state, summed over species, $\delta_\text{O}^\star = 2.9(1) \delta_\text{O}$. In other terms, the asymmetry-commensurate polarisation $\mathcal{I}_z^\star$, averaged over species, exceeds the underlying fractional polarisation $\mathcal{I}_z$ by a factor of $2.9(1)$, well beyond the bound set out by Eq.\,\ref{eq:asympolarbound3} .

\begin{figure*}
	\includegraphics[width=2\columnwidth]{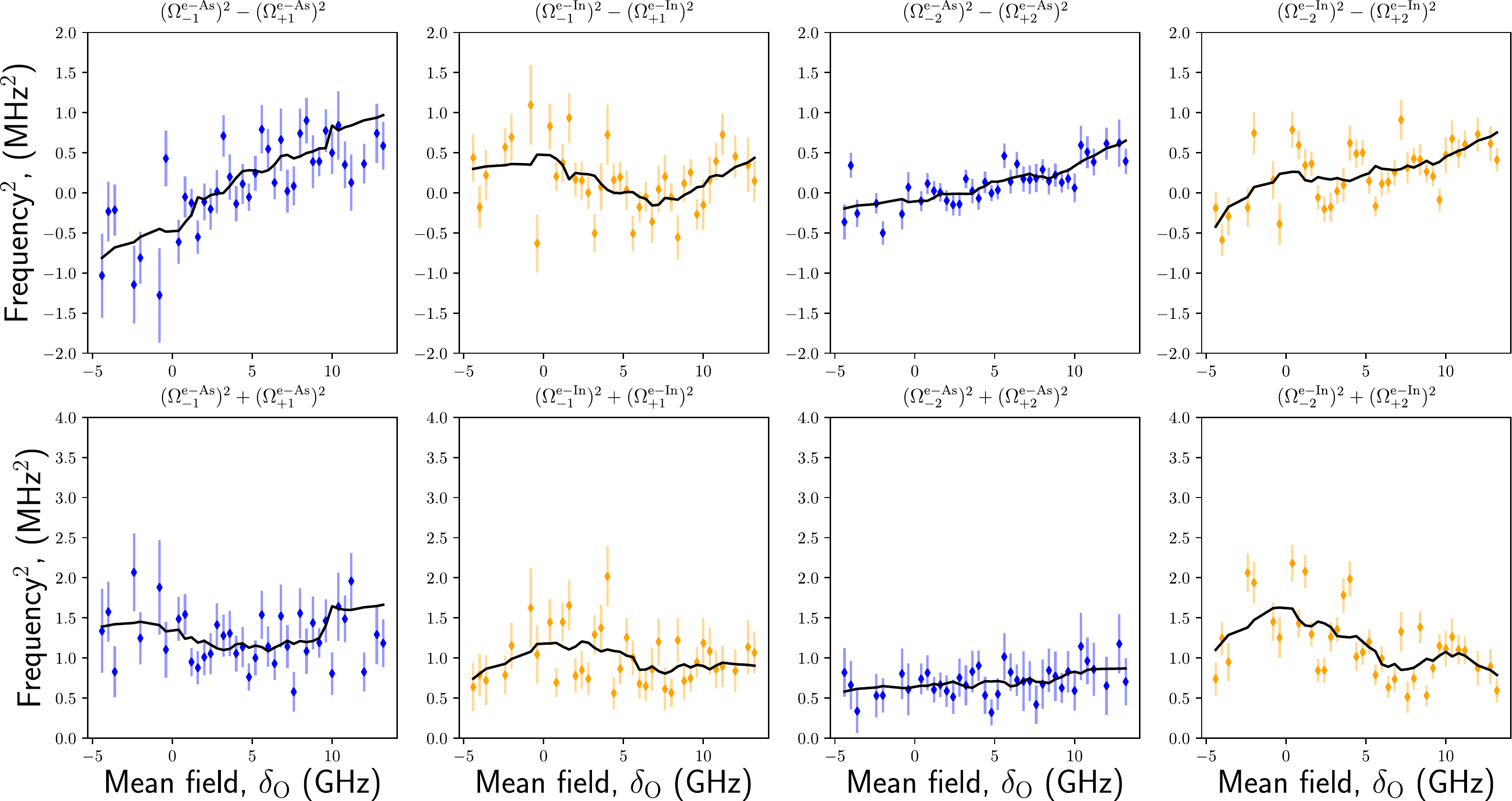}
	\caption{\textbf{Numerators and denominators of the magnon asymmetries $\nu_k$:} The top row shows $(\Omega^{\mathrm{e-j}}_{-k})^2-(\Omega^{\mathrm{e-j}}_{+k})^2$ for $k=1,2$ and $j=\mathrm{As},\mathrm{In}$. The bottom row shows the values of $(\Omega^{\mathrm{e-j}}_{-k})^2+(\Omega^{\mathrm{e-j}}_{+k})^2$. These quantities are straightforwardly calculated from the data analysis presented in Fig. \ref{fig:excfreqs}. The solid black curves are generated by passing the data through a first-order Savitsky-Golay filter with a 3.2-GHz window. The relative magnitudes of the sums in the bottom panels, as well as the gradients of the trends in the top panels, reflect a principal quadrupolar axis close to (further from) the growth axis for indium (arsenic) - see section \ref{sec:model}. }
	\label{fig:constsmotion}
\end{figure*}

The quantum picture introduced in section \ref{subsubsec:asymmetry} prompts us to evaluate separately the numerator and the denominator of our sideband asymmetry parameter (c.f. Eq. \ref{eq:asymmetry}). The numerator is directly proportional to a longitudinal population difference up to a factor that is independent of the nuclear wavefunction symmetry. The denominator, on the other hand, directly reflects the transverse noise. We therefore show these quantities as experimentally measured (Fig. \ref{fig:constsmotion}). These data show that the longitudinal population difference changes linearly with polarisation ($\propto \Omega^2_--\Omega^2_+$), whereas the transverse noise ($\propto \Omega^2_++\Omega^2_-$) remains constant within experimental noise. We thus conclude that the optical preparation of the nuclear ensemble not only reduce longitudinal fluctuations by a factor of $\sim400$, but it also suppresses transverse fluctuations by a factor of $\approx 3$, and that this fluctuation reduction is independent of polarisation.

\subsubsection{Sub-radiance}

We can re-express the asymmetry parameter, Eq.\,\ref{eq:asymmetry}, as function of total angular momentum $\langle \hat{I}^2\rangle = \langle \hat{I}_x^2 + \hat{I}_y^2 + \hat{I}_z^2 \rangle$, and for a reduced fluctuation state for which $\langle \hat{I}_z^2 \rangle \approx \langle \hat{I}_z \rangle^2$:

\begin{equation}
\nu = \frac{\langle \hat{I}_z \rangle}{\langle \hat{I}^2 \rangle-\langle \hat{I}_z \rangle^2}
\end{equation}

The enhancement of asymmetry to $\nu \sim 1$,  i.e. beyond that featured by a low-polarisation product state (which is $\nu=\langle \hat{I}_z \rangle / I^{\mathrm{max}} \ll 1)$, is a signature of: 
\begin{equation}
\langle \hat{I}^2 \rangle\sim \langle \hat{I}_z \rangle(\langle \hat{I}_z \rangle+1)
\end{equation}
characteristic for a dark state, as it implies $I\sim I_z$ even for small $I_z$. It is instructive to consider the opposite limit, where:
\begin{equation}
\langle \hat{I}^2 \rangle\sim I^\mathrm{max}(I^\mathrm{max}+1)
\end{equation}
In such case, for states of $\langle \hat{I}_z \rangle \ll I_z^{\mathrm{max}}$ the asymmetry is decreased close to zero. 

These insights allow the introduction of a phase diagram (Fig.\,4b inset) showing $\nu = \mathcal{I}_z^\star$ vs. $\mathcal{I}_z$ where, in the case of a longitudinally narrow state $\Delta^2 \mathcal{I}_z \approx 0$, deviation from the one-to-one classical line (no coherences),  $\mathcal{I}_z^\star = \mathcal{I}_z$, can indicate the dominant type of interference (constructive or destructive).

%
%

\newpage 

\section{Summary of model parameters}

\begin{table}[h]
\begin{tabular}{|c||c|}
	\hline
	Total hyperfine interaction, arsenic, $A^{\mathrm{As}}$ & $11.1~\mathrm{GHz}$\\ 
	\hline
	Total hyperfine interaction, indium, $A^{\mathrm{In}}$ & $13.5~\mathrm{GHz}$\\ 
	\hline
	Total hyperfine interaction, gallium, $A^{\mathrm{Ga}}$ & $9.2~\mathrm{GHz}$\\ 
	\hline
	Indium concentration $\mathrm{In}_x\mathrm{Ga}_{1-x}\mathrm{As}$, x  & $0.5$\\ 
	\hline

	External magnetic field, $B$   & $3.5~\mathrm{T}$\\ 
	\hline
	Arsenic Zeeman splitting, $\omega^{\mathrm{As}}/B$ & $7.22~\mathrm{MHz}/\mathrm{T}$\\ 
	\hline
	Indium Zeeman splitting, $\omega^{\mathrm{In}}/B$ & $9.33~\mathrm{MHz}/\mathrm{T}$\\ 
	\hline
	Gallium-69 Zeeman splitting, $\omega^{\mathrm{Ga-69}}/B$ & $10.22~\mathrm{MHz}/\mathrm{T}$\\ 
	\hline
	Gallium-71 Zeeman splitting, $\omega^{\mathrm{Ga-71}}/B$ & $12.98~\mathrm{MHz}/\mathrm{T}$\\ 
	\hline
	
	Inhomogeneous dephasing time, $T_2^*$   & $39 ~\mathrm{ns}$\\ 
	\hline
\end{tabular}
\caption{Parameters that were constrained through measurement or taken from the literature.}
\end{table}

\begin{table}[h]
\begin{tabular}{|c||c|}
	\hline
	Electron homogeneous dephasing time, $T_2$ & $4.55~\mu\mathrm{s}$\\ 
	\hline
	\shortstack{Electron Rabi frequency \\ in asymmetry measurement, $\Omega$} & $6.7~\mathrm{MHz}$\\ 
	\hline
	\shortstack{Pure dephasing timescale \\for arsenic sideband transitions, $T_2^{\mathrm{e-As}}$} & $275.4~\mathrm{ns}$\\ 
	\hline
	\shortstack{Pure dephasing timescale \\ for indium sideband transitions, $T_2^{\mathrm{e-In}}$} & $213.8~\mathrm{ns}$\\ 
	\hline

\end{tabular}
\caption{Parameters fitted to the time dependence of the magnon spectrum, used to extract individual exchange frequencies in the asymmetry measurement.}
\end{table}

\begin{table}[h!]
\begin{tabular}{|c||c|}
	\hline
	Arsenic quadrupolar angle, $\theta$ & $18.45^\circ$\\ 
	\hline
	
	
	Rabi frequency, $\Omega$ & $21~\mathrm{MHz}$\\ 
	\hline
	Trion linewidth, $\Gamma_0$ & $150~\mathrm{MHz}$\\ 
	\hline
	Effective optical pumping rate, $\Gamma$ & $20~\mathrm{MHz}$\\ 
	\hline
	Number of arsenic Nuclei, $N^{\mathrm{As}}$ & $40000$\\ 
	\hline
	Electron homogeneous dephasing time, $T_2$ & $1.25 ~\mu \mathrm{s}$\\ 
	\hline
	Arsenic quadrupolar constant, $B_Q$ & $800 ~\mathrm{kHz}$\\ 
	\hline
	Arsenic spin diffusion rate, $N\Gamma_{\mathrm{nuc}}$ & $7.5~\mathrm{kHz}$\\ 
	\hline
	
\end{tabular}
\caption{Parameters fixed in Fokker-Planck modelling of the cooling steady-state.}
\end{table}

\newpage

\bibliography{bib}
\bibliographystyle{naturemagdan.bst}